\documentclass[pdflatex,sn-mathphys-num]{sn-jnl}% Math and Physical Sciences Numbered Reference Style
\usepackage{graphicx}%
\usepackage{multirow}%
\usepackage{amsmath,amssymb,amsfonts}%
\usepackage{amsthm}%
\usepackage{mathrsfs}%
\usepackage{braket}
\usepackage[title]{appendix}%
\usepackage{xcolor}%
\usepackage{textcomp}%
\usepackage{manyfoot}%
\usepackage{booktabs}%
\usepackage{algorithm}%
\usepackage{algorithmicx}%
\usepackage{algpseudocode}%
\usepackage{listings}%
\usepackage{soul}%
\usepackage{placeins}
\usepackage{longtable}
\usepackage{array}
\usepackage{enumitem}
\usepackage{geometry}

\sethlcolor{yellow}

\theoremstyle{thmstyleone}%
\theoremstyle{thmstyletwo}%

\theoremstyle{thmstylethree}%

\begin{document}


%TC:ignore
\author*[1,2]{\fnm{Donald J.} \sur{Jacobs}}\email{djacobs1@charlotte.edu}

\affil*[1]{\orgdiv{Ishwar Aggarwal Department of Physics and Optical Science}, \orgname{University of North Carolina at Charlotte}, 
\orgaddress{\street{9201 University City Blvd.}, \city{Charlotte}, \postcode{28213}, \state{NC}, \country{USA}}}

\affil[2]{\orgdiv{Affiliate of the School of Data Science}, \orgname{University of North Carolina at Charlotte}, 
\orgaddress{\street{9201 University City Blvd.}, \city{Charlotte}, \postcode{28213}, \state{NC}, \country{USA}}}

%TC:endignore

\title{Quantum Interference as a Proposal Mechanism for Combinatorial Optimization}
%\title{Quantum Interference Proposal Search for Non-Variational Binary Optimization}
%\title{Finite-Shot Quantum Interference Proposal Search for Binary Optimization}

%TC:ignore
\abstract{Quantum Interference Proposal Search (QIPS) uses seed-conditioned quantum circuits to generate localized interference patterns as finite-shot proposal distributions for QUBO/Ising optimization. Candidate $n_b$-bit strings are sampled from these distributions, scored classically and used to update an elite frontier of low-energy solutions. QIPS uses a fixed two-layer gate-based circuit architecture with 100 shots per circuit while the Hilbert-space dimension grows as $2^{n_b}$. Across six benchmark families with $18 \le n_b \le 29$, QIPS maintains competitive progress relative to a matched classical control that preserves the same search loop, frontier update rule and proposal budget, with total proposals proportional to $n_b$. Performance is assessed using top-$K$ coverage, hit rate, multiplicity, Hilbert-space coverage and dyadic-rank metrics. The results identify localized quantum interference as a resource-efficient proposal mechanism for computational quantum optimization.}
%TC:endignore

%\keywords{quantum proposal search, localized interference patterns, finite-shot quantum optimization, resource-efficient quantum algorithm, Ising spin models}
\keywords{quantum optimization, QUBO/Ising models, finite-shot sampling, localized quantum interference, quantum-classical benchmarking}


\maketitle

Combinatorial optimization is a central computational problem across science,
engineering and industry. Many such problems can be expressed as quadratic
unconstrained binary optimization (QUBO), which maps naturally to an Ising
energy function and to computational-basis measurements on quantum
devices~\cite{REF09,REF08,REF39}. Despite extensive progress, scalable quantum
advantage for QUBO-like optimization under matched classical resources remains
elusive~\cite{REF13}. Recent work has therefore emphasized the need for
reproducible, resource-aware benchmarking and careful comparison with classical
methods when assessing quantum optimization algorithms~\cite{REF13,refNCS1}.
These considerations are especially important because both annealing and
gate-based approaches are sensitive to modeling choices, embedding overhead,
hardware connectivity, sampling budgets and classical post-processing
resources~\cite{REF08,REF39,REF13,REF04,REF34,REF35,REF36}.

The quantum approximate optimization algorithm (QAOA) is the most widely
studied gate-based framework for QUBO and Ising optimization. QAOA is usually
implemented as a variational quantum algorithm in which circuit parameters are
iteratively updated to improve an objective estimated from measurement
samples~\cite{REF03,REF43}. This approach is flexible, but creates a
variational bottleneck~\cite{REF43,REF14,REF05}. Substantial effort has
therefore focused on warm starts, adaptive ansatz construction, low-shot
optimizers, compressed or constraint-preserving search spaces,
tail-sensitive objectives, low-dimensional parameter maps and
resource-efficient adaptive circuits~\cite{REF05,REF01,REF10,REF23,REF33,REF38}.

Other approaches reduce this bottleneck by reweighting measured bitstrings,
tracking low-energy samples, using comparison-based amplification or adaptively
remapping noisy quantum runs~\cite{REF17,REF28,REF06,REF25,REF26,REF37,REF42}.
Recent multi-objective QAOA work further illustrates that quantum sampling can
be computationally useful when diverse high-quality bitstrings, rather than a
single optimized state, are the relevant output~\cite{refNCS2}. More broadly,
recent non-variational quantum-walk optimization approaches have demonstrated
promising bitstring exploration over combinatorial solution
spaces~\cite{NEW01,REF24,REF27}. Together, these developments shift attention
from variational state preparation alone toward the computational value of
structured bitstring proposal distributions. Nevertheless, persistent
difficulties remain for gate-based optimization when useful progress must be
extracted from shallow circuits and resource-limited measurement
samples~\cite{REF14,REF32,NEW02,NEW03}.

Focusing on finite-shot bitstring proposals motivates Quantum Interference
Proposal Search (QIPS), a non-variational, seed-conditioned proposal mechanism
for binary optimization. Candidate bitstring evaluations become the primitive
objects of the search, and the quantum circuit functions as a structured
finite-shot proposal generator. A seed-conditioned circuit is constructed with 
controlled randomized
parameter variations, and feedback is used to maintain localized quantum
interference patterns that generate measurable candidate bitstrings. As
lower-energy states are discovered, an elite frontier is updated and used to
provide new seeds for subsequent circuits. QIPS therefore differs from
non-variational quantum-walk optimization, where a quantum walk over the
solution graph is designed to concentrate probability globally on high-quality
states~\cite{NEW01,REF24,REF27}. It also differs from quantum-enhanced greedy
solvers, where quantum samples guide variable freezing, and from filtering 
variants of the variational quantum eigensolver, where a variational state is 
trained against a modified objective~\cite{REF06,REF17,REF28,REF42}.

Here, QIPS is evaluated by ideal gate-based quantum simulation using a fixed
two-layer circuit architecture and a matched classical proposal search under
the same outer-loop protocol, frontier update rule and proposal budget. Across
six QUBO/Ising benchmark families, the results show that QIPS accesses
low-energy states through emergent structure in seed-conditioned quantum
proposals. Dyadic-rank measures reveal where this structure is most
competitive as a function of seed rank and system size. Both QIPS and the
matched classical control drive an elite frontier toward low-energy states,
but QIPS resamples near-optimal solutions much more frequently. These results
identify localized quantum interference as a finite-shot proposal resource for
computational quantum optimization.

\section*{Results}

%%%%%%%%%%%%%%%%%%%%%%%%%%%%%%%%%%%%%%%%%%%%%%%%%%%%%%%%%%%%%%%%%%%%%%%%%%%%%%%
%%%%%%%%%%%%%%%%%%%%%%%%%%%%%%%%%%%%%%%%%%%%%%%%%%%%%%%%%%%%%%%%%%%%%%%%%%%%%%%
\subsection*{Seed-conditioned proposal search}

The main analysis evaluates QIPS on six QUBO/Ising benchmark families with
$18\le n_b\le29$ binary variables and 32 matched instances per family and size
(Table~\ref{table:6systems}). The full benchmark scope is described in
Supplementary Note~1. The benchmark set spans sparse constraint problems,
weighted graph optimization, dense mean-field spin glasses and structured
quadratic energy landscapes. This diversity tests whether QIPS is exploiting
general structure in QUBO/Ising objectives rather than special properties of a
single problem family. Randomly shuffling the state labels destroys this
structure and reduces QIPS to blind search. Because all systems studied exhibit
similar qualitative behavior, results are often reported in aggregate form.

%TC:ignore
\begin{table}[hbt]
\caption{QUBO/Ising benchmark families used for systematic size-dependent analysis. RRG
denotes a random regular graph; AFM and FM respectively denote antiferromagnetic and 
ferromagnetic couplings.}
\label{table:6systems}%
\begin{tabular}{@{}ll@{}}
\toprule
Graph topology & Problem or coupling model \\
\midrule
3-RRG    & Ising, 70\% AFM / 30\% FM $J_{ij}$ \\
3-RRG    & MaxCut, uniform weights \\
3-RRG    & MIS, uniform weights and penalties \\
6-RRG    & MaxCut, log-normal weights \\
Complete & exponential weak quenched disorder\\
Complete & Sherrington-Kirkpatrick model \\
\botrule
\end{tabular}
\end{table}

\begin{figure}[hbt]
    \centering
    \includegraphics[
        width=0.98\linewidth,
        trim={0.01in 0.18in 0.00in 0.11in},      % trim={left bottom right top}
        clip
    ]{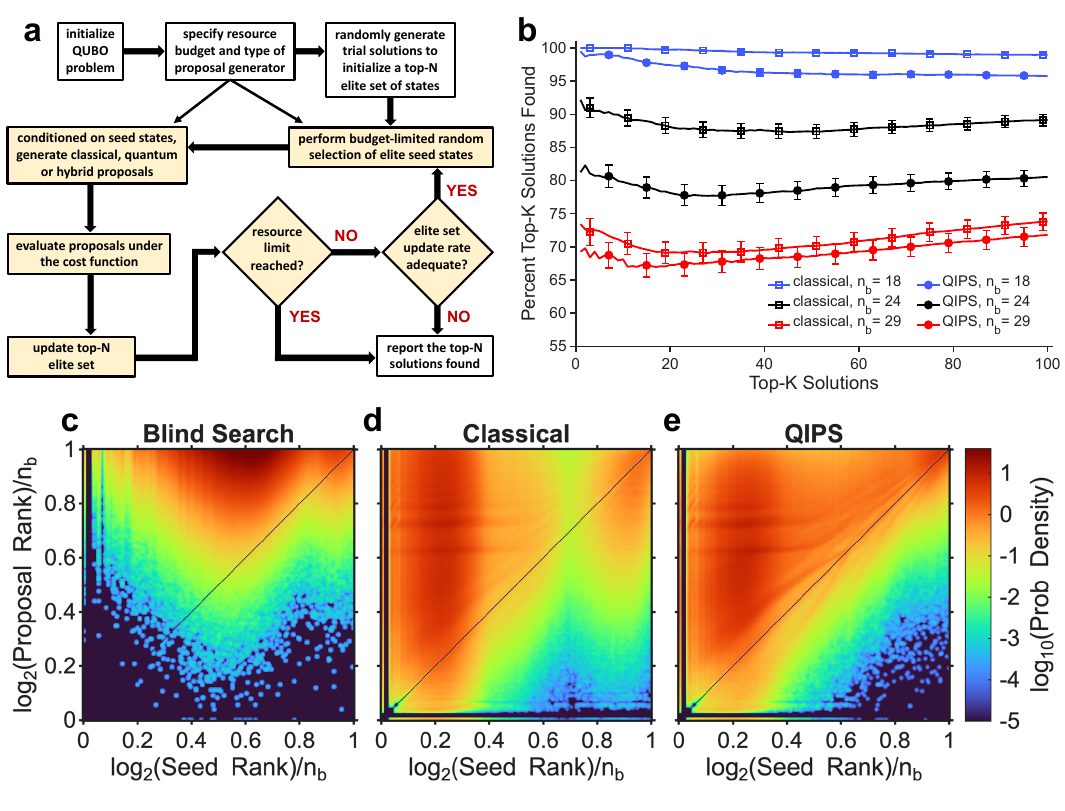}
    \caption{\textbf{QIPS workflow and proposal statistics.}
    \textbf{a}, Schematic of Quantum Interference Proposal Search (QIPS). An initial 
    set of trial solutions is used to construct a top-$N$ elite set. 
    Seed-conditioned classical, quantum, or hybrid proposal generators then
    produce candidate bitstrings, which are evaluated under the QUBO objective
    and used to update the elite set until either the elite-set update rate becomes
    sufficiently small or the allotted proposal budget is exhausted. The fixed-budget
    criterion is used for the results reported here.
    \textbf{b}, Percentage of the top-$K$ lowest-energy solutions recovered by QIPS and 
    by the matched classical proposal search for $n_b=18$, 24, and 29. Classical and 
    quantum searches use the same QUBO instances and equal total proposal budgets. 
    Results are aggregated over the six benchmark families in 
    Table~\ref{table:6systems}, with 32 instances for each system and system size.
    Error bars represent the standard error of the mean over the aggregated
    benchmark ensemble.
    \textbf{c--e}, Empirical conditional probability densities for generating a 
    proposal of normalized logarithmic coordinate $\log_2(r_{\mathrm{p}})/n_b$ from a 
    seed of normalized logarithmic coordinate $\log_2(r_{\mathrm{s}})/n_b$, shown for 
    blind search, the classical proposal generator, and QIPS, respectively. Rank $r=1$ 
    denotes the lowest-energy state. The diagonal marks proposals with the same rank 
    as the seed; points below the diagonal represent rank improvement. Color indicates 
    $\log_{10}$ probability density. The distributions are aggregated over all six 
    benchmark families and all system sizes $18\le n_b\le29$.}
    \label{fig:mainResults}
\end{figure}
%TC:endignore

The iterative QIPS workflow is summarized in
Figure~\ref{fig:mainResults}a. Proposals are generated conditionally from seed
states drawn from an elite outer frontier, and candidate bitstrings are scored
classically under the same QUBO objective used by the matched classical control.
This design isolates the proposal generator as the controlled difference between
the two searches: the outer-loop protocol, seed-selection rule, frontier update
rule and total proposal budget are held fixed. The outer frontier balances search
diversity against computational cost, while a target frontier tracks the best
solutions found. Here, the outer and target frontier sizes are fixed at 100 and
10, respectively.

For benchmarking, the total proposal budget is fixed at $2000n_b$ evaluations
for both classical and quantum searches. In QIPS, this budget is distributed over
$20n_b$ circuits with 100 shots per circuit. Twenty consecutive circuits define
one search round. Although the implementation supports optional early
termination based on frontier stagnation, all matched benchmarks reported here
use the full fixed proposal budget. The 100-shot circuit budget provides a
finite-sampling stress test of whether localized quantum proposals remain useful
as the Hilbert space grows exponentially.

To quantify recovery of the low-energy spectrum, Figure~\ref{fig:mainResults}b
shows the fraction of the ground-truth top-$K$ states found by QIPS and by the
matched classical search. Classical proposals achieve higher top-$K$ recovery,
but QIPS remains competitive under the same budget of $2000n_b$ evaluations.
For both methods, recovery decreases with system size because the proposal
budget grows only linearly with $n_b$, while the number of candidate bitstrings
grows exponentially.

Figures~\ref{fig:mainResults}c--e compare the conditional seed-to-proposal rank
distributions for blind search, classical proposals, and QIPS. Blind proposals
are uniformly distributed over ranks $1$ to $2^{n_b}$, whereas both classical and
quantum proposals retain strong correlations with the seed rank. The QIPS
distribution shows reduced probability in the lower-right region, indicating
that low-rank proposals are less accessible from poorly ranked seeds. This
feature emphasizes the computational role of the elite frontier: progressively
better seeds condition subsequent circuits toward more productive proposal
distributions. The corresponding system-resolved results are provided in
Supplementary Note~2.

%%%%%%%%%%%%%%%%%%%%%%%%%%%%%%%%%%%%%%%%%%%%%%%%%%%%%%%%%%%%%%%%%%%%%%%%%%%%%%%
%%%%%%%%%%%%%%%%%%%%%%%%%%%%%%%%%%%%%%%%%%%%%%%%%%%%%%%%%%%%%%%%%%%%%%%%%%%%%%%
\subsection*{Non-variational feedback-controlled quantum circuits}
Quantum proposals are generated by a fixed two-layer circuit architecture
conditioned on the selected seed state. Feedback regulates randomized circuit
deviations to maintain localized finite-shot interference patterns. For
$n_b$ qubits, the resulting trial state is

\begin{equation}
\ket{\psi_{\mathrm{trial}}}
=
\hat U_2
\bigl(
\boldsymbol{\alpha}_2,
\boldsymbol{\theta}_2,
\boldsymbol{\phi}_2,
\boldsymbol{\delta}_2,
\gamma_2
\bigr)
\hat U_1
\bigl(
\boldsymbol{\alpha}_1,
\boldsymbol{\theta}_1,
\boldsymbol{\phi}_1,
\boldsymbol{\delta}_1,
\gamma_1
\bigr)
\ket{+}^{\otimes n_b},
\label{eq:qips_trial_state}
\end{equation}

\noindent where $\ket{+}^{\otimes n_b}$ is obtained by applying a Hadamard gate
to each qubit. For layer $\ell\in\{1,2\}$,
$\boldsymbol{\alpha}_{\ell}$ and $\boldsymbol{\delta}_{\ell}$ are rotation
parameters, while $\boldsymbol{\theta}_{\ell}$ and
$\boldsymbol{\phi}_{\ell}$ specify polar and azimuthal rotation-axis
orientations. Each bold parameter contains one angle per qubit, whereas
$\gamma_{\ell}$ is a layer-dependent scalar angle for the phase separator.
Suppressing the layer index, define

\begin{equation}
\hat D(\boldsymbol{\delta})
=
\sum_{j=1}^{n_b}
\delta^{j}\hat{\sigma}_{j}^{z},
\label{eq:qips_A_operator}
\end{equation}
and the local rotation axis

\begin{equation}
\hat B_j(\theta^{j},\phi^{j})
=
\sin(\theta^{j})\cos(\phi^{j})\hat{\sigma}_{j}^{x}
+
\sin(\theta^{j})\sin(\phi^{j})\hat{\sigma}_{j}^{y}
+
\cos(\theta^{j})\hat{\sigma}_{j}^{z}.
\label{eq:qips_B_operator}
\end{equation}
A single circuit layer is

\begin{equation}
\hat U
\bigl(
\boldsymbol{\alpha},
\boldsymbol{\theta},
\boldsymbol{\phi},
\boldsymbol{\delta},
\gamma
\bigr)
=
\left[
\prod_{j=1}^{n_b}
\exp\left(
-i\alpha^{j}
\hat B_j(\theta^{j},\phi^{j})
\right)
\right]
\exp\left[
-i\gamma
\left(
\hat K+\hat D
\right)
\right].
\label{eq:qips_layer}
\end{equation}

\noindent where \(\hat K\) is the diagonal QUBO/Ising cost operator
that introduces the problem-dependent multi-qubit phases. In this work, a
circuit layer denotes one algorithmic block containing a diagonal cost-phase
operation followed by parallel single-qubit rotations. The single-qubit
rotations act on distinct qubits and can therefore be executed in parallel.
Increasing $n_b$ increases the number of local rotation angles, but not the
algorithmic depth of the local-rotation block. The physical gate depth required
to implement the diagonal cost operator depends on the problem graph, native
gate set and hardware connectivity.

The complete circuit contains $8n_b+2$ angles. Each qubit-dependent angle is
written as a canonical value plus a feedback-controlled deviation,

\begin{equation}
\xi_{\ell}^{j}
=
\bar{\xi}_{\ell}^{j}
+
\Delta\xi_{\ell}^{j},
\end{equation}

\noindent where $\xi$ denotes $\alpha$, $\theta$, $\phi$, or $\delta$. The
canonical first-layer values are

\begin{equation}
\bar{\gamma}_{1}=0,
\qquad
\bar{\delta}_{1}^{j}=0,
\qquad
\bar{\phi}_{1}^{j}=0,
\qquad
\bar{\theta}_{1}^{j}=\frac{\pi}{4},
\qquad
\bar{\alpha}_{1}^{j}=\frac{\pi}{2}.
\label{eq:qips_canonical_layer1}
\end{equation}

For the second layer,

\begin{equation}
\bar{\gamma}_{2}=0,
\qquad
\bar{\delta}_{2}^{j}=0,
\qquad
\bar{\phi}_{2}^{j}=0,
\qquad
\bar{\theta}_{2}^{j}
=
\frac{\pi}{2}b_{j}^{(\mathrm{seed})},
\qquad
\bar{\alpha}_{2}^{j}=\frac{\pi}{2},
\label{eq:qips_canonical_layer2}
\end{equation}

\noindent where $b_{j}^{(\mathrm{seed})}\in\{0,1\}$ is the $j$th bit of the
seed bitstring.

Early implementations framed the seed-conditioned circuit within a variational
optimization paradigm, treating the circuit parameters as degrees of freedom to
be optimized directly. The focus shifted when this approach revealed that the
sampled bitstring distribution itself could be used as a proposal-search 
resource, where progress is measured not by the mean of a sampled cost
distribution but by whether the sampled bitstrings provide useful search
proposals. QIPS uses this resource by generating finite-shot proposal
distributions from which improved bitstrings can be detected, harvested and used
as new seeds, rather than by training a single circuit to concentrate probability
on a fixed low-energy state. QIPS therefore adopts a stochastic proposal
framework in which deviations from the seed-conditioned canonical angles are
sampled and updated by feedback to maintain detectable localization while
preserving enough diversity to discover improved bitstrings.

The feedback controller regulates the scale of the angle deviations. Smaller
deviations generally strengthen localization, whereas larger deviations broaden
the proposal distribution. Localization is assessed directly from the bitstrings
obtained in 100 shots. A useful quantum interference pattern must return the
seed repeatedly while also assigning statistically accessible probability to a
limited set of additional outcomes. This requirement makes localization an
operational finite-shot condition: useful proposal states must occur often enough
to be detected, harvested and passed forward by the search. Localization
therefore concentrates the proposal distribution strongly enough that informative
candidate states remain statistically significant within a small measurement
sample.

Operationally, the feedback step is implemented as a stochastic search over
angle deviations. Each trial circuit is sampled with 100 shots, and the observed
seed multiplicity and number of distinct measured bitstrings define a
pseudo-energy that scores whether the resulting interference pattern is
sufficiently localized while retaining non-seed diversity. Trial deviations are
accepted or rejected using a Monte Carlo rule on this pseudo-energy. The
pseudo-energy is not the QUBO objective; it is an internal controller objective
used to select circuits that generate statistically useful finite-shot proposal
distributions.

The seed multiplicity and number of distinct measured states therefore define
the feedback objective, which determines whether the deviation parameters are
retained and how broadly subsequent trials are sampled. The controller
maintains a prescribed average qualification rate for localized patterns while
preserving stochastic diversity among accepted circuits. Because the
localization criteria are defined relative to the fixed shot count, neither the
feedback objective nor its target qualification rate depends explicitly on
system size. This makes the controller a size-independent finite-shot rule for
selecting proposal circuits.

%%%%%%%%%%%%%%%%%%%%%%%%%%%%%%%%%%%%%%%%%%%%%%%%%%%%%%%%%%%%%%%%%%%%%%%%%%%%%%%
%%%%%%%%%%%%%%%%%%%%%%%%%%%%%%%%%%%%%%%%%%%%%%%%%%%%%%%%%%%%%%%%%%%%%%%%%%%%%%%
\subsection*{Emergent structure of localized quantum interference}
\begin{figure}[!hbt]
    \centering
    \includegraphics[
        width=0.98\linewidth,
        trim={0.00in 0.00in 0.00in 0.00in},  % trim={left bottom right top}
        clip
    ]{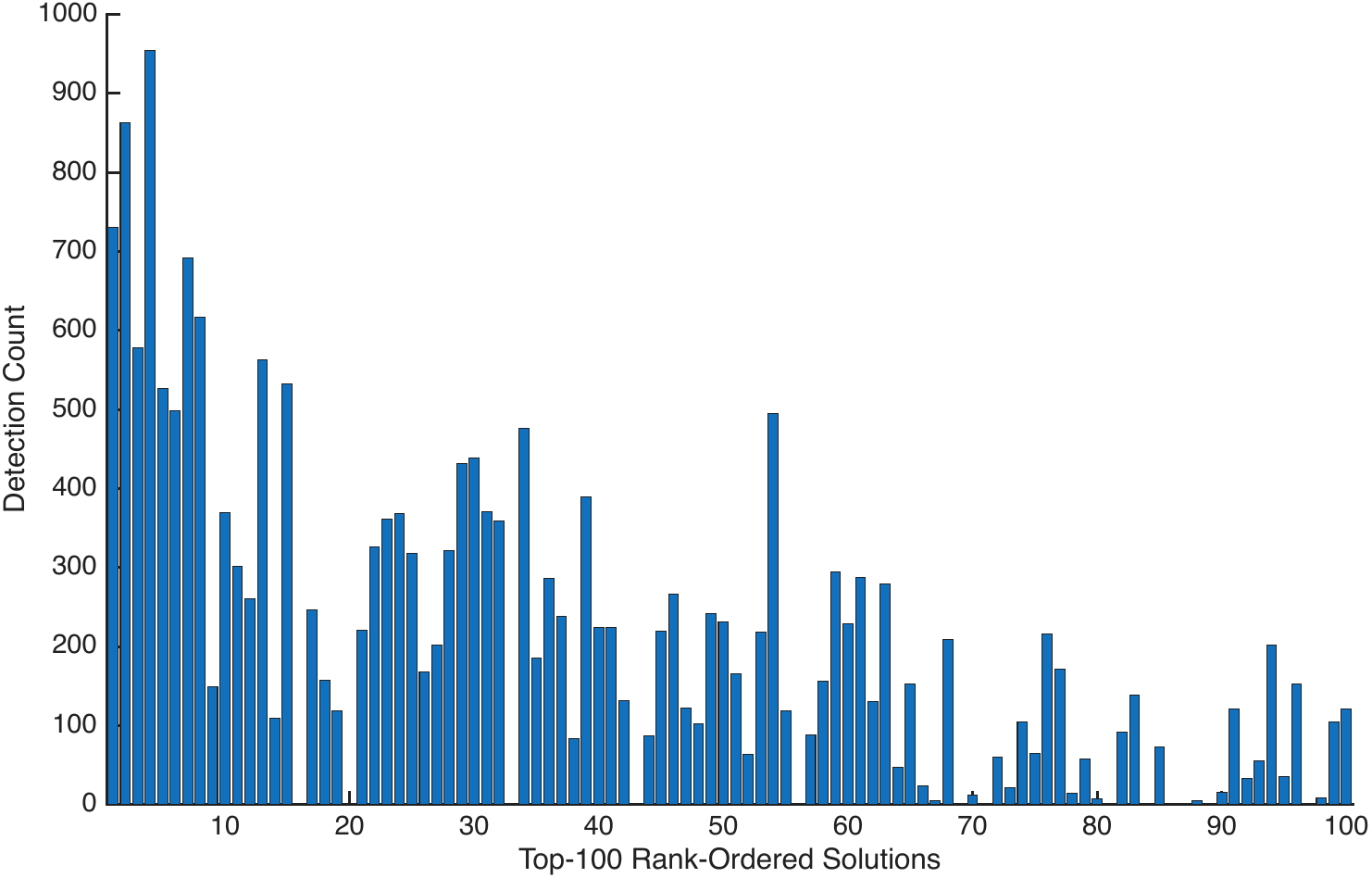}
\caption{\textbf{Finite-shot detection of top-100 low-energy states.}
Detection counts for the top-100 rank-ordered solutions in a representative
QIPS run on the SK model for $n_b=29$. Rank 1 denotes the lowest-energy state. 
Counts are accumulated over the fixed proposal budget used in the search. The 
lowest-rank states are detected most frequently, but detections extend across 
the top-100 set. This pattern shows that localized finite-shot proposals can
repeatedly access near-optimal states while maintaining diversity within the
elite frontier.}
    \label{fig:DetectionCounts}
\end{figure}

What does finite-shot access to the low-energy tail look like when it is
generated by localized quantum interference? Figure~\ref{fig:DetectionCounts}
answers this question for a representative Sherrington--Kirkpatrick (SK) run.
QIPS produces broad coverage across the top-100 states, but this coverage is
not uniform: repeated detections fluctuate strongly, and some low-energy states
are missed altogether. This pattern reflects the ensemble nature of the
proposal process. Different seed-conditioned circuits expose different
statistically accessible states, and the search harvests useful bitstrings from
this ensemble rather than requiring one circuit to resolve the entire
low-energy tail. A detailed finite-shot search trajectory analysis is provided
in Supplementary Note~3. The same qualitative behavior is observed across the
benchmark study and sharpens the mechanistic question: what probability
structure in the seed-conditioned quantum circuits makes such finite-shot
access possible?

%TC:ignore
%TC:endignore

%TC:ignore
\begin{figure}[!ht]
    \centering
    \includegraphics[
        width=0.98\linewidth,
        trim={0.00in 0.00in 0.00in 0.00 in},      % trim={left bottom right top}
        clip
    ]{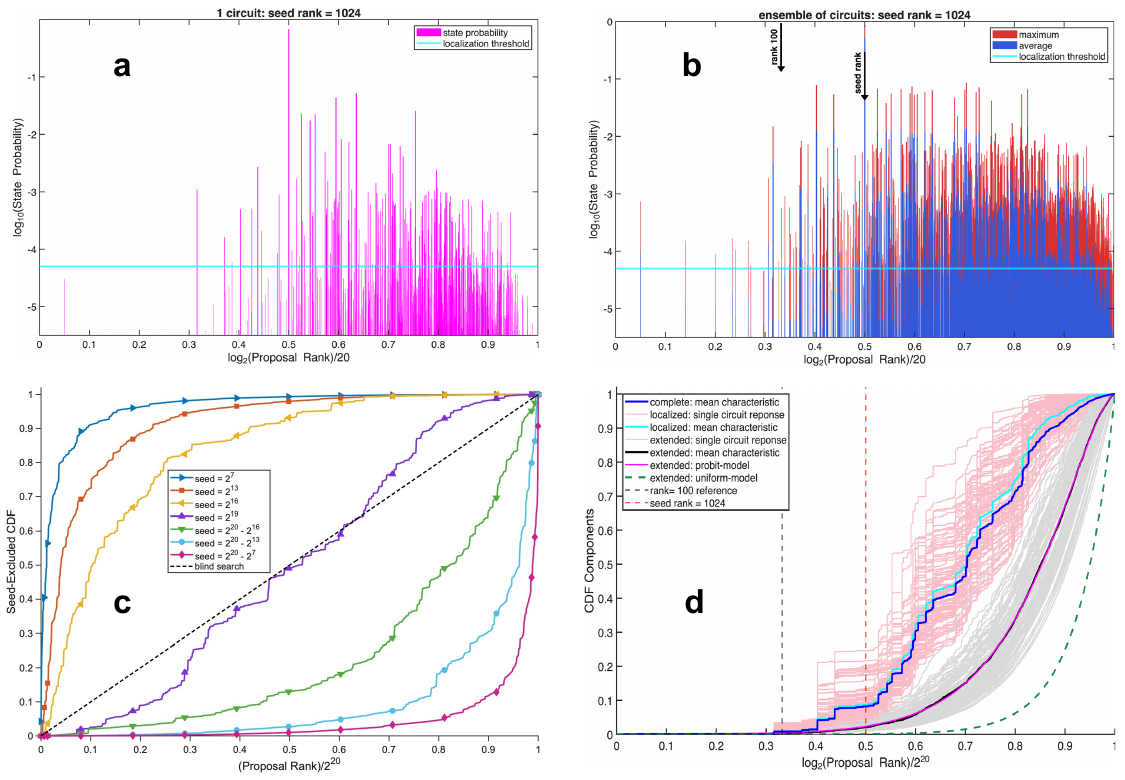}
\caption{\textbf{Emergent structure of QIPS proposal distributions for the SK model.}
\textbf{a}, State probabilities from a single localized circuit for
$n_b=20$ and seed rank $r_s=1024$, plotted against the normalized logarithmic
rank coordinate $\log_2(r)/n_b$. This coordinate resolves probability
concentration near the low-energy tail more clearly than the linear coordinate
$r/2^{n_b}$. The horizontal line is the localization threshold
$p_{\mathrm{loc}}=0.005/N_{\mathrm{shots}}=5\times10^{-5}$ for
$N_{\mathrm{shots}}=100$, which estimates the probability needed for 
statistical detectability in a 100-shot sample.
\textbf{b}, Maximum and mean state probabilities obtained from the 96 localized
circuits retained from an ensemble of 120 circuits generated from the same
seed. The maximum curve identifies states that become highly probable in at
least one circuit, whereas the mean curve identifies probability structure
that recurs across the circuit ensemble.
\textbf{c}, Seed-excluded cumulative distribution functions (CDFs) for
representative seed ranks. The CDF gives the probability that a proposal,
excluding return to the seed, has rank less than or equal to $r$. Seeds near
the low-energy edge favor low-rank proposals, seeds near the high-energy edge
favor high-rank proposals, and a seed near the spectral midpoint produces a
CDF close to blind search.
\textbf{d}, Decomposition of the conditional CDF for $n_b=20$ and
$r_s=1024$. The seed contribution is treated separately. Non-seed states above
$p_{\mathrm{loc}}$ define the localized component, and the remaining
probability defines the extended component. Thin curves show localized and
extended components from individual retained circuits, while thick curves show
the corresponding ensemble-averaged characteristics. The CDF for a uniform
distribution is shown as the blind-search reference. The extended component is
well described by a probit CDF and remains more structured than blind search,
showing that probability structure persists even outside the localized peaks.}
    \label{fig:QuantumEfficent}
\end{figure}
%TC:endignore 

Figure~\ref{fig:QuantumEfficent}a shows the state probabilities generated by a
representative circuit that produces a localized quantum interference pattern.
A total of 213 states lie above the localization threshold, which estimates the
statistically significant detection level for 100 shots, and together account
for 0.98695 of the probability mass. Excluding the seed state, six states have
probability greater than 0.01 and together carry 0.18342 of the probability
mass. More generally, finite-shot localization is quantified by sorting the state
probabilities from largest to smallest and asking whether most of the total
probability accumulates over a small leading subset of computational-basis
states. Localization therefore makes finite-shot proposal search possible by 
assigning statistically detectable probability to a restricted set of candidate
bitstrings. Localization alone, however, does not ensure a productive search.
For robust downhill movement, statistically significant peaks must occur at
energies below the seed energy. The circuit in this example produces a highly
localized quantum interference pattern, but only 0.00652 of the total 
probability mass lies below the seed energy and it is therefore relatively 
ineffective for advancing the search.

Figure~\ref{fig:QuantumEfficent}b considers an ensemble of 120 circuits
generated from the same seed. Supplementary Note~4 evaluates the localization
quality of each circuit. Across this ensemble, Hilbert-space coverage increases
from 0.020\% for a single circuit to 0.347\%, while the number of states with
probability greater than 0.01 increases from 7 to 31. Many statistically
significant peaks recur across the circuit ensemble, whereas others appear only
intermittently. Thus, QIPS does not rely on a single reproducible interference
pattern. Instead, it harvests useful low-energy proposals from an ensemble of
circuits whose localized interference peaks vary across realizations. The mean
behavior of these peaks within the top-100 energy-ranked states is shown in
Supplementary Note~5, and their effectiveness in driving downhill search
dynamics is illustrated in Supplementary Note~6. The cumulative distribution
function (CDF) of proposal ranks provides a quantitative characterization of
this emergent ensemble behavior.

The seed dependence of the proposal distribution is summarized by the
seed-excluded CDFs in Figure~\ref{fig:QuantumEfficent}c, which quantify how
proposal ranks change when the return-to-seed event is removed.
Supplementary Note~7 shows how this CDF is constructed and decomposed into
seed, localized, and extended contributions. Probability concentrates toward
the low- or high-energy edge as the seed approaches the corresponding end of
the spectrum, whereas a seed near the midpoint, $r_m=2^{n_b-1}$, produces a CDF
close to blind search. An approximate symmetry,
$\mathrm{CDF}(u)\approx1-\mathrm{CDF}(1-u)$ with $u=r/2^{n_b}$, is observed
across the sizes examined for all six systems listed in
Table~\ref{table:6systems}. Supplementary Note~8 compares the quantum proposal
CDFs with matched classical proposal CDFs for the SK system at $n_b=20$ across 
varying seeds, as well as over different system sizes at fixed seed rank and 
fixed normalized seed rank. No corresponding symmetry is present in the 
classical proposal distributions, which are explicitly biased toward
lower-energy states.

Figure~\ref{fig:QuantumEfficent}d shows that the localized and extended
components vary across circuit realizations but have stable ensemble-averaged
forms. When the localization threshold is chosen so that $q_E<0.1$, the
localized component typically contains hundreds to thousands of states,
increasing to tens of thousands and beyond only when localization is lost.
Thus, low-rank probability concentration emerges not from a single optimized 
circuit, but from an ensemble of localized finite-shot proposal distributions 
generated by stochastic feedback.

%%%%%%%%%%%%%%%%%%%%%%%%%%%%%%%%%%%%%%%%%%%%%%%%%%%%%%%%%%%%%%%%%%%%%%%%%%%%%%%
%%%%%%%%%%%%%%%%%%%%%%%%%%%%%%%%%%%%%%%%%%%%%%%%%%%%%%%%%%%%%%%%%%%%%%%%%%%%%%%
\subsection*{Matched classical and quantum proposal benchmarks}
\begin{figure}[!hbt]
    \centering
    \includegraphics[
        width=0.98\linewidth,
        trim={0.07in 0.05in 0.00in 0.05in},
        clip
    ]{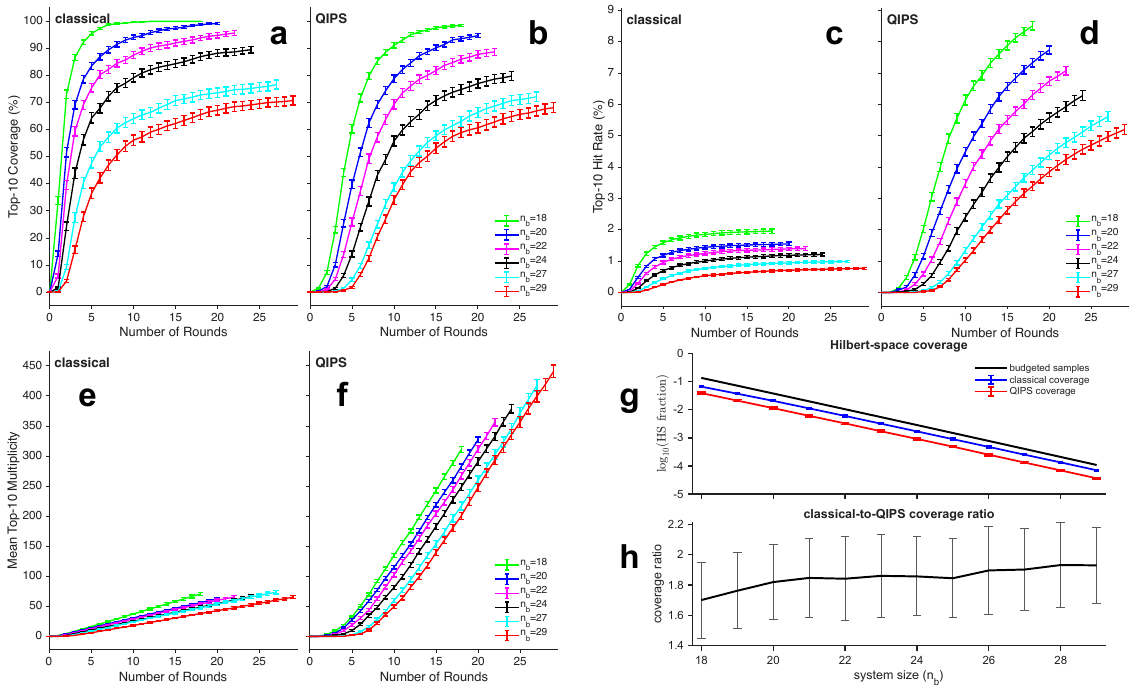}
    \caption{\textbf{Matched classical and QIPS proposal-search benchmarks 
    aggregated over six systems.}
    \textbf{a,b}, Top-10 coverage for the classical and QIPS searches, defined as the
    fraction of distinct ground-truth top-10 states found.
    \textbf{c,d}, Top-10 hit rate, defined as the fraction of all proposals that
    correspond to a ground-truth top-10 state.
    \textbf{e,f}, Mean top-10 multiplicity, defined as the number of top-10 hits
    divided by the number of distinct top-10 states found.
    \textbf{g}, Hilbert-space coverage, defined as the number of distinct sampled
    states divided by $2^{n_b}$, compared with the total sampling budget.
    \textbf{h}, Ratio of classical to QIPS Hilbert-space coverage. Results are shown as
    functions of search round or system size, and error bars denote the standard error
    of the mean.}
    \label{fig:classicalVsQIPS}
\end{figure}

Figure~\ref{fig:classicalVsQIPS} compares QIPS with the matched classical
proposal search under identical outer-loop protocols, seed-selection rules,
frontier update rules and proposal budgets. Panels a--f show the evolution of
top-10 coverage, hit rate and multiplicity. The classical search reaches high
coverage more rapidly, whereas QIPS exhibits greater repeated sampling of
low-energy states as the frontier matures. Both methods lose coverage as
$n_b$ increases, but their qualitative behavior remains consistent across the
six benchmark families. Corresponding top-1 and top-100 results, together with
system-resolved data, are provided in Supplementary Note~9.

%TC:ignore
%TC:endignore

The difference between the two searches is clearest in the hit-rate and
multiplicity curves. Classical proposals produce rapid early progress followed
by strong diminishing returns. QIPS advances more gradually but continues to
revisit low-energy states at high rates after the elite frontier begins to
saturate. This repeated sampling is consistent with localized quantum
interference concentrating finite-shot probability on a restricted low-energy
subset, and provides an empirical signal that the search has entered a
productive low-energy region. A declining rate of novel frontier updates
therefore provides a natural fixed-budget or early-stopping diagnostic.

Figure~\ref{fig:classicalVsQIPS}g compares the fraction of Hilbert space explored.
QIPS samples fewer distinct states than the classical search because substantial
probability remains concentrated on the seed and a small set of recurring
low-energy proposals. Figure~\ref{fig:classicalVsQIPS}h shows that the classical
search covers roughly twice as many distinct states over the range studied.
This lower coverage is not itself a disadvantage: QIPS trades broad exploration
for repeated access to a statistically significant low-energy subset. The
relevant diagnostic is therefore not Hilbert-space coverage alone, but how
effectively the sampled probability is concentrated within the low-energy tail.

%TC:ignore
\begin{figure}[!ht]
    \centering
    \includegraphics[
        width=0.98\linewidth,
        trim={0.07in 0.05in 0.08in 0.05in},      % trim={left bottom right top}
        clip
    ]{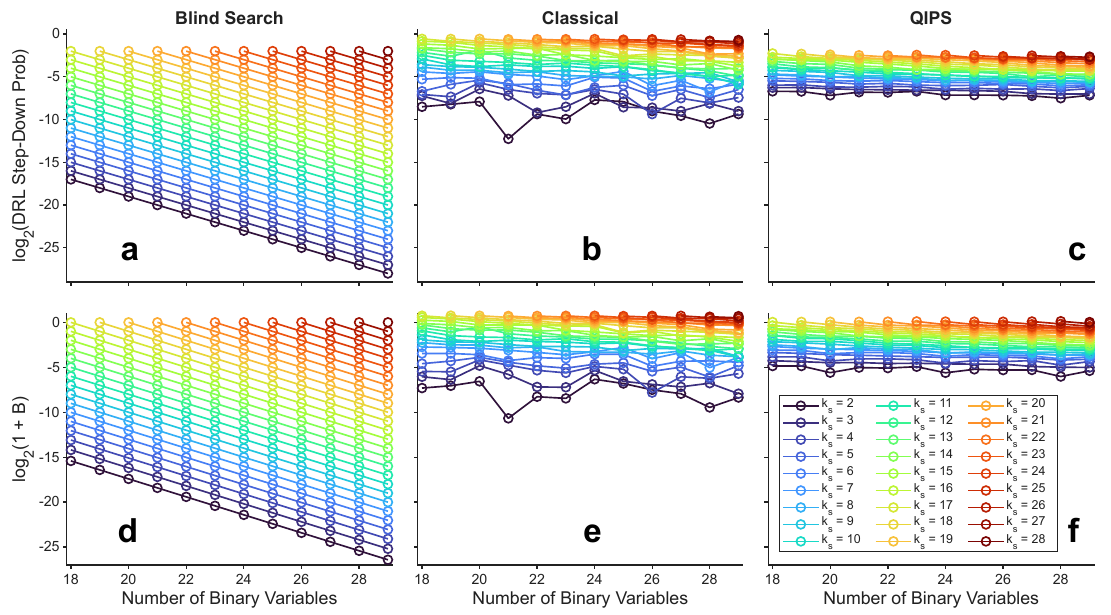}
    \caption{\textbf{Dyadic-rank progression metrics for the six benchmark systems.}
    \textbf{a--c}, Dyadic-rank landmark step-down probability for blind search,
    classical proposals and QIPS, respectively. For dyadic rank level $k_s$, the 
    reference landmark is $r_s=2^{k_s}$. The step-down probability is the conditional 
    probability that a proposal has rank $r_p\leq 2^{k_s-1}$, corresponding to an 
    improvement of at least one dyadic rank level. To increase statistical sampling, 
    seed states with ranks within $\pm20\%$ of each reference landmark are included.
    \textbf{d--f}, Rank-directional bias for the same proposal models. The bias is
    defined as $B=(p_L-p_R)/(p_L+p_R)$, where $p_L$ and $p_R$ are the probabilities
    of proposing ranks below and above the reference landmark, respectively; proposals
    at the reference rank are excluded. The plotted quantity, $\log_2(1+B)$, resolves
    the approach to the limiting bias $B=-1$. Curves are shown as functions of system
    size, with color indicating dyadic seed-rank level $k_s$. Results are aggregated
    over all six benchmark systems at each size. Error bars, corresponding to the 
    standard error of the mean over the aggregated benchmark ensemble, are omitted 
    for visual clarity.}
    \label{fig:QIPSmetaData}
\end{figure}
%TC:endignore

The rank dependence of this concentration is quantified in
Figure~\ref{fig:QIPSmetaData}. Dyadic-rank metrics ask whether a proposal moves
the search across logarithmic rank scales, which is the relevant scale when the
number of candidate bitstrings grows exponentially with $n_b$. Panels a--c show
the probability that a proposal improves by at least one dyadic rank level. Blind
search decays exponentially with system size. Both classical proposals
and QIPS retain substantially larger step-down probabilities. QIPS becomes
increasingly competitive as the seed moves toward the ground state, where
localized proposal structure is most consequential.

Panels d--f show the corresponding rank-directional bias. As the seed
approaches the ground state, proposal mechanisms become increasingly biased
toward larger ranks because fewer improving states remain available. Blind
search approaches this limiting behavior most rapidly. Classical proposals and
QIPS preserve a substantially larger probability of moving toward lower ranks,
with QIPS showing the strongest relative behavior for the best-ranked seeds.
Together, the dyadic metrics show that the classical search is strongest during
early descent from poorly ranked seeds, whereas QIPS is most competitive after
the frontier has moved into the low-energy tail. Thus, the main QIPS signature
is not uniform dominance over the matched classical proposal generator, but a
distinct finite-shot proposal profile: localized interference repeatedly
samples low-energy states while preserving enough rank-improving probability to
support continued frontier refinement.

%%%%%%%%%%%%%%%%%%%%%%%%%%%%%%%%%%%%%%%%%%%%%%%%%%%%%%%%%%%%%%%%%%%%%%%%%%%%%%%%%%%%%%
%%%%%%%%%%%%%%%%%%%%%%%%%%%%%%%%%%%%%%%%%%%%%%%%%%%%%%%%%%%%%%%%%%%%%%%%%%%%%%%%%%%%%%
\section*{Discussion}

\subsection*{Localized interference as a proposal resource}
The defining feature of QIPS is that finite-shot randomness is not treated as
a nuisance to be averaged away, but as a proposal resource shaped by localized
quantum interference. This framing differs from variational optimization, where 
progress is usually measured through improvement of an expectation value, and 
removes the requirement that a single circuit place high probability on a target 
low-energy state. Instead, progress is made through a sequence of seed-conditioned 
circuits whose localized interference patterns generate finite-shot proposal
distributions. Candidate bitstrings are detected, harvested, scored classically
and used to update an elite frontier. Across the benchmark range studied, this
structure preserves repeated access to low-energy states with a fixed
two-layer circuit architecture and a fixed per-circuit shot budget, making small
measurement samples statistically productive.

The results raise mechanistic questions about how localized quantum
interference patterns emerge and acquire reproducible structure. A useful
theoretical direction is to characterize how randomized deviations scatter
probability away from the canonical seed-centered pattern, and how the resulting
peaks depend on the cost operator, deviation statistics and seed rank. Future
work should determine why such patterns arise across QUBO families, how their
support and rank bias scale with $n_b$, and whether the observed dyadic-rank
progression approaches a non-vanishing asymptotic limit.

\subsection*{Algorithm refinement and future benchmarks}
These findings do not establish quantum advantage or the asymptotic complexity
of QIPS. They do, however, identify finite-shot signatures that warrant further
testing beyond ideal simulation. Across all six benchmark families, QIPS
exhibits high detection probability in the low-energy tail, repeated recovery
of near-optimal states, and rank-dependent proposal structure that becomes most
competitive after the frontier has moved toward lower-ranked seeds. Whether
these trends persist, saturate or cross over remains open. The decisive next
step is experimental: QIPS should be implemented on current gate-based quantum
computers, tested on larger systems and diverse problem classes, and benchmarked
under matched end-to-end resources. If localized finite-shot behavior survives
realistic hardware noise, QIPS could provide a practical route toward useful
gate-based quantum optimization.

The algorithm also leaves substantial room for refinement. Ablation studies
should identify which circuit elements, feedback rules, frontier sizes,
seed-selection strategies, localization thresholds and proposal budgets are
essential, which can be simplified, and which can be optimized for hardware
execution. Hybrid proposal strategies provide another direction: the quantum
circuit could supply a non-local proposal step, while classical optimization or
repair methods refine the proposed bitstrings before frontier updates. Simple
alternating quantum and classical proposals did not improve
performance in the limited tests performed here, but more structured hybrid
schedules remain open. Future studies should also test whether the same 
seed-conditioned proposal principle extends to more general discrete binary 
optimization problems.

\subsection*{Quantum proposal generation on hardware}
QIPS is a software-level algorithm, but its central hardware-facing unit is the
quantum proposal generator: a seed-conditioned circuit whose finite-shot
measurement record supplies candidate bitstrings for classical evaluation. The
feedback controller selects circuit deviations using quantities estimated from
repeated measurements. This controller functionality could be incorporated into 
quantum-computing hardware to support efficient circuit-parameter updates, 
including calibrated perturbations around seed-conditioned canonical angles. The
resulting finite-shot proposal distributions could then be benchmarked against
ideal or calibrated reference distributions to inform hardware-specific
calibration, noise mitigation, measurement-error correction and qubit-mapping
choices around the proposal stage. Such distribution-level diagnostics may also
be useful for calibrating and mapping gate-based quantum circuits
whose computational value depends on structured finite-shot sampling.

%%%%%%%%%%%%%%%%%%%%%%%%%%%%%%%%%%%%%%%%%%%%%%%%%%%%%%%%%%%%%%%%%%%%%%%%%%%%%%%
%%%%%%%%%%%%%%%%%%%%%%%%%%%%%%%%%%%%%%%%%%%%%%%%%%%%%%%%%%%%%%%%%%%%%%%%%%%%%%%
%TC:ignore
\section*{Methods}

\subsection*{Benchmark problems and rank representation}
Each benchmark problem is represented as a quadratic unconstrained binary
optimization (QUBO) problem or equivalently as an Ising cost function on
$n_b$ binary variables. A computational-basis state is a bitstring
$b=(b_1,\ldots,b_{n_b})$, with $b_j\in\{0,1\}$. The corresponding Ising
spin convention is $s_j=2b_j-1$. For every problem instance, all
computational-basis states are assigned a cost and then sorted by energy.
The energy rank $r$ runs from $1$ to $2^{n_b}$, with $r=1$ denoting the
lowest-energy state.

Rank ordering is used throughout the analysis because it provides a common
coordinate for comparing distinct QUBO families. The numerical energy scale,
density of states and degeneracy structure differ across benchmark systems,
but a proposal with low rank always denotes a low-energy candidate relative
to that instance. The normalized logarithmic coordinate
$\log_2(r)/n_b$ is used when resolving the low-energy tail, because it
spreads exponentially small rank fractions over a visible range.

The systematic benchmarks use six QUBO/Ising families, 32 matched instances
per family and size, and system sizes $18\le n_b\le29$. Classical and QIPS
proposal searches are applied to the same problem instances using the same
outer search protocol and the same total proposal budget.

\subsection*{Outer search loop and resource accounting}
QIPS combines a classical outer search with a quantum proposal generator.
The outer search maintains an elite frontier containing the best states
found so far. Seed states are drawn from this frontier, and each seed
conditions either a quantum or classical proposal generator. Proposed
bitstrings are evaluated under the QUBO/Ising cost function and merged into
the elite set. The frontier is then truncated back to a fixed size.

The production calculations use an outer frontier of 100 states and a target
frontier of 10 states. Each search round selects 20 seed states from the
frontier. In QIPS, each selected seed defines one quantum circuit measured
with $N_{\mathrm{shots}}=100$ shots. The benchmark uses $n_b$ rounds, giving
$N_{\mathrm{circ}} = 20n_b$ quantum circuits and
$N_{\mathrm{prop}} = 100 \times 20n_b = 2000n_b$
proposal evaluations. The matched classical search is given the same total
number of proposal evaluations. Repeated measurements, repeated bitstrings
and non-improving proposals all count against this budget. This convention
treats sampling effort as the shared resource.

Seed states are selected from the outer frontier using a rank-biased
distribution that favors lower-energy frontier states while retaining a small
exploration tail. This allows the search to exploit the best available seeds
without losing diversity. For the classical proposal generator, diversity
corresponds to exploring different basins of the cost landscape. For QIPS,
diversity corresponds to generating different high-probability peaks in the
seed-conditioned quantum interference patterns. The reported matched benchmarks
use the full fixed proposal budget.

\subsection*{Seed-conditioned quantum proposal circuit}
The quantum proposal generator is a fixed two-layer circuit initialized in the
uniform superposition $\ket{+}^{\otimes n_b}$. The seed bitstring is encoded
through the canonical circuit angles rather than by preparing the register as
the computational-basis seed state. These canonical angles define a
seed-centered localization pattern. Stochastic deviations then dress this
reference pattern into a family of localized quantum interference patterns by
perturbing the constructive and destructive interference among
computational-basis amplitudes.

This construction is used as a proposal generator rather than as a variational
ansatz. The two layers are not intended to form an expressive approximation to
an unknown optimum state. Instead, they provide a compact interference
architecture that combines seed-centered localization, cost-dependent phase
structure, and feedback-regulated stochastic deviations. The deviations may be
viewed as controlled scatter around the canonical pattern: if they are too
small, the distribution remains seed dominated; if they are too large,
localization is lost. QIPS operates in the intermediate regime, where
localization is preserved but a finite set of statistically significant
non-seed peaks is generated, with peak locations shaped by the QUBO/Ising cost
structure.

The search therefore does not optimize a single circuit or minimize an
expectation value with respect to circuit parameters. It samples an ensemble of
related circuits whose localized interference patterns generate different
finite-shot proposals. The useful object is this ensemble of proposals, not an
individually optimized quantum state with probability concentrated on the
ground state or other low-energy states.

The phase separator uses a normalized cost operator so that different problem
instances and sizes can be treated on a comparable angular scale. A negligible
static perturbation defines an operational ordering through exact degeneracies.
Some circuit realizations also use dynamic cost-operator jitter to diversify
recurring interference peaks. In all cases, measured bitstrings are evaluated
and ranked using the operational search energy, not the instantaneous jittered
spectrum used to generate the circuit phase. Additional implementation details
are provided in Supplementary Note~10.

\subsection*{Feedback control of localized interference}
The purpose of feedback is to maintain localized but diverse quantum
interference patterns. A useful circuit should return the seed repeatedly,
showing that the distribution remains localized, but it should also produce
a finite set of additional measurable outcomes that can update the frontier.
A very broad probability distribution gives too many distinct outcomes with
negligible individual probability. Excessive localization on the seed gives
too few new proposals.

The feedback controller is based only on the 100-shot measurement record.
For a circuit conditioned on seed state $s_{\mathrm{seed}}$, define
$n_{\mathrm{seed}}$ as the number of measured shots equal to the seed and
$n_{\mathrm{unique}}$ as the number of distinct measured bitstrings. These
two quantities define a pseudo-energy objective that favors intermediate
seed multiplicity and a finite number of distinct outcomes. A Metropolis
rule accepts or rejects trial deviation parameters based on this
pseudo-energy. A separate smoothed qualification-rate controller adjusts the
scale of subsequent trials so that a prescribed fraction of circuits remain
localized while stochastic diversity is preserved.

Because the objective is defined relative to the fixed shot count, the
feedback target does not explicitly depend on the Hilbert-space dimension.
This is important for finite-shot scaling: the circuit is not required to
resolve the full $2^{n_b}$-state distribution. It is only required to
concentrate enough probability into a small set of outcomes that informative
states can be detected with 100 shots.

\subsection*{Statistically significant states and localization threshold}
Localization is defined here as concentration of probability mass over a small
effective support, independent of the energy ordering or semantic interpretation of
the computational-basis states. Let $P(b)$, with $b \in \{0,1\}^{n_b}$, denote 
an output probability distribution over computational-basis bitstrings, and let
\[
p_{(1)} \ge p_{(2)} \ge \cdots \ge p_{(2^{n_b})}
\]
denote the same probabilities sorted in descending order. The cumulative mass
\[
C(m)=\sum_{i=1}^{m}p_{(i)}
\]
characterizes localization. The relevant localization condition is that the leading 
probability masses must accumulate on a scale comparable to, and preferably smaller 
than, the number of measurement shots:
\[
m = O(N_{\mathrm{shots}}), \qquad m < N_{\mathrm{shots}}
\]
for the statistically significant portion of the distribution. This finite-shot condition
ensures that repeated detections of some high-probability states can occur within a
single circuit measurement record. If the effective support becomes comparable to or
larger than \(N_{\mathrm{shots}}\), the measurement record becomes too sparse to verify
statistically significant peaks reliably.

This definition depends only on the sorted probability masses, not on whether the
corresponding bitstrings are close to a seed, low in energy, feasible, or favorable under
the objective function. Those task-dependent relationships determine whether a
localized interference pattern is productive for optimization.

For post hoc analysis of the exact simulated probability distribution, a state is treated
as statistically significant when its probability exceeds
\[
p_{\mathrm{loc}}=\frac{0.005}{N_{\mathrm{shots}}}.
\]
For \(N_{\mathrm{shots}}=100\), this gives
\[
p_{\mathrm{loc}}=5\times 10^{-5}.
\]
This threshold provides a finite-shot operational approximation to the leading
probability masses in the sorted distribution: states far below it are unlikely to be
observed reproducibly in a 100-shot record, whereas states above it can contribute
measurable peaks. The threshold is used for analysis and decomposition of exact
probability distributions. The feedback controller itself uses only the measured values
of \(n_{\mathrm{seed}}\) and \(n_{\mathrm{unique}}\).

Localization is therefore operationally connected to finite-shot detection.
A localized interference pattern concentrates probability into states that
can appear with statistical significance. However, localization alone is not
sufficient for optimization. Productive search also requires statistically
significant probability mass at energies below the seed energy.

\subsection*{Detection probability from intermittent peaks}
A low-energy state need not have high probability in every circuit to be
found reliably. QIPS often accesses low-energy states through intermittent
probability peaks: a state may have negligible probability for many circuits
and then become highly probable for one or a few seed-conditioned circuit
realizations. The relevant quantity is therefore the accumulated detection
probability over the circuit ensemble.

If state $s$ has measurement probability $p_s(k)$ in circuit $k$, then after
100 shots from each circuit its accumulated detection probability is

\begin{equation}
P_D(s)
=
1
-
\prod_k
\left[1-p_s(k)\right]^{100}.
\end{equation}
This expression distinguishes finite-shot search from inspection of a single
probability distribution. A state with modest ensemble-mean probability may
still be detected frequently if it appears as a large intermittent peak.
Conversely, a broad distribution with no statistically significant peaks will
result in poor finite-shot utility. Detection counts and detection probabilities
therefore provide direct diagnostics of whether localized interference
patterns produce operationally useful proposals.

\subsection*{Seed-excluded cumulative proposal-rank distributions}
QIPS circuits generally return the seed state with substantial probability.
Repeated recovery of the seed confirms localization but does not produce a
new candidate bitstring. To compare the quality of non-seed proposals, the
seed contribution is removed from the proposal distribution. For a seed rank
$r_s$, the seed-excluded conditional distribution is

\begin{equation}
P(r_p \mid r_p\ne r_s, r_s)
=
\frac{P(r_p\mid r_s)}{1-p_s},
\qquad r_p\ne r_s ,
\end{equation}
where $p_s=P(r_p=r_s\mid r_s)$ is the probability of returning to the seed.
The corresponding seed-excluded cumulative distribution function (CDF)
measures the probability that a non-seed proposal has rank less than or
equal to $r$.

The seed-excluded CDF isolates the rank structure of proposals that can
advance the search. It is used to compare QIPS with the matched classical
proposal generator, and to examine how proposal quality changes with seed
rank and system size. Because all systems are compared by rank rather than
raw energy, the same CDF framework applies across sparse graphs, dense spin
glasses and structured QUBO instances.

\subsection*{Localized and extended CDF components}
The exact proposal distribution from a localized circuit contains three
conceptually distinct parts: the seed probability, a localized set of
statistically significant non-seed peaks, and an extended low-probability
background. The conditional CDF can therefore be decomposed as

\begin{equation}
\mathrm{CDF}(r_p\mid r_s)
=
p_s\,\delta_{r_p,r_s}
+
q_L(r_s)\,\mathrm{CDF}_L(r_p\mid r_s)
+
q_E(r_s)\,\mathrm{CDF}_E(r_p\mid r_s).
\end{equation}

Here, $p_s$ is the seed probability, $q_L$ is the total probability in
non-seed states above $p_{\mathrm{loc}}$, and $q_E$ is the remaining
extended probability. The localized CDF describes the statistically
significant peaks, whereas the extended CDF describes the background after
those peaks are removed.

The extended component is not structureless. Across the systems and sizes
examined, the empirical $\mathrm{CDF}_E(r_p\mid r_s)$ fits markedly well to a
two-parameter probit CDF. In normalized rank coordinate $u=r/2^{n_b}$, the probit form is

\begin{equation}
F_E(u;m,s)
=
\Phi\left[
\frac{\Phi^{-1}(u)-m}{s}
\right],
\qquad 0<u<1,
\end{equation}
where $\Phi$ is the standard normal CDF. A uniform blind search would give
$F(u)=u$. The fitted probit form typically remains more concentrated toward
low-energy ranks than this blind-search reference, showing that rank
structure persists even outside the localized peaks. This decomposition also
provides a compact representation of large probability distributions without 
storing every basis-state probability explicitly. High reconstruction 
accuracy can be achieved with very low memory requirements for post hoc
analysis of simulation data.

\subsection*{Classical proposal generator}
The matched classical baseline uses the same outer frontier, seed-selection
rule and proposal budget as QIPS, but replaces the quantum circuit with a
classical kick-and-repair proposal generator. Starting from a selected seed,
a random number of bits is flipped. Most kicks are short range in Hamming
distance, while a smaller fraction are longer range to preserve exploration.
The kicked state is then subjected to a small number of greedy one-bit repair
moves when these moves improve the cost.

This baseline is intentionally strong but generic. It uses local information
from the QUBO cost function to bias proposals toward lower-energy states,
while still allowing nonlocal moves. All attempted proposals and repair
neighbors count against the classical proposal budget, with the same total
budget used for QIPS. A blind-search control is obtained by drawing bitstrings
uniformly from the full Hilbert space.

For CDF comparisons, classical proposal probabilities are defined empirically
from the finite proposal batch: a state proposed $n_s$ times in
$N_{\mathrm{shots}}=100$ attempts is assigned probability
$n_s/N_{\mathrm{shots}}$, while unobserved states are assigned zero probability.

\subsection*{Aggregate performance measures}
Search performance is quantified using top-$K$ coverage, hit rate and
multiplicity. Top-$K$ coverage is the fraction of the true lowest-energy
top-$K$ states found at least once. Top-$K$ hit rate is the fraction of all
proposal evaluations that belong to the true top-$K$ set. Mean top-$K$
multiplicity is the number of top-$K$ hits divided by the number of distinct
top-$K$ states recovered. These measures separate three effects: whether the
search finds the low-energy set, how often proposals land in that set, and
whether the search repeatedly samples the same near-optimal states.

Hilbert-space coverage is the fraction of distinct computational-basis states
sampled during the search. An ideal proposal mechanism would recover all states
in the target frontier while visiting as little of Hilbert space as possible.
Thus, the critical issue is the quality of coverage rather than its quantity:
when exploration is not accompanied by improved low-energy recovery, it
represents inefficient cost rather than search progress. Some exploration is
nevertheless unavoidable, because viable paths through a rugged energy landscape
must be discovered. The objective is therefore directed exploration toward the
low-energy tail, not maximal coverage. QIPS trades broad coverage
for repeated access to a structured low-energy subset. For this reason, coverage
is interpreted together with hit rate, multiplicity and rank-based progression
metrics.

\subsection*{Dyadic-rank progression metrics}
Two dyadic-rank measures are used to quantify proposal quality as a function
of seed position and system size. For dyadic level $k_s$, the reference seed
rank is $r_s=2^{k_s}$. Because the sampled seeds occupy only a sparse subset
of the full Hilbert space, proposal statistics are accumulated in a finite
rank window around each dyadic landmark. Specifically, seeds with ranks within
$\pm 20\%$ of $2^{k_s}$ are included in the bin. The dyadic step-down
probability is then the conditional probability that a proposal improves by at
least one dyadic rank level,

\begin{equation}
P_{\mathrm{step}}(k_s)
=
P(r_p \le 2^{k_s-1}\mid r_s\approx 2^{k_s}),
\end{equation}

\noindent where $r_s\approx 2^{k_s}$ denotes this $\pm20\%$ seed-rank bin.
The measure asks whether proposals generated from seeds near one logarithmic
rank scale can move to the next lower scale. Blind search decays exponentially
with system size under this measure, whereas structured proposal mechanisms
can retain much larger step-down probabilities.

The second measure is a rank-directional bias. Let $p_L$ be the probability
of proposing a rank below the reference landmark and $p_R$ the probability
of proposing a rank above it, excluding proposals exactly at the landmark.
The bias is

\begin{equation}
B
=
\frac{p_L-p_R}{p_L+p_R}.
\end{equation}

Values near $B=1$ indicate predominantly downhill proposals, values near
$B=-1$ indicate predominantly uphill proposals, and $B=0$ indicates balance.
The plotted quantity $\log_2(1+B)$ resolves the approach to the limiting
uphill bias that occurs near the ground state, where few improving states
remain. Together, the dyadic step-down probability and rank-directional bias
show where a proposal generator is most effective along the search trajectory.

\subsection*{Simulation and statistical aggregation}
All quantum results reported here are obtained by ideal state-vector
simulation of the two-layer circuit. No hardware noise model is included.
For large Hilbert spaces, finite-shot sampling is accelerated by explicitly
retaining the statistically significant part of the probability distribution
and treating the remaining low-probability mass collectively. This
acceleration is used only to generate samples efficiently in simulation and
does not change how proposed bitstrings are evaluated. Additional
probability-storage compression is used only for post hoc analysis.

Main-text benchmark curves are aggregated over the six systematic benchmark
families and 32 matched instances per family and size. Error bars denote the
standard error of the mean over the corresponding ensemble. System-resolved
results, fixed-seed proposal distributions, finite-shot trajectory analyses,
and implementation details are provided in the Supplementary Information.

\subsection*{Large language model usage}
ChatGPT (OpenAI) was used during the research and manuscript-preparation process
to assist with literature assessment, code review, refactoring, drafting,
organization, language refinement, and presentation of technical material.

%%%%%%%%%%%%%%%%%%%%%%%%%%%%%%%%%%%%%%%%%%%%%%%%%%%%%%%%%%%%%%%%%%%%%%%%%%%%%%%

\bmhead{Supplementary information}

Supplementary information accompanies this article
and includes benchmark definitions, system-resolved proposal statistics,
finite-shot search trajectory analysis, localized-interference diagnostics,
proposal-rank CDF decompositions, matched classical comparisons, top-K
benchmarks, and circuit-construction and feedback-control details.

\bmhead{Acknowledgments}

The author thanks the University Research Computing group at the University of 
North Carolina at Charlotte for providing the computational resources used for 
the large-scale simulations, especially for dedicated resources that were 
required to obtain the 29-qubit results.

\bmhead{Author contributions}

D.J.J. conceived the QIPS framework, developed the quantum and classical
algorithms, implemented the numerical simulations, performed the data analysis,
interpreted the results, prepared the figures, and wrote the manuscript.

\bmhead{Competing interests}

The author declares no competing interests.

\bmhead{Data availability}
The numerical results needed to support the conclusions of this study are 
contained in the article and Supplementary Information. Additional 
processed data are available from the corresponding author upon reasonable 
request.

\bmhead{Code availability}
The custom implementation used in this study is not publicly released 
with this preprint. A documented public release of the code is planned 
following peer-reviewed publication. Inquiries may be directed to the 
corresponding author.

\newpage
%\bibliographystyle{naturemag}
%\FloatBarrier
\bibliography{NPqips}

\clearpage
\onecolumn

\newgeometry{
   left=0.8in,
    right=0.8in,
    top=0.8in,
    bottom=0.8in
}

\begingroup

\renewcommand{\figurename}{Supplementary Figure}
\renewcommand{\tablename}{Supplementary Table}

\setcounter{figure}{0}
\setcounter{table}{0}
\setcounter{equation}{0}

%%%%\documentclass[11pt]{article}

% Paragraph formatting
\setlength{\parindent}{1.5em}
\setlength{\parskip}{0pt}

%\captionsetup[figure]{name=Supplementary Figure}
%\captionsetup[table]{name=Supplementary Table}

% Supplementary Note command:
% \suppnote{number}{title}{label}
\newcommand{\suppnote}[3]{%
    \clearpage
    \phantomsection
    \section*{Supplementary Note #1: #2}
    \addcontentsline{toc}{section}{Supplementary Note #1: #2}
    \label{#3}
}

%%%\sethlcolor{yellow}

\renewcommand{\figurename}{Supplementary Figure}
\renewcommand{\thefigure}{\arabic{figure}}

\renewcommand{\tablename}{Supplementary Table}
\renewcommand{\thetable}{\arabic{table}}

%%%%%\begin{document}

\begin{center}

{\Large\bfseries Supplementary Information}

\vspace{1.2em}

{\LARGE\bfseries
Quantum Interference as a Proposal Mechanism\\
for Combinatorial Optimization
}

\vspace{1.5em}

{\large Donald J. Jacobs$^{1,2,*}$}

\vspace{0.8em}

\begin{minipage}{0.90\textwidth}
\centering
$^{1}$Ishwar Aggarwal Department of Physics and Optical Science, \\
University of North Carolina at Charlotte,
Charlotte, North Carolina, USA

\vspace{0.3em}

$^{2}$Affiliate of the School of Data Science,
University of North Carolina at Charlotte, \\
Charlotte, North Carolina, USA

\vspace{0.5em}

$^{*}$Correspondence:
\href{mailto:djacobs1@charlotte.edu}{djacobs1@charlotte.edu}
\end{minipage}

\end{center}

\vspace{1.5em}

\tableofcontents

\clearpage

%%%%%%%%%%%%%%%%%%%%%%%%%%%%%%%%%%%%%%%%%%%%%%%%%%%%%%%%%%%%%%%%%%%%%%%%%%%%%%%
%%%%%%%%%%%%%%%%%%%%%%%%%%%%%%%%%%%%%%%%%%%%%%%%%%%%%%%%%%%%%%%%%%%%%%%%%%%%%%%
\suppnote{1}{Benchmark systems and instance generation}
{supp:BenchmarkSystems}

%+++++++++++++++++++++++++++++++++++++++++++++++++++++++++++++++++++++++++++++++++++++++++
\subsection*{QUBO and Ising conventions}
A QUBO problem is written as
\begin{equation}
C(\mathbf{b})
=
\sum_i g_i b_i
+
\sum_{i<j} Q_{ij} b_i b_j,
\qquad
b_i\in\{0,1\}.
\end{equation}
Binary variables are mapped to Ising spins using
$s_i=2b_i-1$, so that $b_i=1$ corresponds to spin up and
$b_i=0$ to spin down. Computational-basis states are indexed by the
integer represented by the bitstring, from $0$ to $2^{n_b}-1$. 
Benchmark energies were generated in Ising form,
\begin{equation}
E(\mathbf{s})
=
-\sum_{k<n} J_{nk}\, s_n s_k
-\sum_{k} h_k\, s_k ,
\qquad s_k \in \{-1,+1\}.
\end{equation}
Thus, $J_{nk}>0$ denotes a ferromagnetic bond and $J_{nk}<0$ denotes an
antiferromagnetic bond. Energy ranks are indexed separately from $1$ to 
$2^{n_b}$, with rank $r=1$ denoting the lowest-energy state.

%+++++++++++++++++++++++++++++++++++++++++++++++++++++++++++++++++++++++++++++++++++++++++
\subsection*{Scope of the benchmark study}
The benchmark study was designed to evaluate QIPS across a diverse set of
QUBO/Ising structure. The systems include sparse constraint problems, 
mean-field spin glasses, various spin models with intrinsic frustration, 
MaxCut problems, and maximum independent set (MIS) 
problems. Couplings and weights include uniform, bimodal, Gaussian-like, 
exponentially distributed, and log-normal forms. Graph topologies span 
sparse graphs, complete graphs and 1D and 2D structured lattices.

Three complementary data collections were generated. First, systematic
size-dependent searches were performed for six benchmark families over
$18 \leq n_b \leq 29$. These data form the basis of the aggregate results in the
main text. Second, additional graph and coupling combinations were evaluated
at selected sizes to test whether the observed behavior depended strongly on
the benchmark family. Third, fixed-seed
ensembles were generated to characterize conditional proposal-rank
distributions and their localized and extended components. 

%+++++++++++++++++++++++++++++++++++++++++++++++++++++++++++++++++++++++++++++++++++++++++
\subsection*{Coupling and problem labels}
The coupling and problem models used in the benchmark study are summarized in
Supplementary Table~\ref{tab:supp_labels}. Most of these models can be paired
with different graph topologies, including sparse random graphs, structured
lattices and complete graphs. The SK model is the main exception: it is used
either on the complete graph or, in reduced-SK form, restricted to the edges of
a specified sparse graph.

\begin{table}[ht]
\centering
\caption{\textbf{Problem and coupling labels used in the benchmark archive.}}
\label{tab:supp_labels}
\begin{tabular}{@{}p{0.24\textwidth}p{0.68\textwidth}@{}}
\toprule
Archive label & Description \\
\midrule
70A/30F
& Bimodal Ising couplings, drawn i.i.d. on each bond with 70\%
antiferromagnetic ($J_A=-1$) and 30\% ferromagnetic ($J_B=1$) probabilities. \\

70A/30F+GQD
& Same 70A/30F coupling model supplemented by Gaussian quenched
disorder with $\sigma_J = 0.20$ for any $J$ and $\sigma_h = 0.02$ for any spin.  \\

EWQD & All i.i.d. antiferromagnetic Ising couplings with
$J_{ij}=-X_{ij}$, with random variable $X_{ij}\sim\mathrm{Exp}(1)$, clipped at
$X_{ij}=100$ for numerical stability. \\

reduced-SK
& Sparse SK-type model with i.i.d. Gaussian couplings and fields:
$J_{ij}\sim\mathcal{N}(0,1)$ on graph edges and
$h_i\sim\mathcal{N}(0,1)$ on spins. \\

maxcut-unf
& MaxCut with $W_{ij}=1$, mapped to Ising couplings
$J_{ij}=-W_{ij}/2=-0.5$ and fields $h_i=0$. \\

maxcut-LN
%& MaxCut with log-normal edge weights. \\
& MaxCut with log-normal edge weights
$W_{ij}=\exp(\sigma_W Z_{ij})$, where
$Z_{ij}\sim\mathcal{N}(0,1)$ i.i.d.; mapped to Ising couplings
$J_{ij}=-W_{ij}/2$ and fields $h_i=0$. \\

MIS-unf
& Maximum independent set with uniform vertex weights $a_i=1$ and penalty
$\lambda=6$; mapped to Ising fields
$h_i=a_i/2-\lambda d_{\mathrm{RRG}}/4$ and edge couplings
$J_{ij}=-\lambda/4$ on graph edges. \\

MIS-LN
& Maximum independent set with log-normal vertex weights
$a_i=\exp(\sigma_A Z_i)>0$, where $Z_i\sim\mathcal{N}(0,1)$ i.i.d.,
and penalty $\lambda=\max_i a_i+5$; mapped to Ising fields
$h_i=a_i/2-\lambda d_{\mathrm{RRG}}/4$ and edge couplings
$J_{ij}=-\lambda/4$ on graph edges. \\

SK
& The Sherrington--Kirkpatrick model on a complete graph 
with the $J$-couplings defined the same as for the reduced SK model.\\
\bottomrule
\end{tabular}
\end{table}

%+++++++++++++++++++++++++++++++++++++++++++++++++++++++++++++++++++++++++++++++++++++++++
\subsection*{Systematic size-dependent benchmark families}
Six benchmark families were simulated for every integer system size from
$n_b=18$ through $29$. For each family and size, 32 independent problem
instances were generated. Classical and QIPS proposal searches were applied
to the same instances using identical outer-loop search protocols and equal
proposal budgets. These six families are listed in
Supplementary Table~\ref{tab:supp_systematic_families} and correspond to the
systems summarized in Table~1 of the main text.

\begin{table}[hb!]
\centering
\caption{\textbf{Systematic benchmark families used for size-dependent
analysis.} Each family was simulated for $n_b=18,\ldots,29$ using
32 matched instances per size. Both classical and quantum proposals 
are considered for direct comparisons on identical problem cases for all 
these system types and sizes.}
\label{tab:supp_systematic_families}
\begin{tabular}{@{}clll@{}}
\toprule
Index & Graph topology & Problem or coupling model & Size range \\
\midrule
1 & 3-RRG    & Ising, 70\% AFM / 30\% FM & $18$--$29$ \\
2 & 3-RRG    & MaxCut, uniform weights    & $18$--$29$ \\
3 & 3-RRG    & MIS, uniform weights       & $18$--$29$ \\
4 & 6-RRG    & MaxCut, log-normal weights & $18$--$29$ \\
5 & Complete & Exponential weak quenched disorder & $18$--$29$ \\
6 & Complete & Sherrington--Kirkpatrick model & $18$--$29$ \\
\bottomrule
\end{tabular}
\end{table}

Here, $k$-RRG denotes a random regular graph of degree $k$. These six
families were selected for systematic analysis because they jointly sample
sparse constraint problems, weighted graph partitioning, and dense frustrated
spin systems while remaining suitable for matched classical and quantum
calculations over the full size range.

%+++++++++++++++++++++++++++++++++++++++++++++++++++++++++++++++++++++++++++++++++++++++++
\FloatBarrier
\subsection*{Additional benchmark families}
The breadth of the systematic results was tested using 30 additional
graph--coupling combinations. Sparse random graphs were examined with degrees
3, 5, and 6; structured graphs included a one-dimensional ring, square lattice
and triangular lattice; and complete graphs supplied dense controls. Most
additional random-graph and complete-graph systems were simulated at
$n_b=24$ and $26$, whereas the structured lattices and 5-RRG systems were
evaluated at $n_b=24$. This size allowed several graph and lattice topologies
with different coordination structures to be compared at the same number of
binary variables.

For the one-dimensional ring, periodic boundary conditions were used. For the
square lattice, periodic boundary conditions were applied in the $y$-direction
with $L_y=3$, and fixed boundary conditions were applied in the $x$-direction
with $L_x=8$, forming a $3\times 8$ grid. For the triangular lattice,
periodic boundary conditions were applied in the $y$-direction with $L_y=4$,
and fixed boundary conditions were applied in the $x$-direction with $L_x=6$,
forming a $4\times 6$ grid. Each case used 32 matched instances for the
classical and QIPS proposal searches.

\begin{longtable}{@{}p{0.20\textwidth}p{0.56\textwidth}p{0.13\textwidth}@{}}
\caption{\textbf{Additional benchmark families used to test system
dependence.}}
\label{tab:supp_additional_families}\\
\toprule
Graph topology & Coupling or problem variants & System sizes \\
\midrule
\endfirsthead

\multicolumn{3}{l}{\textit{Supplementary Table
\ref{tab:supp_additional_families} continued}}\\
\toprule
Graph topology & Coupling or problem variants & System sizes \\
\midrule
\endhead

3-RRG
& 70A/30F+GQD; EWQD; reduced-SK; MaxCut with log-normal weights;
MIS with log-normal weights
& 24, 26 \\

6-RRG
& 70A/30F+GQD; EWQD; reduced-SK; 70A/30F;
MaxCut with uniform weights; MIS with uniform weights;
MIS with log-normal weights
& 24, 26 \\

Complete graph
& 70A/30F+GQD; 70A/30F
& 24, 26 \\

1D ring
& 70A/30F+GQD; EWQD; reduced-SK; 70A/30F
& 24 \\

5-RRG
& 70A/30F+GQD; EWQD; reduced-SK; 70A/30F
& 24 \\

Triangular lattice
& 70A/30F+GQD; EWQD; reduced-SK; 70A/30F
& 24 \\

Square lattice
& 70A/30F+GQD; EWQD; reduced-SK; 70A/30F
& 24 \\
\bottomrule
\end{longtable}

The additional systems were used as robustness checks and were not included 
in the aggregated main-text statistics. Their purpose was to test whether 
benchmark families outside the six systematic systems exhibited qualitatively 
different QIPS behavior. No such differences were observed: all simulated 
systems showed the same general search characteristics. Thus, references in 
the main text to behavior shared across the benchmark set also include the 
systems and sizes listed in Supplementary Table~\ref{tab:supp_additional_families}.

%+++++++++++++++++++++++++++++++++++++++++++++++++++++++++++++++++++++++++++++++++++++++++
\subsection*{Matched search protocol}

For the systematic and additional search datasets, each run contained
$n_b$ rounds with 20 proposal steps per round. A QIPS proposal step consisted
of one circuit sampled with 100 shots, giving
\begin{equation}
N_{\mathrm{prop}}
=
20n_b\times100
=
2000n_b
\end{equation}
proposal evaluations per instance. The matched classical search used the
same total number of proposal evaluations. Classical and QIPS searches were
performed on identical QUBO instances so that differences arose from the
proposal mechanism rather than the problem ensemble.

Blind-search controls used uniformly sampled bitstrings. Limited hybrid
controls, in which classical and QIPS proposal steps were alternated under
the same total budget, were also generated for selected systems. These
controls were used diagnostically and were not included in the principal
matched classical--QIPS aggregates.

%+++++++++++++++++++++++++++++++++++++++++++++++++++++++++++++++++++++++++++++++++++++++++
\subsection*{Fixed-seed proposal-rank ensembles}

A separate data collection was generated to characterize the conditional
proposal distribution $P(r_p\mid r_s)$ independently of the evolving search
frontier. The same six systematic benchmark families were examined at
\begin{equation}
n_b\in\{12,16,20,24,28\},
\end{equation}
using the seven reference seed ranks
\begin{equation}
r_s\in\{1,4,15,64,261,1024,4087\}.
\end{equation}
For every combination of benchmark family, system size, seed rank, and
proposal mechanism, 20 matched problem instances were used. Each instance
was sampled using 100 independently generated proposal distributions with
2,000 samples per distribution. Thus, each system--size--seed case contained
\begin{equation}
20\times100\times2000
=
4.0\times10^6
\end{equation}
proposal samples for each proposal mechanism. Across the six systems,
seven seed ranks, and five sizes, this produced $8.4\times10^8$ classical
and $8.4\times10^8$ QIPS proposal samples.

The reference ranks are fixed while $n_b$ varies. Consequently,
$\log_2(r_s)/n_b$ decreases with increasing system size, as shown in
Supplementary Table~\ref{tab:supp_seed_coordinates}. This allows the data to
probe the low-rank ``end-game'' regime in which a fixed-rank seed occupies an
exponentially smaller fraction of Hilbert space.

\begin{table}[ht]
\centering
\caption{\textbf{Normalized logarithmic coordinates of the fixed reference
seed ranks.}}
\label{tab:supp_seed_coordinates}
\begin{tabular}{@{}rrrrrr@{}}
\toprule
$r_s$ & $n_b=12$ & $n_b=16$ & $n_b=20$ & $n_b=24$ & $n_b=28$ \\
\midrule
1    & 0      & 0      & 0      & 0      & 0      \\
4    & 0.1667 & 0.1250 & 0.1000 & 0.0833 & 0.0714 \\
15   & 0.3256 & 0.2442 & 0.1953 & 0.1628 & 0.1395 \\
64   & 0.5000 & 0.3750 & 0.3000 & 0.2500 & 0.2143 \\
261  & 0.6690 & 0.5017 & 0.4014 & 0.3345 & 0.2867 \\
1024 & 0.8333 & 0.6250 & 0.5000 & 0.4167 & 0.3571 \\
4087 & 0.9997 & 0.7498 & 0.5998 & 0.4999 & 0.4285 \\
\bottomrule
\end{tabular}
\end{table}

\paragraph{Remark on the normalized logarithmic rank scale.}
The normalized logarithmic coordinate provides a useful intuition for the
difficulty of reaching the low-energy tail as $n_b$ increases. For example,
under blind search with $N_{\mathrm{draw}}=10{,}000$ independent draws, the
typical best rank is on the order of $2^{n_b}/N_{\mathrm{draw}}$. Thus the
corresponding normalized logarithmic coordinate is approximately
\begin{equation}
\frac{\log_2(2^{n_b}/N_{\mathrm{draw}})}{n_b}
=
1-\frac{\log_2 N_{\mathrm{draw}}}{n_b}.
\end{equation}
For $n_b=15$, this gives
$1-\log_2(10{,}000)/15 \simeq 0.114$, whereas for $n_b=29$ it gives
$1-\log_2(10{,}000)/29 \simeq 0.542$. Thus the same finite sampling budget
moves from being close to the ground-state end of the spectrum at small
system size to being much farther from it at larger system size. This simple
scaling illustrates why algorithms can appear disproportionately effective for
small binary optimization problems: random chance, or weakly structured
exploration, can still reach very low ranks. As the Hilbert space grows
exponentially, that accidental productivity rapidly decreases, making
rank-resolved scaling essential for interpreting performance on combinatorial
optimization problems.

%+++++++++++++++++++++++++++++++++++++++++++++++++++++++++++++++++++++++++++++++++++++++++
\subsection*{Localized--extended CDF decomposition dataset}

A third collection was generated for detailed analysis of the localized and
extended components of QIPS proposal distributions. These calculations used
the SK model and 120 independently randomized circuits for each selected
system-size and seed-rank combination. Circuits were retained for decomposition
only when they satisfied the localization criterion defined in 
Supplementary Note~10. The retained distributions were used to construct 
the seed, localized,
and extended components discussed in Supplementary Note~7.

%%%%%%%%%%%%%%%%%%%%%%%%%%%%%%%%%%%%%%%%%%%%%%%%%%%%%%%%%%%%%%%%%%%%%%%%%%%%%%%% 
%\clearpage
\vspace{0.6cm}
\refstepcounter{table}
\label{tab:supp_decomposition_scope}

\noindent
\textbf{\tablename~\thetable. 
Datasets used for the proposal-distribution decomposition.}
The listed seed-rank and system-size combinations were analyzed for the SK model.
For $n_b=20$, the same seed-rank cases were also checked across all six systematic
benchmark families to test whether the qualitative CDF structure and approximate
spectral symmetry observed for SK persist across problem classes.

\vspace{1em}

\begin{center}
\begin{tabular}{ll}
\toprule
Seed rank & System size(s) \\
\midrule
$2^7$              & 20 \\
$2^8$              & 18 \\
$2^{10}=1024$      & 18, 20, 24, 28 \\
$2^{13}$           & 20, 23 \\
$2^{16}$           & 20 \\
$2^{18}$           & 28 \\
$2^{19}$           & 20 \\
$2^{20}-2^{16}$    & 20 \\
$2^{20}-2^{13}$    & 20 \\
$2^{20}-2^{10}$    & 20 \\
$2^{20}-2^7$       & 20 \\
\bottomrule
\end{tabular}
\end{center}
%%%%%%%%%%%%%%%%%%%%%%%%%%%%%%%%%%%%%%%%%%%%%%%%%%%%%%%%%%%%%%%%%%%%%%%%%%%%%%%%

The collection includes seeds on both sides of the energy spectrum and several
system sizes at fixed seed rank. It therefore supports the analyses of
seed-rank dependence, approximate spectral symmetry, system-size dependence,
and the compact representation of the extended proposal component. The detailed
figures are shown primarily for the SK model, but the $n_b=20$ cases in
Supplementary Table~\ref{tab:supp_decomposition_scope} were also checked
across all six systematic benchmark families. These comparisons showed the
same qualitative proposal-rank structure, supporting the use of SK as the
representative system for the detailed decomposition analysis.

%%%%%%%%%%%%%%%%%%%%%%%%%%%%%%%%%%%%%%%%%%%%%%%%%%%%%%%%%%%%%%%%%%%%%%%%%%%%%%%
%%%%%%%%%%%%%%%%%%%%%%%%%%%%%%%%%%%%%%%%%%%%%%%%%%%%%%%%%%%%%%%%%%%%%%%%%%%%%%%
\FloatBarrier
\suppnote{2}{System-resolved proposal-search results}
{supp:SystemResolvedSearch}

Supplementary Figure~\ref{fig:topK6systems} resolves the aggregate top-$K$ recovery results
from the main text by benchmark family. For each system, the curves are averaged
over 32 matched problem instances. The system-resolved results reproduce the same 
qualitative trends observed in the aggregate data,
while revealing differences in recovery across benchmark families.

\begin{figure}[hbt]
    \centering
    \includegraphics[
        width=0.90\linewidth,
        trim={0.01in 0.00in 0.00in 0.00in},      % trim={left bottom right top}
        clip
    ]{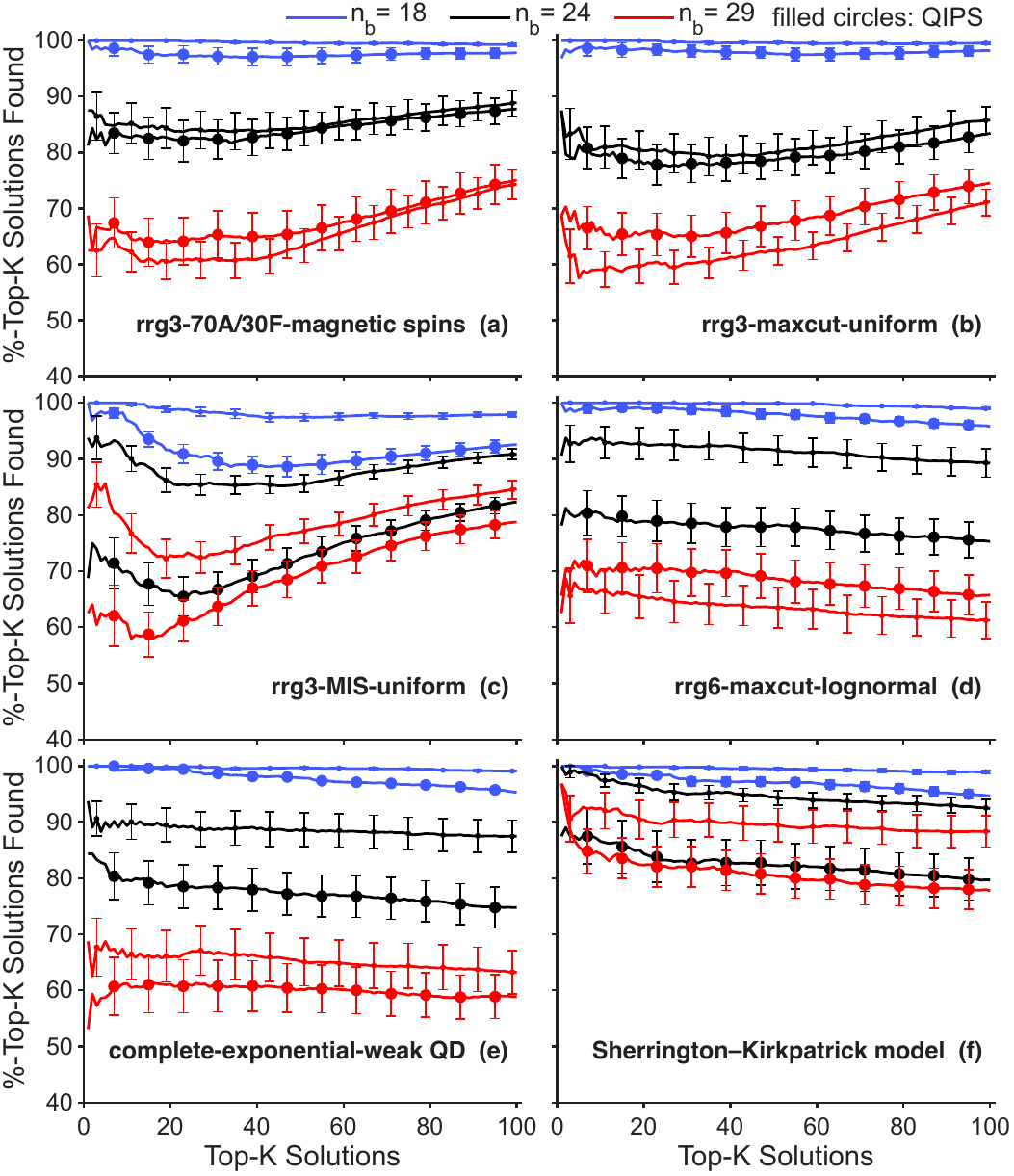}
    \caption{\textbf{System-resolved recovery of the top-$K$ solutions.}
    Percentage of the true top-$K$ lowest-energy states recovered by the
    matched classical and QIPS searches for $n_b=18$, 24, and 29.
    Results are averaged over 32 matched instances for each benchmark family
    and system size. Classical and QIPS searches use identical QUBO instances,
    outer-loop search protocols, and total proposal budgets. Open symbols denote
    the classical search, and filled circles denote QIPS. Panels correspond to
    \textbf{a}, 3-RRG Ising with 70\% antiferromagnetic and 30\% ferromagnetic
    couplings;
    \textbf{b}, 3-RRG MaxCut with uniform weights;
    \textbf{c}, 3-RRG maximum independent set with uniform weights and penalties;
    \textbf{d}, 6-RRG MaxCut with log-normal weights;
    \textbf{e}, complete-graph Ising with exponential weak quenched disorder; and
    \textbf{f}, the Sherrington--Kirkpatrick model.
    Error bars denote the standard error of the mean.}
    \label{fig:topK6systems}
\end{figure}

The following heat maps show the 
classical and quantum conditional proposal-rank densities for the six benchmark 
families listed in Supplementary Table~2. The blind-search heat map is not shown
again because uniform sampling is independent of the system and seed state.

\begin{figure}[p]
    \centering

    \begin{minipage}{0.90\textwidth}
        \centering
        \includegraphics[width=\linewidth]
        {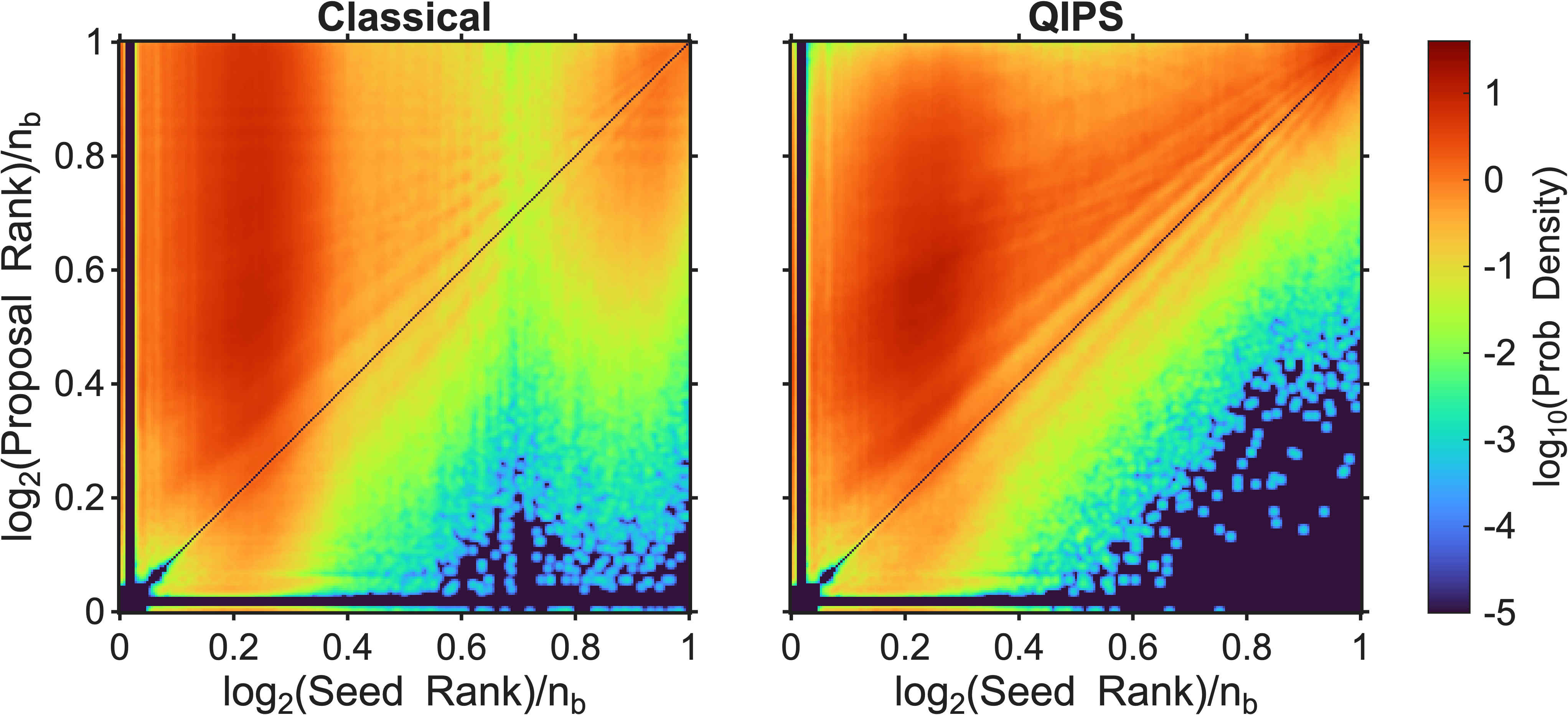}

        \smallskip
        \textbf{(a)} 3-RRG Ising model with 70\% antiferromagnetic
        and 30\% ferromagnetic couplings.
    \end{minipage}

    \vspace{0.5em}

    \begin{minipage}{0.90\textwidth}
        \centering
        \includegraphics[width=\linewidth]
        {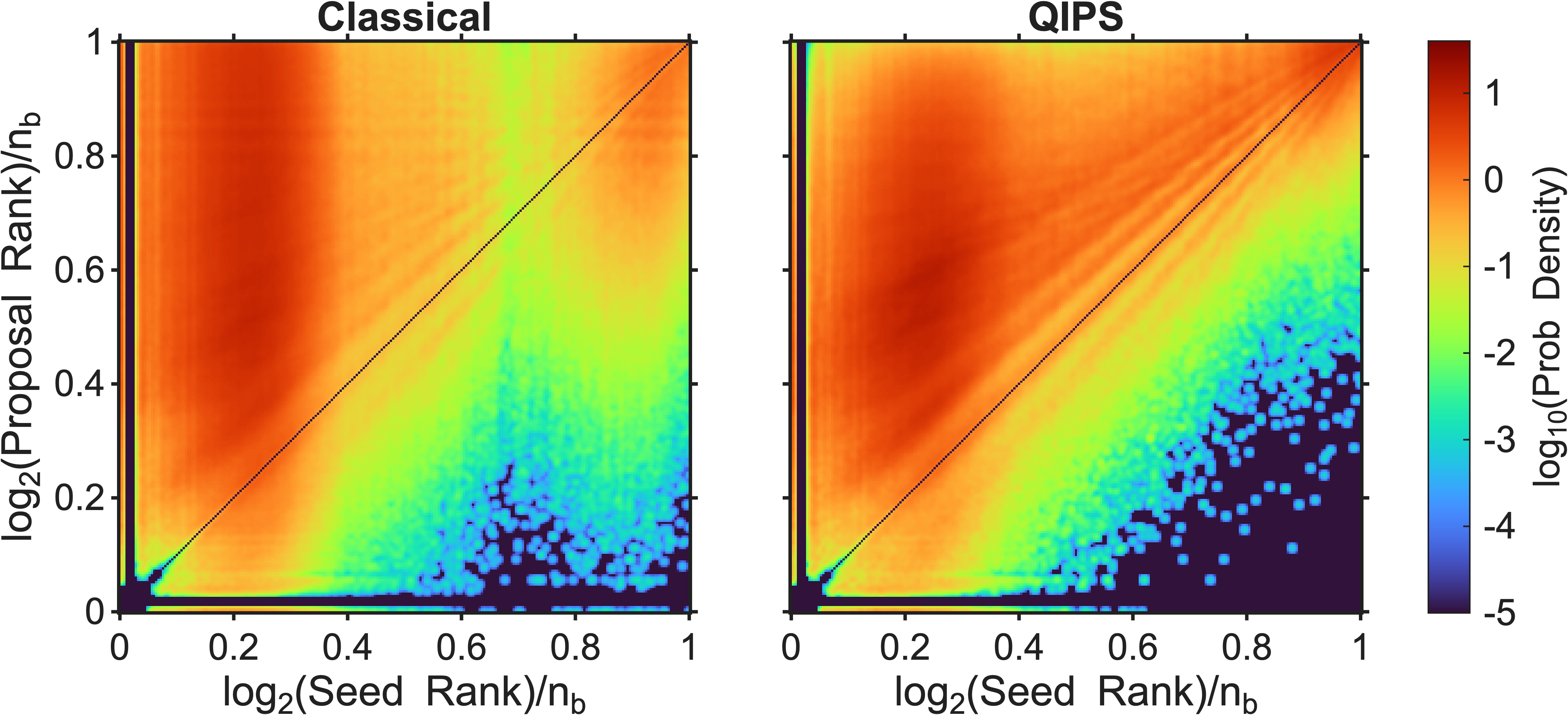}

        \smallskip
        \textbf{(b)} 3-RRG MaxCut with uniform edge weights.
    \end{minipage}

    \vspace{0.5em}

    \begin{minipage}{0.90\textwidth}
        \centering
        \includegraphics[width=\linewidth]
        {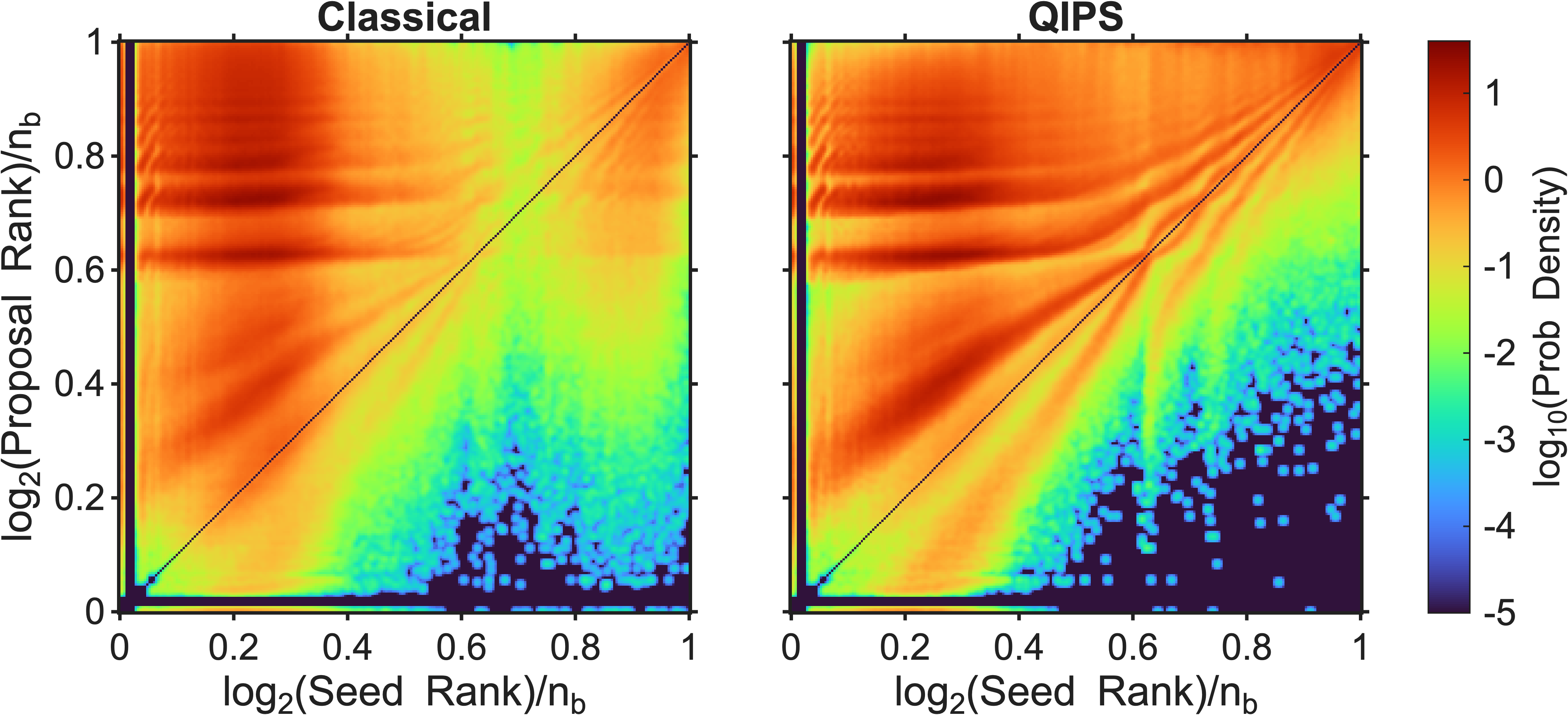}

        \smallskip
        \textbf{(c)} 3-RRG maximum independent set with uniform
        weights and penalties.
    \end{minipage}

    \caption{\textbf{System-resolved conditional proposal-rank densities
    for benchmark families 1--3.}
    Each row compares the classical proposal generator with QIPS.
    The horizontal and vertical coordinates are the normalized logarithmic
    seed and proposal ranks, respectively. Color indicates the base-10
    logarithm of the empirical conditional probability density.
    The diagonal corresponds to proposals having the same rank as the seed;
    points below the diagonal represent rank improvement.}
    \label{fig:condRankMaps_1to3}
\end{figure}

\begin{figure}[p]
    \centering

    \begin{minipage}{0.90\textwidth}
        \centering
        \includegraphics[width=\linewidth]
        {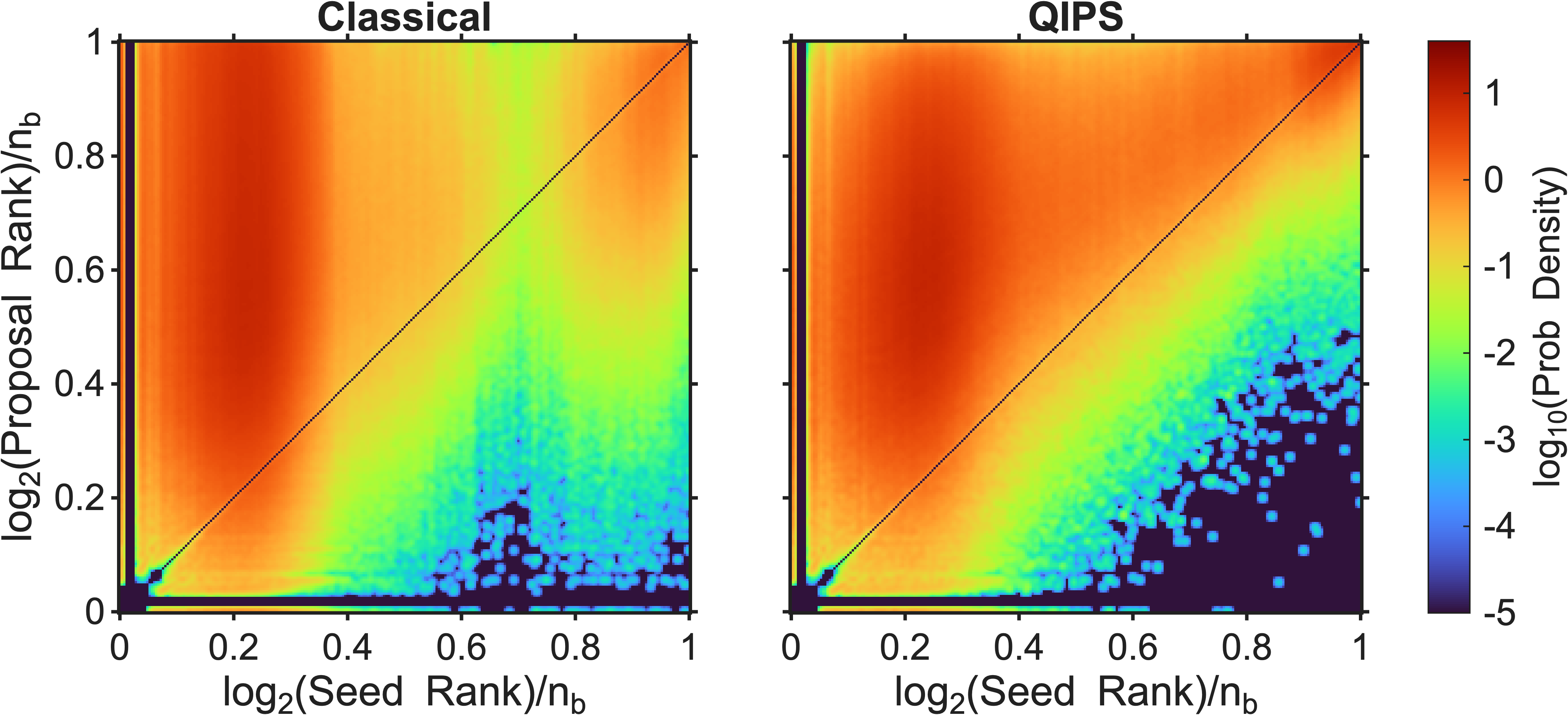}

        \smallskip
        \textbf{(a)} 3-RRG Ising model with 70\% antiferromagnetic
        and 30\% ferromagnetic couplings.
    \end{minipage}

    \vspace{0.5em}

    \begin{minipage}{0.90\textwidth}
        \centering
        \includegraphics[width=\linewidth]
        {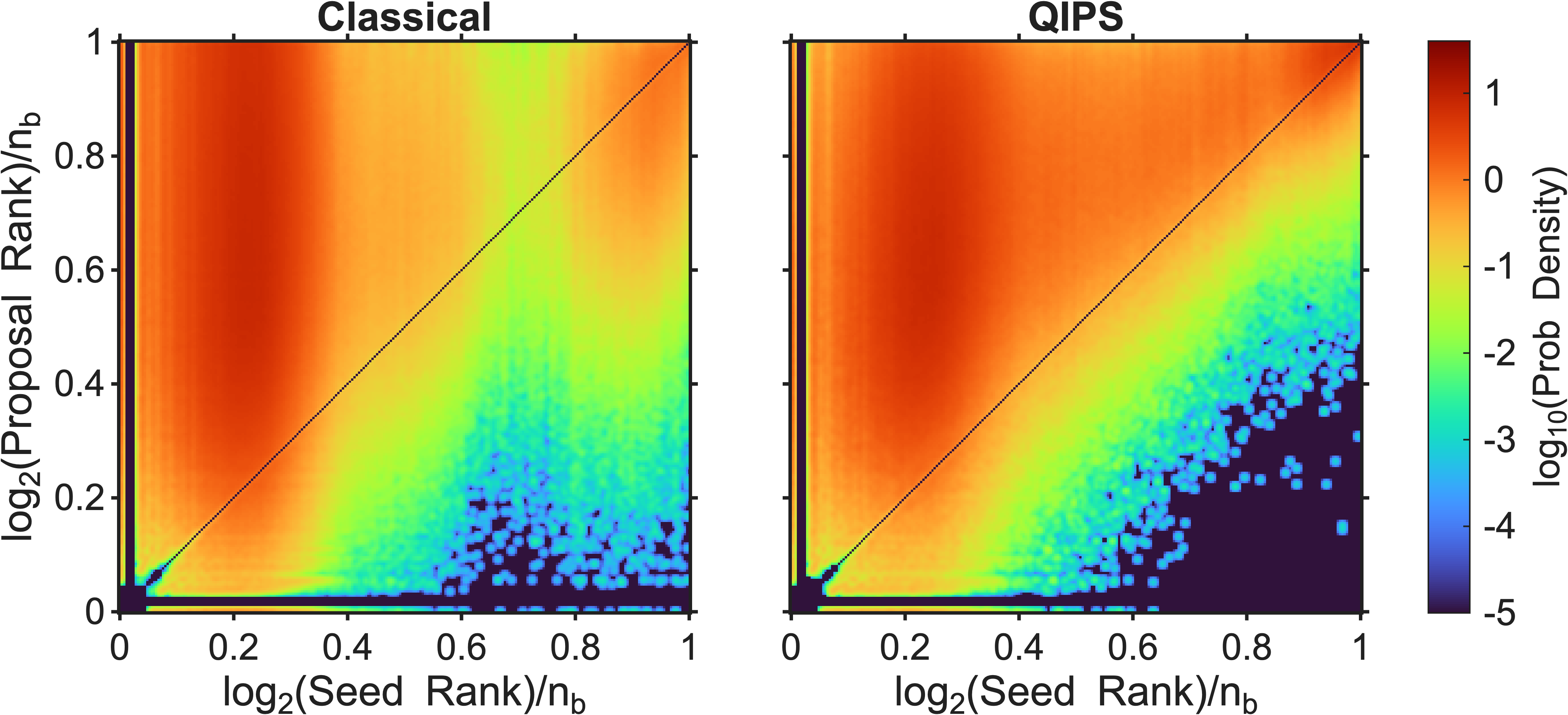}

        \smallskip
        \textbf{(b)} 3-RRG MaxCut with uniform edge weights.
    \end{minipage}

    \vspace{0.5em}

    \begin{minipage}{0.90\textwidth}
        \centering
        \includegraphics[width=\linewidth]
        {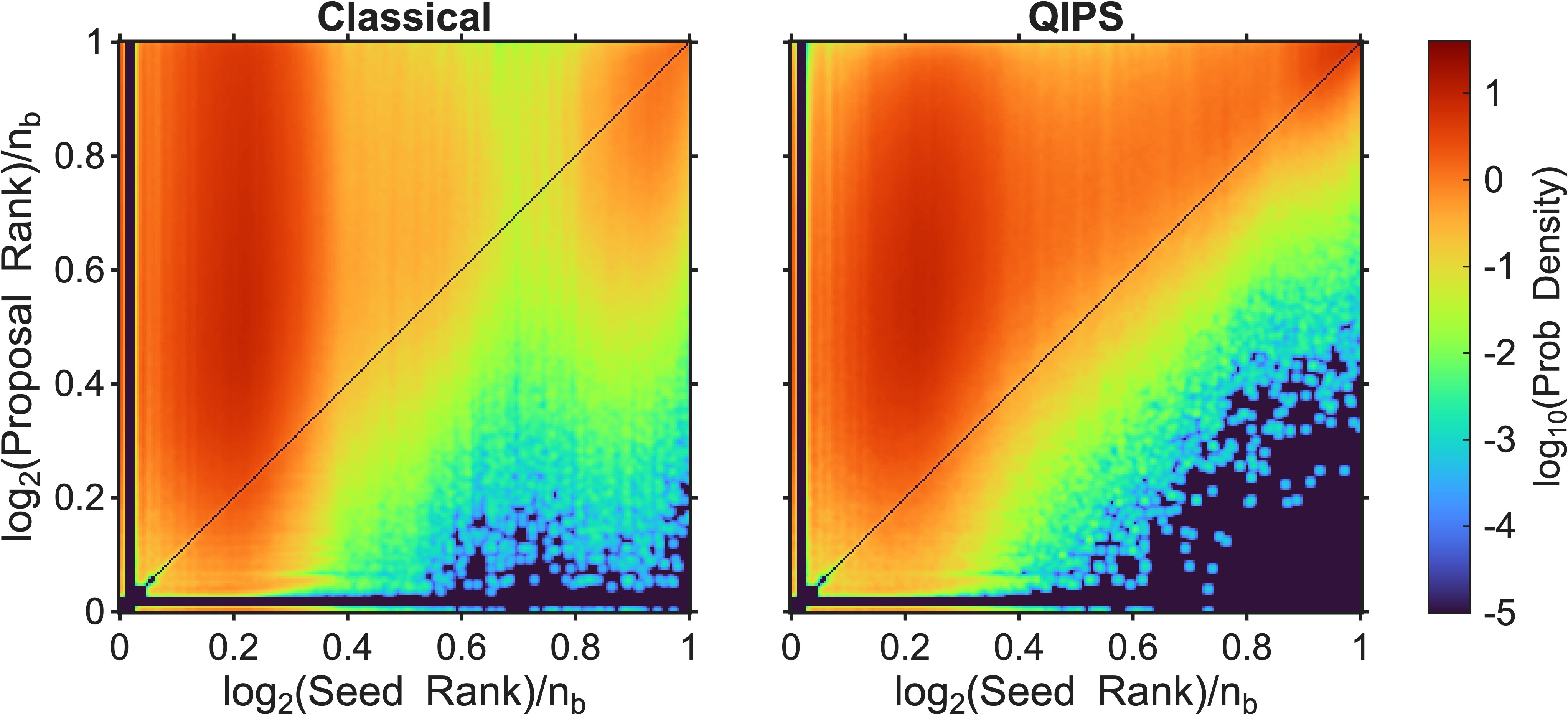}

        \smallskip
        \textbf{(c)} 3-RRG maximum independent set with uniform
        weights and penalties.
    \end{minipage}

    \caption{\textbf{System-resolved conditional proposal-rank densities
    for benchmark families 4--6.}
    Each row compares the classical proposal generator with QIPS.
    Coordinates, color scale, and diagonal interpretation are the same as in
    Supplementary Figure~\ref{fig:condRankMaps_1to3}.}
    \label{fig:condRankMaps_1to3}
\end{figure}

%%%%%%%%%%%%%%%%%%%%%%%%%%%%%%%%%%%%%%%%%%%%%%%%%%%%%%%%%%%%%%%%%%%%%%%%%%%%%%%
%%%%%%%%%%%%%%%%%%%%%%%%%%%%%%%%%%%%%%%%%%%%%%%%%%%%%%%%%%%%%%%%%%%%%%%%%%%%%%%
\FloatBarrier
\suppnote{3}{Finite-shot search trajectory analysis}
{supp:FiniteShotSearchDynamics}
Results for a representative QIPS run on the $n_b=29$ Sherrington--Kirkpatrick 
(SK) model are shown in Supplementary Figure~\ref{fig:SKexample}. The qualitative 
behavior is consistent across systems and sizes, although extensive degeneracies 
produce visible plateau structure in some problem classes. The discussion therefore
focuses on generic features shared across the benchmark set. Also note that
Supplementary Figure~\ref{fig:SKexample}d is copied from Figure~2 of the main
text to facilitate convenient comparisons. 

Supplementary Figure~\ref{fig:SKexample}a shows that the mean excess energy 
saturates rather than approaching zero. This is expected because QIPS does not 
optimize the mean energy. 
Progress instead depends on the lower-energy tail of the proposal distribution: 
even a small but finite probability mass below the seed energy can generate improved
states reliably with a modest number of shots. Localized quantum interference 
patterns are therefore essential, because they concentrate enough probability on a 
limited set of measurable outcomes for these low-energy proposals to remain 
statistically accessible.

\begin{figure}[!hbt]
    \centering
    \includegraphics[
        width=0.98\linewidth,
        trim={0.00in 0.00in 0.00in 0.00in},      % trim={left bottom right top}
        clip
    ]{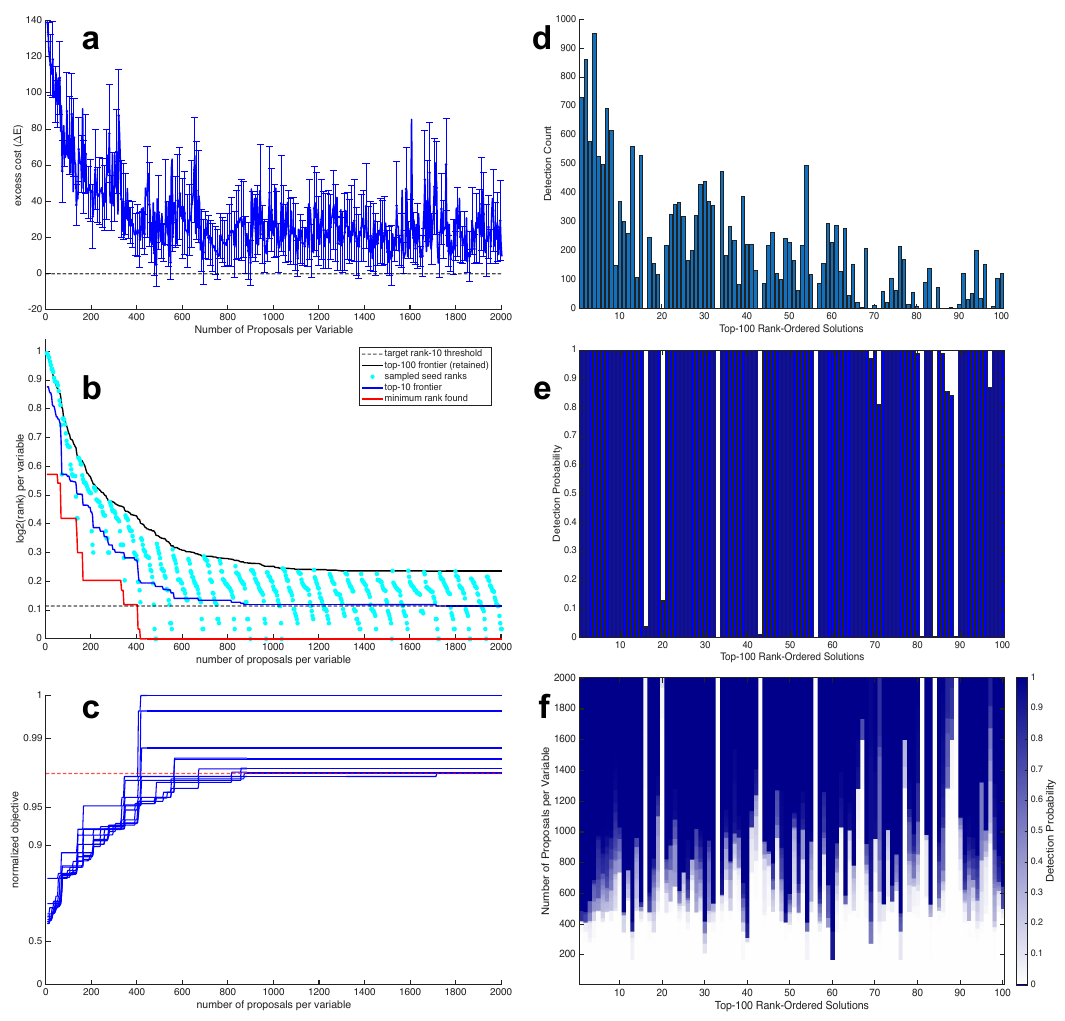}
     \caption{\textbf{Typical single-run behavior for the 29-spin SK model.}
     \textbf{a}, The evolution of the mean excess energy and standard error per 100-shots is shown. 
      \textbf{b}, The evolution of the top-100, top-10 and top-1 elite frontiers are shown. 
       The ground-truth top-10 threshold and the ranks of seed states 
       selected from the top-100 frontier are also shown. Outer-frontier seeds are sampled 
       using an exponentially decaying rank weight favoring lower ranks, with the decay scaled 
       to place a fixed probability $q_{\mathrm{tail}}=0.05$ beyond the frontier; this overflow 
       probability is redistributed uniformly over the frontier.
       \textbf{c}, Evolution of the normalized objective values for the ten states in the top-10 
       frontier; the dashed line marks the ground-truth threshold.
       \textbf{d}, Number of times each of the 100 lowest-energy states is proposed during the search.
       \textbf{e}, Final detection probabilities for these states after accumulating contributions
       from all sampled circuits.
       \textbf{f}, Evolution of the detection probabilities with proposal budget, showing the 
       intermittent acquisition of low-energy states during the search.}
    \label{fig:SKexample}
\end{figure}

Supplementary Figure~\ref{fig:SKexample}b shows the evolution of the elite frontier 
and the best rank found. Progress is rapid initially and then slows as the frontier 
approaches the low-energy tail, where new improvements become increasingly rare. The 
frontier retains the 100 best states found, so perfect recovery would make it 
coincide with the true top-100 spectrum.

The outer frontier collects productive states, including rare intermittent
low-energy proposals, which can become seeds for subsequent circuits. As the
frontier improves, lower-ranked seeds condition the search toward proposals
with statistically significant peaks at still lower energies. Only a subset of
the retained states is sampled as seeds, preserving diversity without spending
resources on every frontier member. The frontier size therefore controls the
exploration--exploitation trade-off: larger frontiers improve coverage, whereas
smaller frontiers concentrate effort on the best-known region.

A key scaling question is whether this mechanism continues to produce localized
interference patterns with statistically significant low-energy peaks after the
frontier accumulates near the ground state, or whether progress eventually
stalls. The present system sizes provide evidence against immediate stalling,
but are not large enough to determine the asymptotic behavior. Supplementary
Notes~5 and~6 therefore examine the proposal structure generated
by low-ranked seeds in more detail.

Supplementary Figure~\ref{fig:SKexample}c shows the evolution of the top-10 frontier 
on a nonlinear objective scale that resolves proximity to the ground state. In this 
representative run, all top-10 states are recovered. Although complete recovery 
is not typical, QIPS shows consistent low-energy performance across systems and 
sizes (shown in Figure~1b, main text). This consistency is supported by repeated 
sampling of low-energy states, as demonstrated by the empirical detection counts 
across the true top-100 states in Supplementary Figure~\ref{fig:SKexample}d.

Supplementary Figure~\ref{fig:SKexample}e gives the cumulative detection probabilities 
associated with these states. Most approach unity, whereas a small number remain effectively 
inaccessible during this run, producing transient blind spots. Detection probability 
differs from the probability assigned by any single circuit. If $p_s(k)$ is the 
probability of measuring state $s$ from circuit $k$, then after 100 shots per circuit,

\begin{equation}
P_D(s)=1-\prod_k\left[1-p_s(k)\right]^{100}.
\end{equation}
As Supplementary Figure~\ref{fig:SKexample}f shows, $P_D(s)$ can remain near zero for many 
circuits and then rise sharply when one seed-conditioned localized interference pattern assigns 
appreciable probability to that state. QIPS therefore succeeds through intermittent but 
statistically significant low-energy proposals accumulated across the circuit ensemble.

%%%%%%%%%%%%%%%%%%%%%%%%%%%%%%%%%%%%%%%%%%%%%%%%%%%%%%%%%%%%%%%%%%%%%%%%%%%%%%%
%%%%%%%%%%%%%%%%%%%%%%%%%%%%%%%%%%%%%%%%%%%%%%%%%%%%%%%%%%%%%%%%%%%%%%%%%%%%%%%
\FloatBarrier
\suppnote{4}{Evaluation of quantum interference patterns}
{supp:LocalizationCriteria}
Supplementary Figure~\ref{fig:QIPfiltering} summarizes the classification of
120 quantum interference patterns generated for the fixed seed considered in
the main text. Each circuit is labeled as acceptable or poor according to the
two-parameter localization objective function (also called pseudo-energy 
landscape) used by the feedback 
controller, which is defined in Supplementary Note~10. This classification is 
based on the empirical 100-shot measurement record associated with each circuit.

\begin{figure}[!htb]
    \centering
    \includegraphics[
        width=0.85\linewidth,
        trim={0.01in 0.00in 0.00in 0.00in},
        clip
    ]{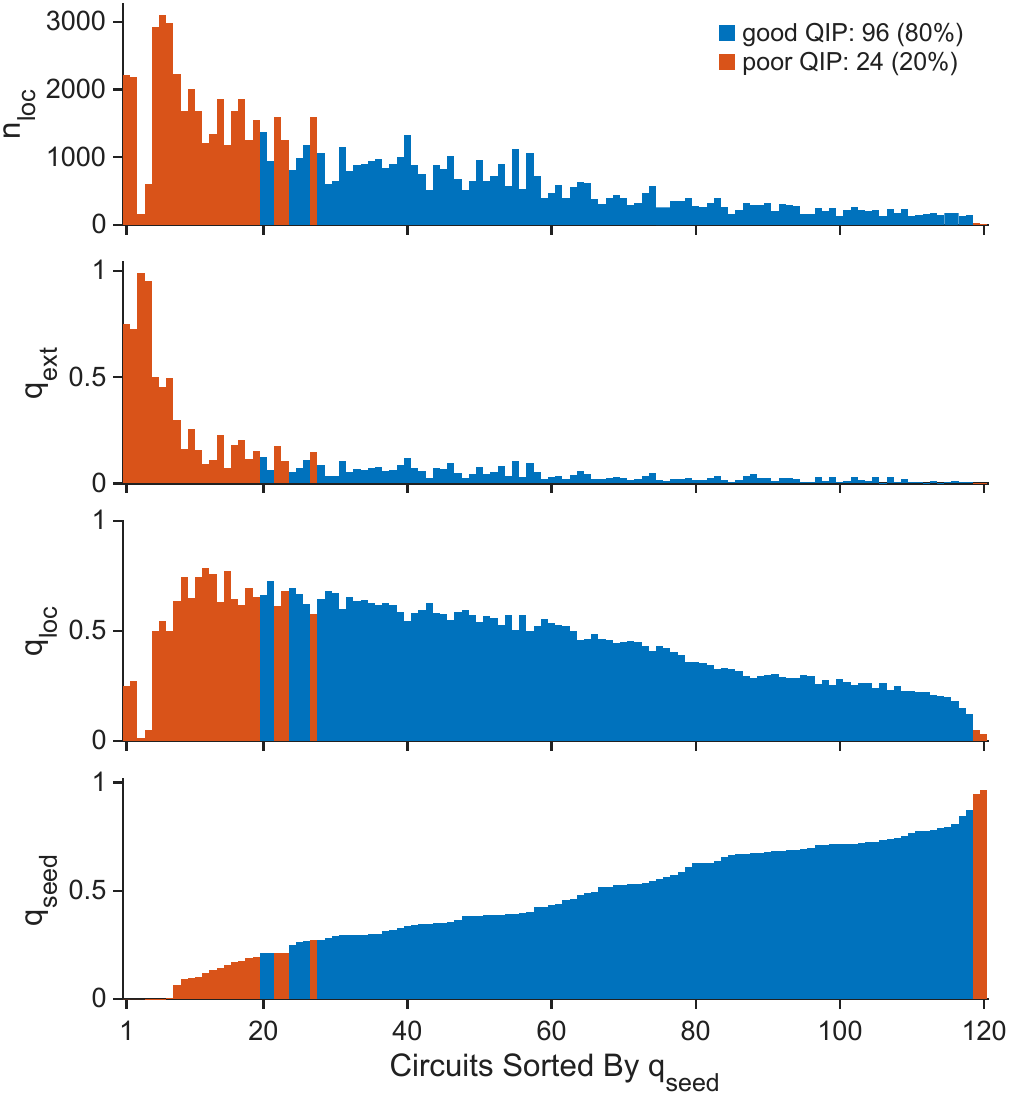}
    \caption{\textbf{Selection of localized quantum interference patterns.}
    Properties of 120 circuits sorted by increasing seed-state probability
    $q_{\mathrm{seed}}$. From bottom to top, the panels show
    $q_{\mathrm{seed}}$, the localized probability $q_{\mathrm{loc}}$, the
    extended probability $q_{\mathrm{ext}}$, and the number of localized states
    $n_{\mathrm{loc}}$. States are classified as localized when their exact
    computational-basis probability exceeds
    $p_{\mathrm{loc}}=5\times10^{-5}$. Blue bars denote circuits accepted by the
    feedback localization criterion, and orange bars denote rejected circuits.
    Of the 120 circuits, 96 are accepted and 24 are rejected. Rejected circuits
    occur both when the distribution is too extended and when it is excessively
    concentrated on the seed state.}
    \label{fig:QIPfiltering}
\end{figure}

For the present analysis, the exact computational-basis probabilities are 
used to calculate the probability assigned to the seed state, $q_{\mathrm{seed}}$,
and to the localized and extended components, $q_{\mathrm{loc}}$ and
$q_{\mathrm{ext}}$. States with probability
\begin{equation}
p_s > p_{\mathrm{loc}},
\qquad
p_{\mathrm{loc}}
=
\frac{0.005}{N_{\mathrm{shots}}}
=
5\times10^{-5},
\end{equation}
for $N_{\mathrm{shots}}=100$, are assigned to the localized component. The
quantity $n_{\mathrm{loc}}$ denotes the number of states above this threshold.
The same threshold is used in the decomposition shown in
Figure~3a,b,d.
It is worth emphasizing that the exact quantities $q_{\mathrm{loc}}$, $q_{\mathrm{ext}}$, 
and $n_{\mathrm{loc}}$ are used only for post hoc characterization and do not 
enter the feedback-based circuit classification.

The circuits are ordered by increasing $q_{\mathrm{seed}}$. Small values of
$q_{\mathrm{seed}}$ generally correspond to extended patterns, whereas larger
values indicate stronger localization around the seed. Excessive localization
is also undesirable because the limiting case
$q_{\mathrm{seed}}\rightarrow1$ suppresses exploration of new states.
Consequently, the feedback objective selects an intermediate regime that
retains substantial seed localization while preserving a finite set of
additional measurable outcomes.

%%%%%%%%%%%%%%%%%%%%%%%%%%%%%%%%%%%%%%%%%%%%%%%%%%%%%%%%%%%%%%%%%%%%%%%%%%%%%%%%%%%%%%%%
%%%%%%%%%%%%%%%%%%%%%%%%%%%%%%%%%%%%%%%%%%%%%%%%%%%%%%%%%%%%%%%%%%%%%%%%%%%%%%%%%%%%%%%%
\FloatBarrier
\suppnote{5}{Illustrating top-100 mean state probabilities}
{supp:Top100AveStateProb}

The following eight figures examine how QIPS distributes probability among the true
top-100 energy-ranked states for individual Sherrington--Kirkpatrick problem
instances. Each panel shows the base-10 logarithm of the mean seed-excluded
state probability, averaged over the indicated ensemble of circuits. No
averaging over problem instances is performed. All generated circuits are
included, irrespective of whether their interference patterns satisfy the
localization criterion, thereby reproducing the circuit ensemble encountered
during an actual search. The number of contributing circuits, $N_c$, and the
total mean probability assigned to the displayed top-100 states are reported
beside each panel. When a single seed rank is used, it is also indicated.
In summary, the strongest conclusion supported by the full series of eight
figures shown below is that low-ranked seed diversity over an ensemble
of circuits, not requiring any privileged seed or circuit, produces broad 
top-100 coverage.

%$$$$$$$$$$$$$$$$$$$$$$$$$$$$$$$$$$$$$$$$$$$$$$$$$$$$$$$$$$$$$$
\begin{figure}[!h]
\centering
\includegraphics[
width=0.82\linewidth,
trim={0.01in 0.00in 0.00in 0.00in},
clip
]{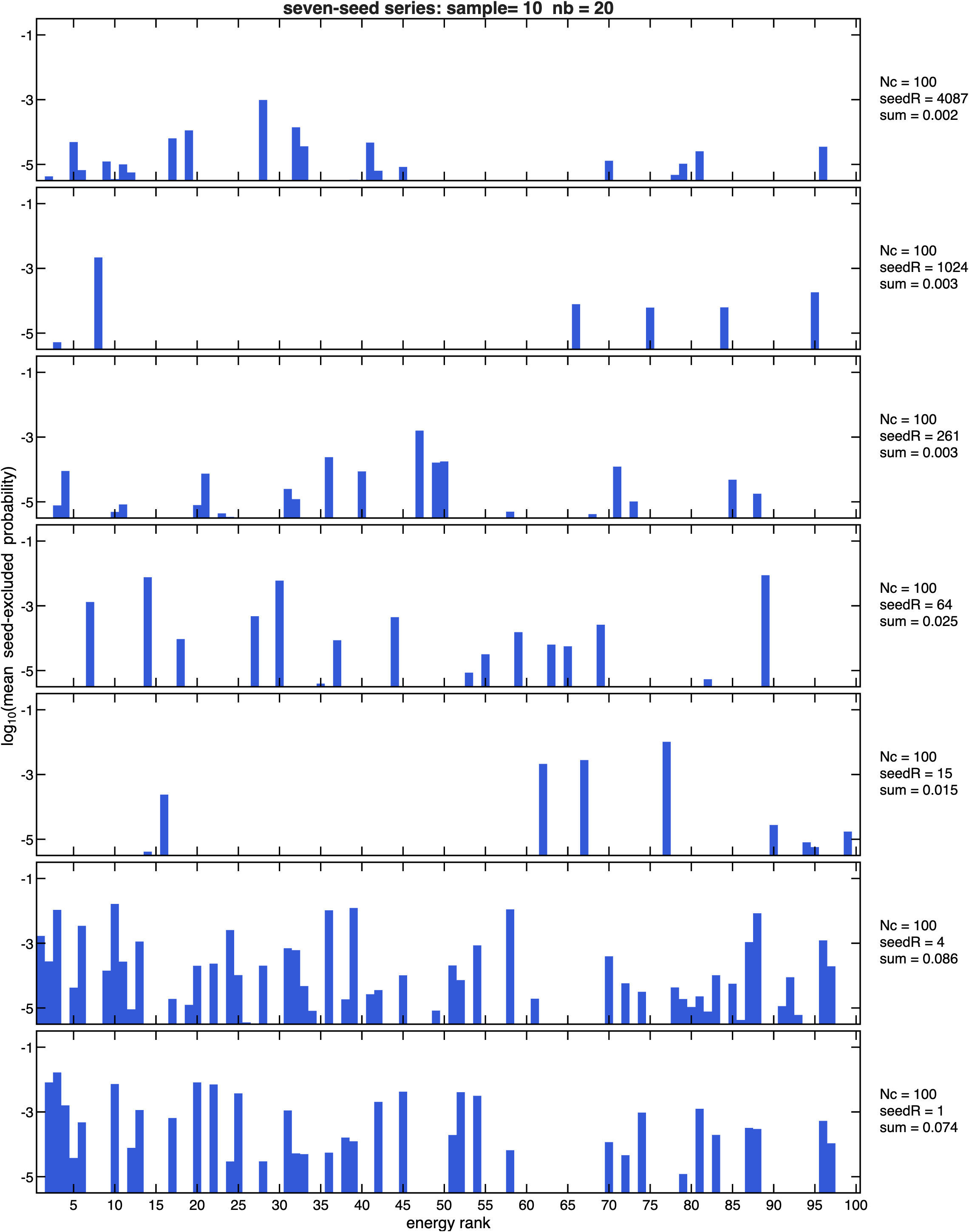}
\caption{\textbf{Dependence of top-100 mean state probabilities on seed rank
for fixed system size.} 
Results are shown for one SK problem instance with $n_b=20$ using seven fixed 
seed ranks, ranging from rank 1 near the ground state to rank 4087. Each panel 
averages over 100 circuits conditioned on the indicated seed. Lower-ranked seeds
generally assign greater total probability to the true top-100 states and
populate a larger fraction of them, whereas higher-ranked seeds produce
sparser and weaker low-energy support. The locations of the dominant peaks
remain irregular, showing that no single low-energy state is favored
systematically.}
\label{fig:aveTop100a1}
\end{figure}

%$$$$$$$$$$$$$$$$$$$$$$$$$$$$$$$$$$$$$$$$$$$$$$$$$$$$$$$$$$$$$$
\begin{figure}[p]
\centering
\includegraphics[
width=0.85\linewidth,
trim={0.01in 0.00in 0.00in 0.00in},
clip
]{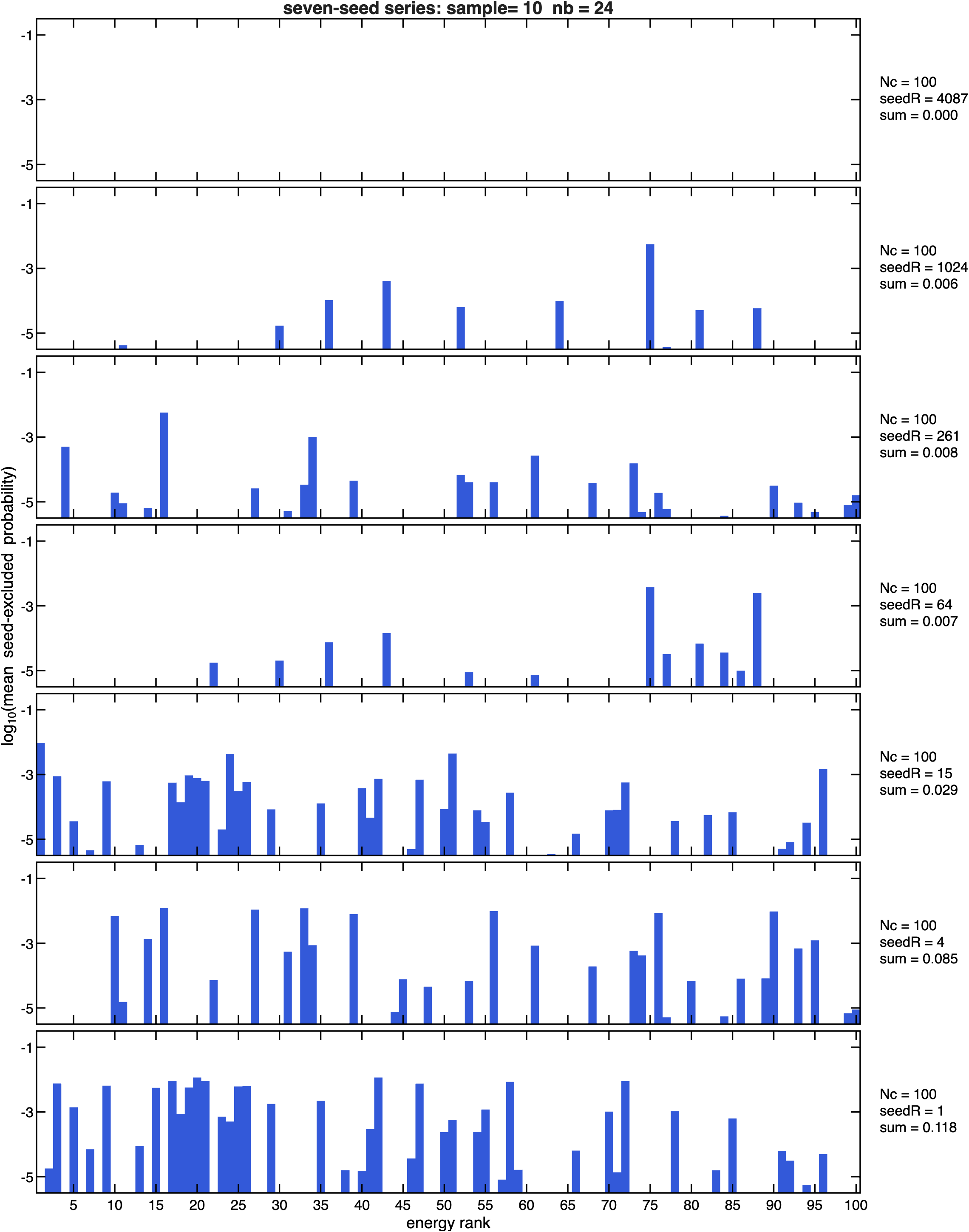}
\caption{\textbf{Dependence of top-100 mean state probabilities on seed rank
for fixed system size.} 
The same problem instance index and seven fixed seed ranks used in
Figure~\ref{fig:aveTop100a1} are examined at larger system size of $n_b=24$. 
Probability within the true top-100 generally decreases and becomes more sparsely
distributed than for $n_b=20$, although lower-ranked seeds continue to produce
broader and stronger low-energy support.}
\label{fig:aveTop100a2}
\end{figure}

%$$$$$$$$$$$$$$$$$$$$$$$$$$$$$$$$$$$$$$$$$$$$$$$$$$$$$$$$$$$$$$
\begin{figure}[p]
\centering
\includegraphics[
width=0.85\linewidth,
trim={0.01in 0.00in 0.00in 0.00in},
clip
]{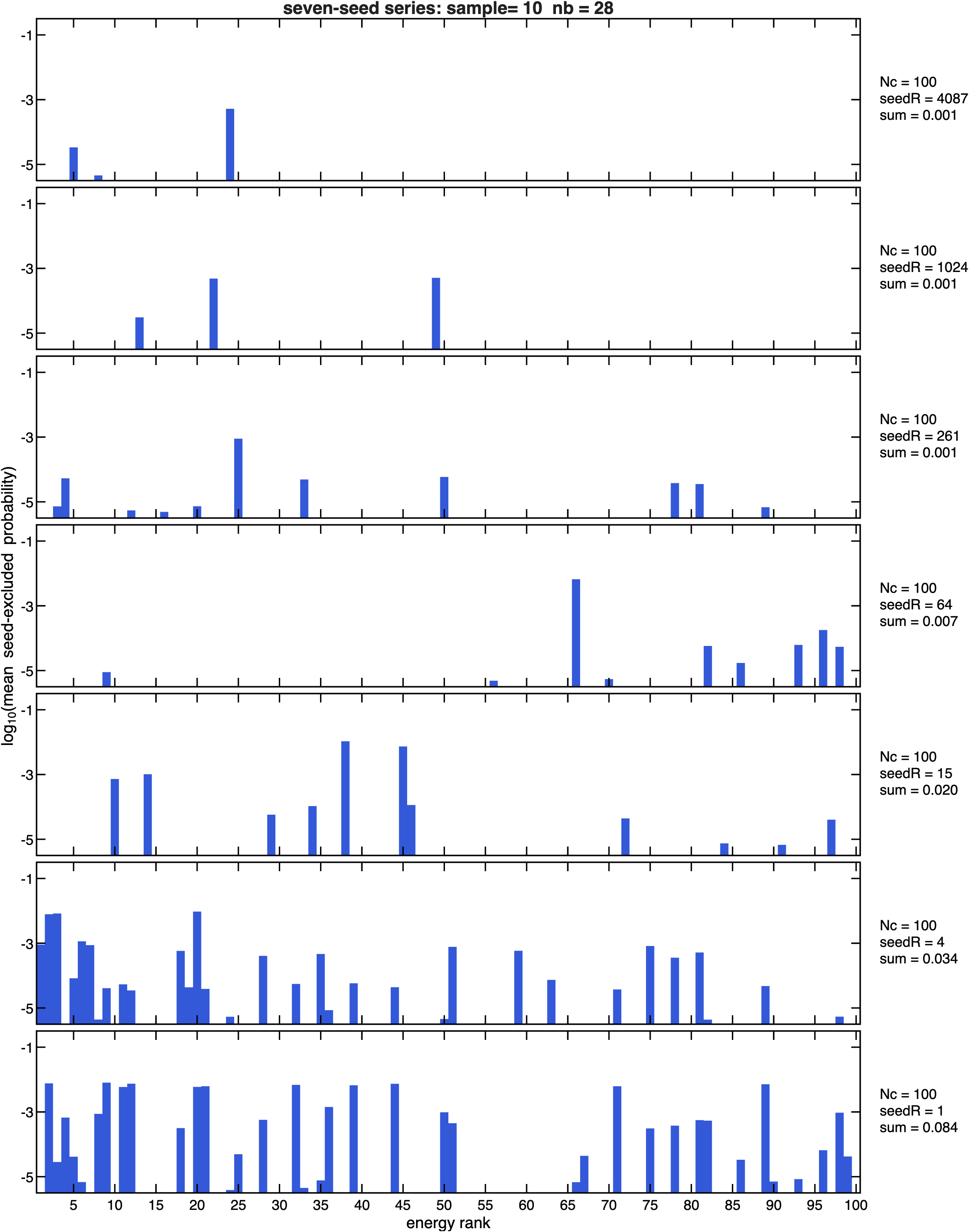}
\caption{\textbf{Dependence of top-100 mean state probabilities on seed rank
for fixed system size.} 
For the same problem-instance index, increasing the system size to $n_b=28$
further reduces the total probability assigned to the top-100 states for fixed
seed rank. Nevertheless, seeds nearer the ground state continue to populate
more of the low-energy spectrum. The sparse and strongly fluctuating peak
locations emphasize that low-energy access is stochastic even when the seed
rank is fixed.}
\label{fig:aveTop100a3}
\end{figure}

%$$$$$$$$$$$$$$$$$$$$$$$$$$$$$$$$$$$$$$$$$$$$$$$$$$$$$$$$$$$$$$
\begin{figure}[p]
\centering
\includegraphics[
width=0.85\linewidth,
trim={0.01in 0.00in 0.00in 0.00in},
clip
]{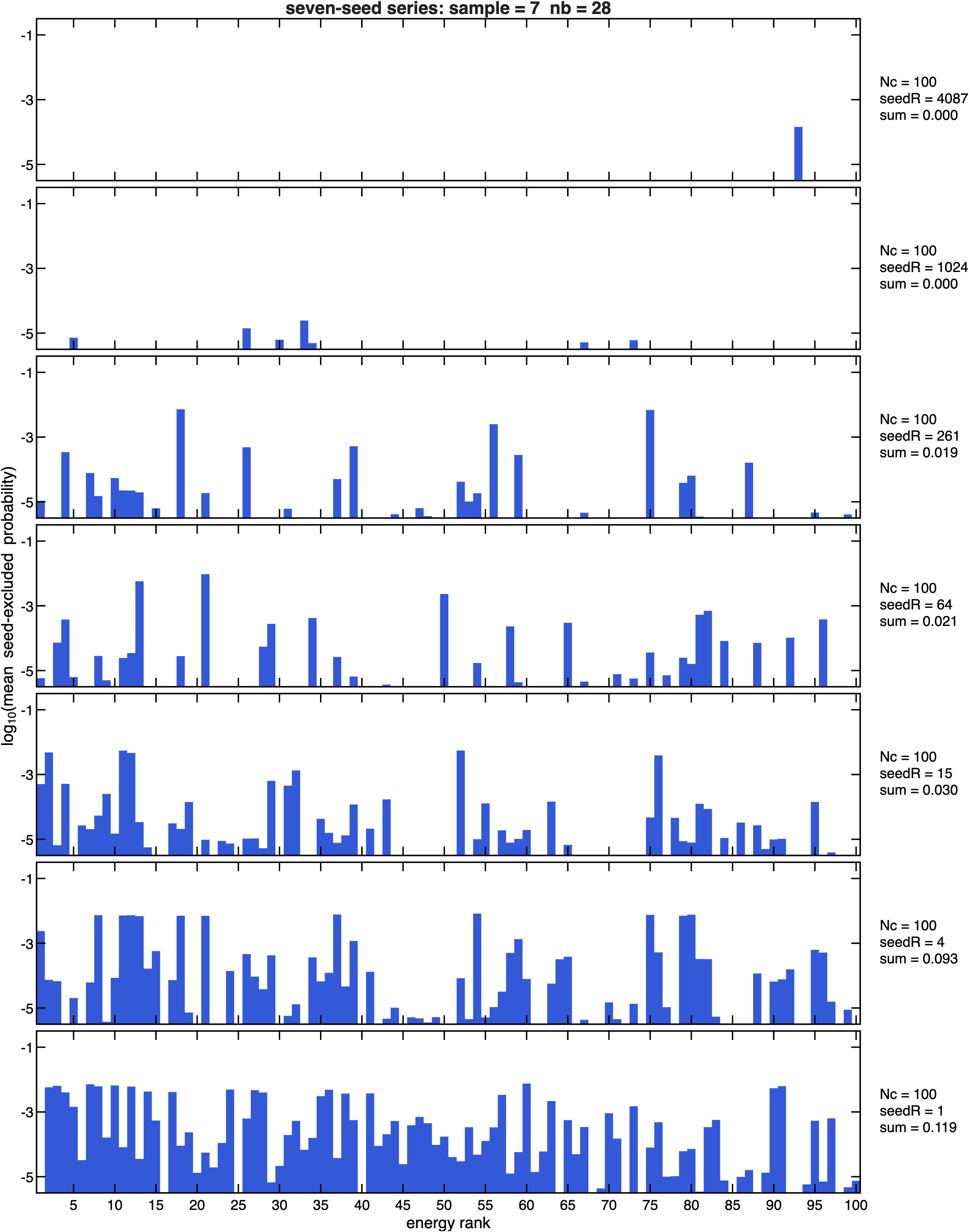}
\caption{\textbf{Seed-rank dependence for a second $n_b=28$ SK instance.}
The seven fixed-seed analysis is repeated for a different problem instance.
Both the locations of the populated top-100 states and the total probability
assigned to them differ substantially from
Figure~\ref{fig:aveTop100a3}. The general improvement obtained from
lower-ranked seeds remains, but no individual seed rank or set of proposal
states is universally favored.}
\label{fig:aveTop100b1}
\end{figure}

%$$$$$$$$$$$$$$$$$$$$$$$$$$$$$$$$$$$$$$$$$$$$$$$$$$$$$$$$$$$$$$
\begin{figure}[p]
\centering
\includegraphics[
width=0.85\linewidth,
trim={0.01in 0.00in 0.00in 0.00in},
clip
]{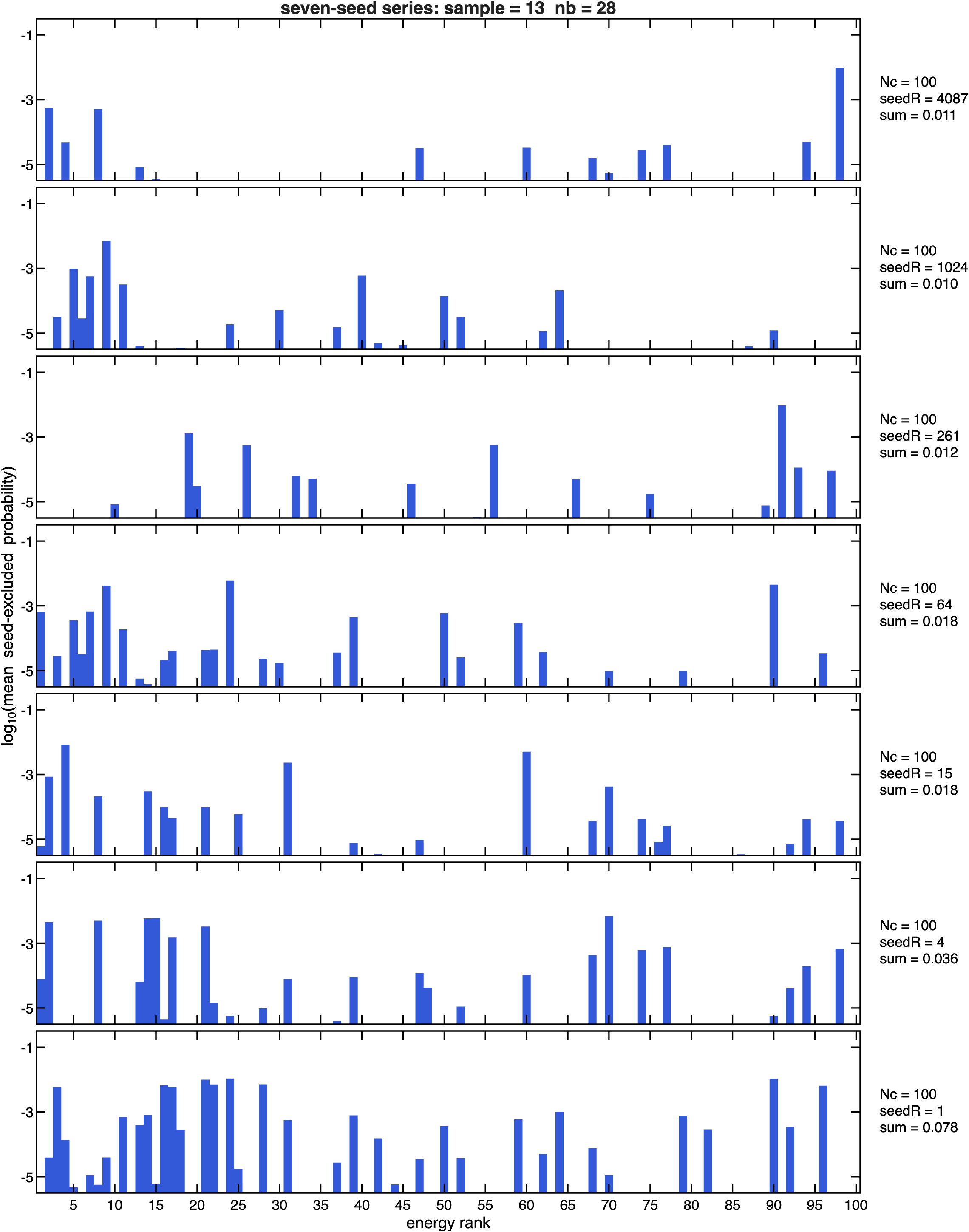}
\caption{\textbf{Seed-rank dependence for a third $n_b=28$ SK instance.}
A third problem instance again exhibits distinct peak locations and substantial
variation in the total top-100 probability. Together with
Figures~\ref{fig:aveTop100a3} and~\ref{fig:aveTop100b1}, this comparison shows
that instance-to-instance fluctuations are large, while the broad dependence
on seed rank remains reproducible. Robust search therefore requires sampling
multiple seeds and randomized circuit realizations rather than relying on a
preferred seed position.}
\label{fig:aveTop100b2}
\end{figure}

%$$$$$$$$$$$$$$$$$$$$$$$$$$$$$$$$$$$$$$$$$$$$$$$$$$$$$$$$$$$$$$
\begin{figure}[p]
\centering
\includegraphics[
width=0.85\linewidth,
trim={0.01in 0.00in 0.00in 0.00in},
clip
]{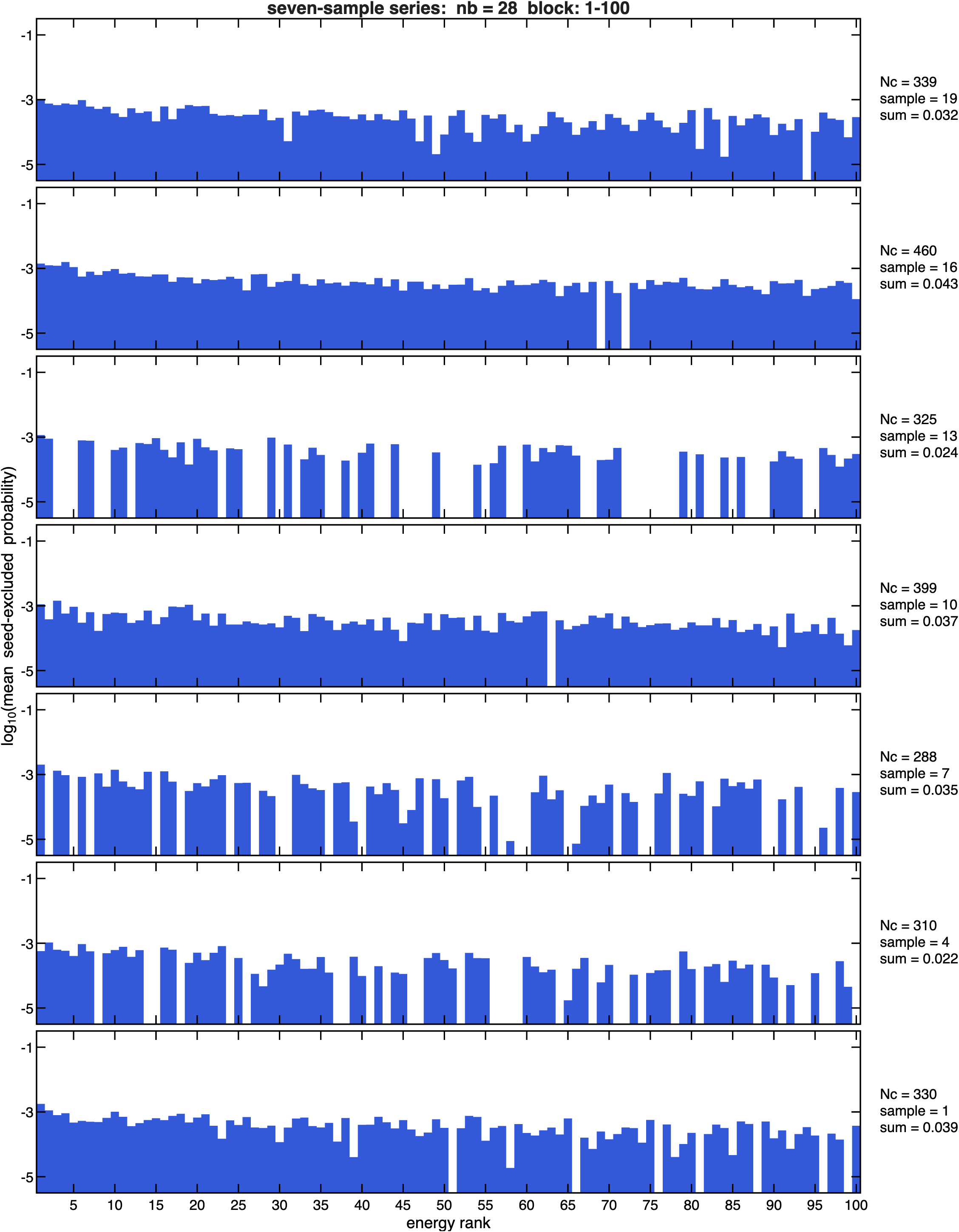}
\caption{\textbf{Top-100 mean state probabilities using all recovered
top-100 seeds.}
Seven $n_b=28$ SK instances are shown. For each instance, the circuit ensemble
contains all search-generated seeds whose true ranks lie between 1 and 100.
Compared with fixed-seed ensembles, the resulting probability is distributed
far more uniformly across the top-100 states. Instances with fewer contributing
circuits exhibit more unpopulated ranks, supporting the use of continued
randomized sampling to reduce, although not necessarily eliminate, such
coverage gaps.}
\label{fig:aveTop100c1}
\end{figure}

%$$$$$$$$$$$$$$$$$$$$$$$$$$$$$$$$$$$$$$$$$$$$$$$$$$$$$$$$$$$$$$
\begin{figure}[p]
\centering
\includegraphics[
width=0.85\linewidth,
trim={0.01in 0.00in 0.00in 0.00in},
clip
]{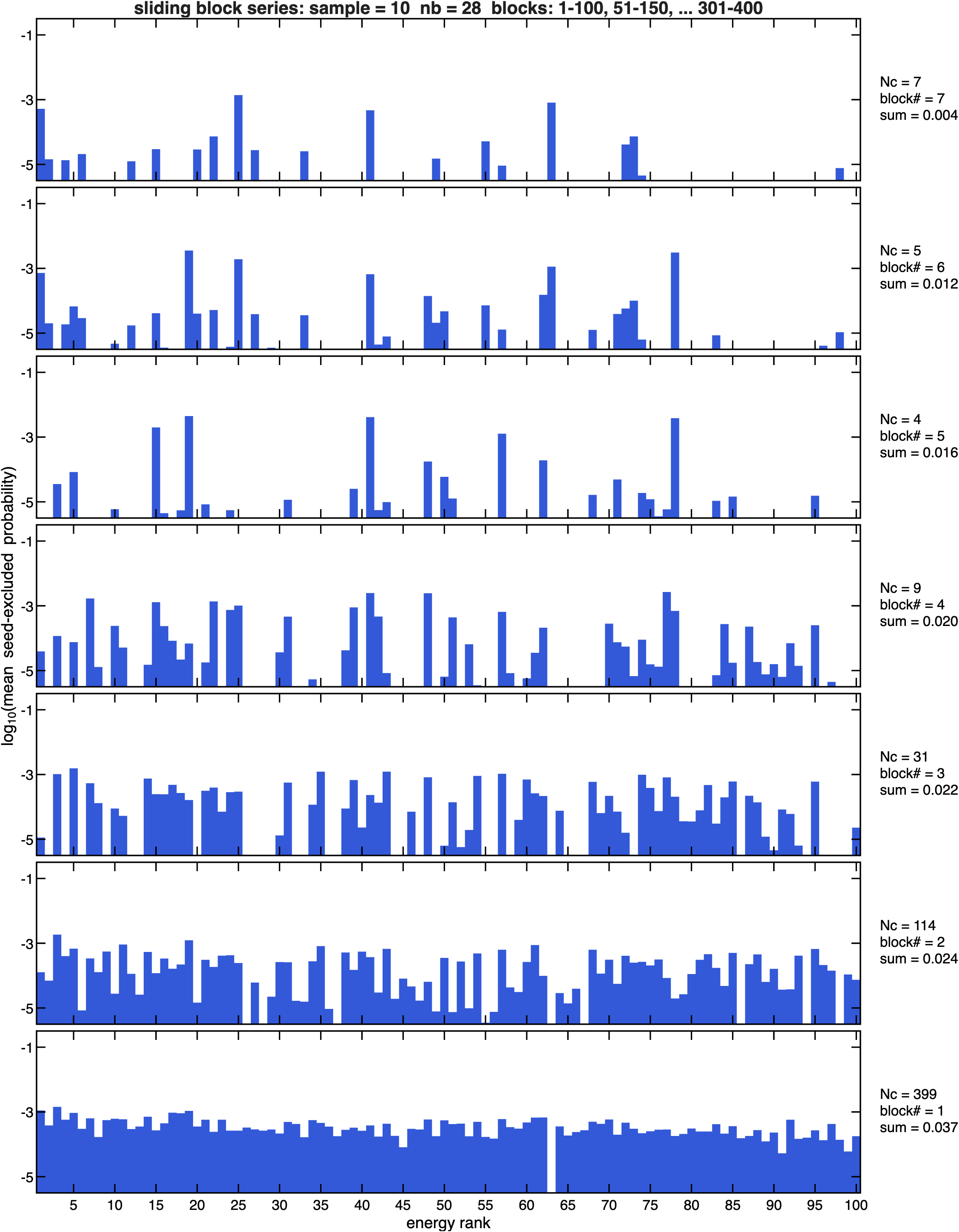}
\caption{\textbf{Dependence of top-100 probability on the seed-rank window.}
For one $n_b=28$ SK instance, circuit ensembles are grouped by overlapping
100-rank seed windows: 1--100, 51--150, \ldots, 301--400. Probability assigned
to the true top-100 states decreases rapidly as the seed window moves away
from the ground state, and the number of contributing circuits also declines.
Thus, broad coverage of the top-100 spectrum is generated primarily after the
elite frontier has accumulated a dense set of low-ranked seeds. Enlarging the
outer frontier to include substantially poorer seeds would contribute little
to top-100 recovery under the present search protocol.}
\label{fig:aveTop100d1}
\end{figure}

%$$$$$$$$$$$$$$$$$$$$$$$$$$$$$$$$$$$$$$$$$$$$$$$$$$$$$$$$$$$$$$
\begin{figure}[p]
\centering
\includegraphics[
width=0.85\linewidth,
trim={0.01in 0.00in 0.00in 0.00in},
clip
]{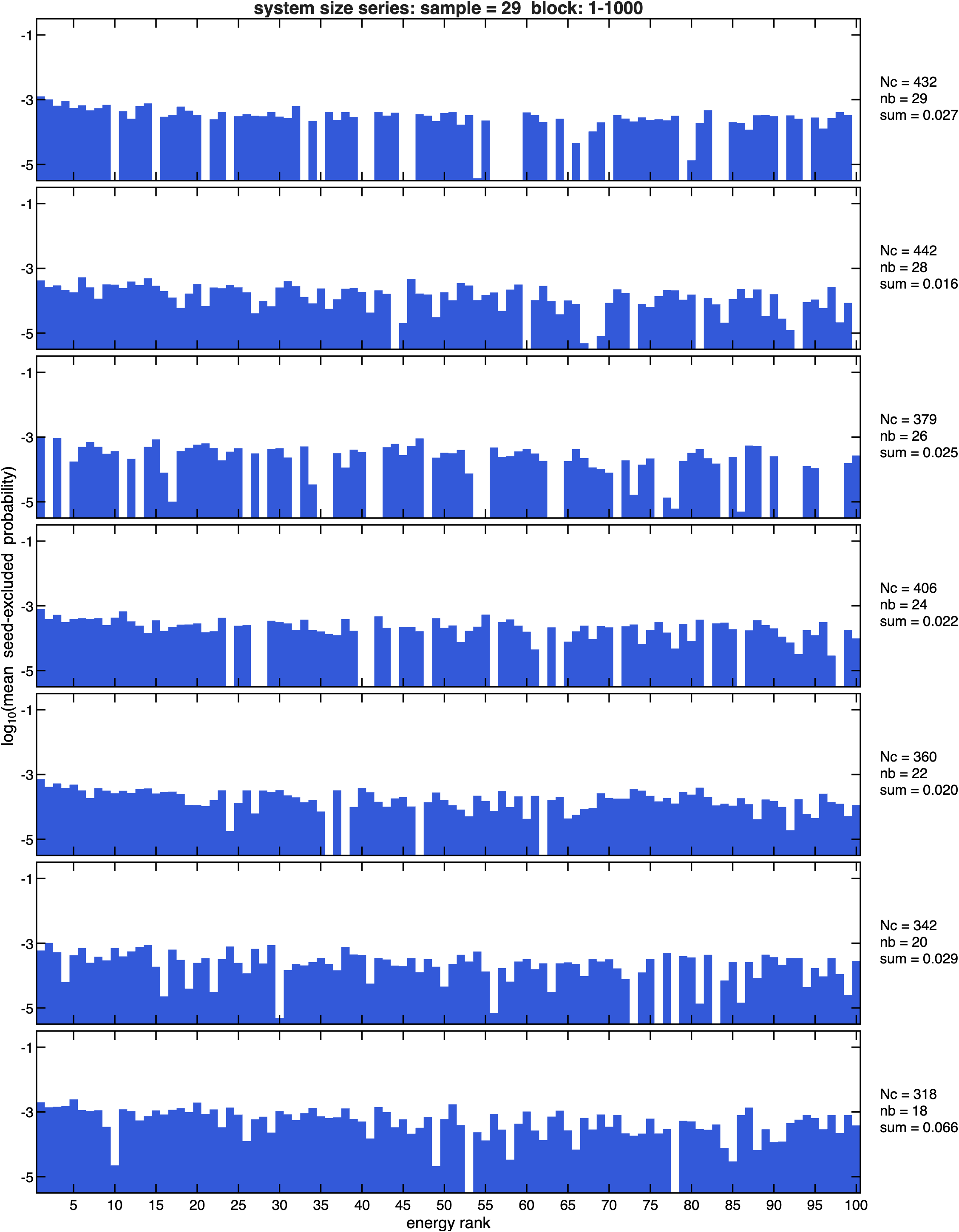}
\caption{\textbf{System-size dependence using seeds from ranks 1--1000.}
For one SK instance at each size, the circuit ensemble includes all
search-generated seeds with true ranks between 1 and 1000. The mean
probability remains broadly distributed across the top-100 states from
$n_b=18$ to 29, although unpopulated ranks become more frequent as the system
grows. The number of contributing circuits changes only modestly across this
range. These results show that QIPS retains broad low-energy support with
increasing system size, while also indicating that larger systems may require
a greater circuit budget to achieve comparable coverage.}
\label{fig:aveTop100e1}
\end{figure}

%%%%%%%%%%%%%%%%%%%%%%%%%%%%%%%%%%%%%%%%%%%%%%%%%%%%%%%%%%%%%%%%%%%%%%%%%%%%%%%%%%%%%%%%
%%%%%%%%%%%%%%%%%%%%%%%%%%%%%%%%%%%%%%%%%%%%%%%%%%%%%%%%%%%%%%%%%%%%%%%%%%%%%%%%%%%%%%%%
\FloatBarrier
\suppnote{6}{Illustrating top-100 maximum state probabilities}
{supp:Top100MaxStateProb}

The following eight figures examine the largest probability assigned to each
of the true top-100 energy-ranked states by any circuit in a specified
ensemble. The underlying Sherrington--Kirkpatrick instances and circuit
ensembles are identical to those analyzed using ensemble-mean probabilities
in Supplementary Note~5. Here, however, the plotted quantity for state $s$ is
\begin{equation}
p_{\max}(s)=\max_k p_s(k),
\end{equation}
after excluding the seed contribution, where $p_s(k)$ is the probability
assigned to state $s$ by circuit $k$.

This statistic isolates the intermittent high-probability events that
dominate finite-shot detection. A state need not carry appreciable probability
across most circuits. It may be found when a single randomized,
seed-conditioned circuit produces a sufficiently large peak for its 100-shot
measurement record to sample that state. Recurrent peaks across multiple
circuits then produce the repeated low-energy hits that accumulate during the 
search with counts far exceeding those possible from a single 100-shot record.
Thus, recurrent hits arise from different circuits.

Compared with the ensemble means in Supplementary Note~5, the maximum
probabilities reveal substantially stronger and broader access to the top-100
states. Their locations remain irregular and instance dependent, consistent
with the stochastic structure of the localized interference patterns.
Randomizing the circuit parameters and seed states spreads these intermittent
peaks across the low-energy spectrum rather than relying on one circuit to
cover all relevant states.

The peak strengths generally decrease as the system grows and as the seed
moves away from the ground state. Thus, maintaining comparable top-100
detection at larger $n_b$ may require more than the $20n_b$ circuits used in
the present benchmark. The results nevertheless support a central design
principle of QIPS: for a fixed measurement budget, sampling many randomized
circuits is likely to provide broader low-energy access than concentrating all
shots on a single circuit. The optimal allocation between circuit diversity
and shots per circuit remains to be determined.

%$$$$$$$$$$$$$$$$$$$$$$$$$$$$$$$$$$$$$$$$$$$$$$$$$$$$$$$$$$$$$$
\begin{figure}[p]
\centering
\includegraphics[
width=0.85\linewidth,
trim={0.01in 0.00in 0.00in 0.00in},
clip
]{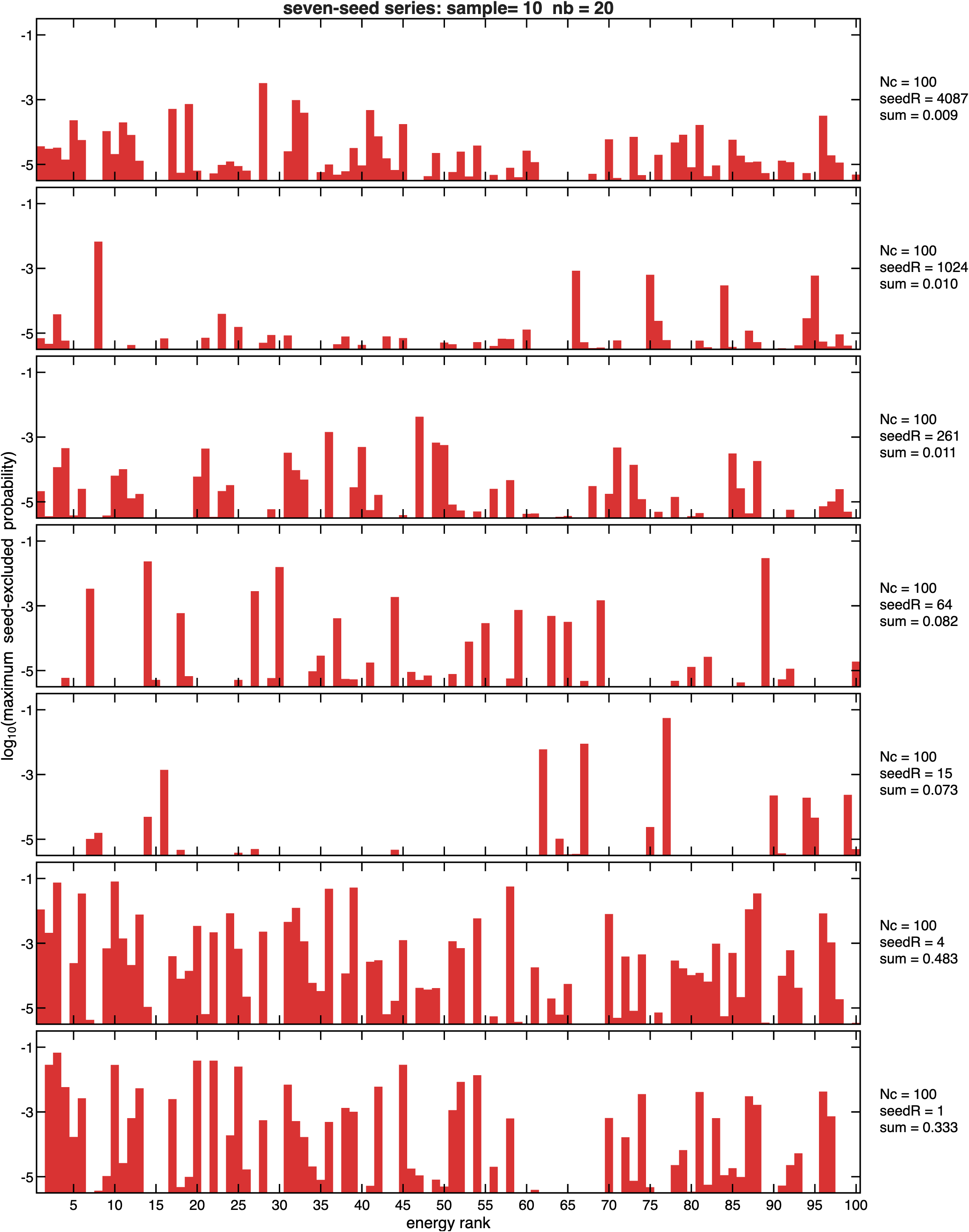}
\caption{\textbf{Dependence of top-100 maximum state probabilities on seed
rank for fixed system size.}
For one SK instance at $n_b=20$, each panel shows the maximum seed-excluded probability
assigned to the true top-100 states across 100 circuits conditioned on the
indicated seed rank. Lower-ranked seeds produce stronger and more broadly
distributed peaks, with several probabilities large enough to be readily
sampled in 100 shots. Higher-ranked seeds retain intermittent access to
low-energy states, but with smaller total probability and more incomplete
coverage.}
\label{fig:maxTop100a1}
\end{figure}

%$$$$$$$$$$$$$$$$$$$$$$$$$$$$$$$$$$$$$$$$$$$$$$$$$$$$$$$$$$$$$$
\begin{figure}[p]
\centering
\includegraphics[
width=0.85\linewidth,
trim={0.01in 0.00in 0.00in 0.00in},
clip
]{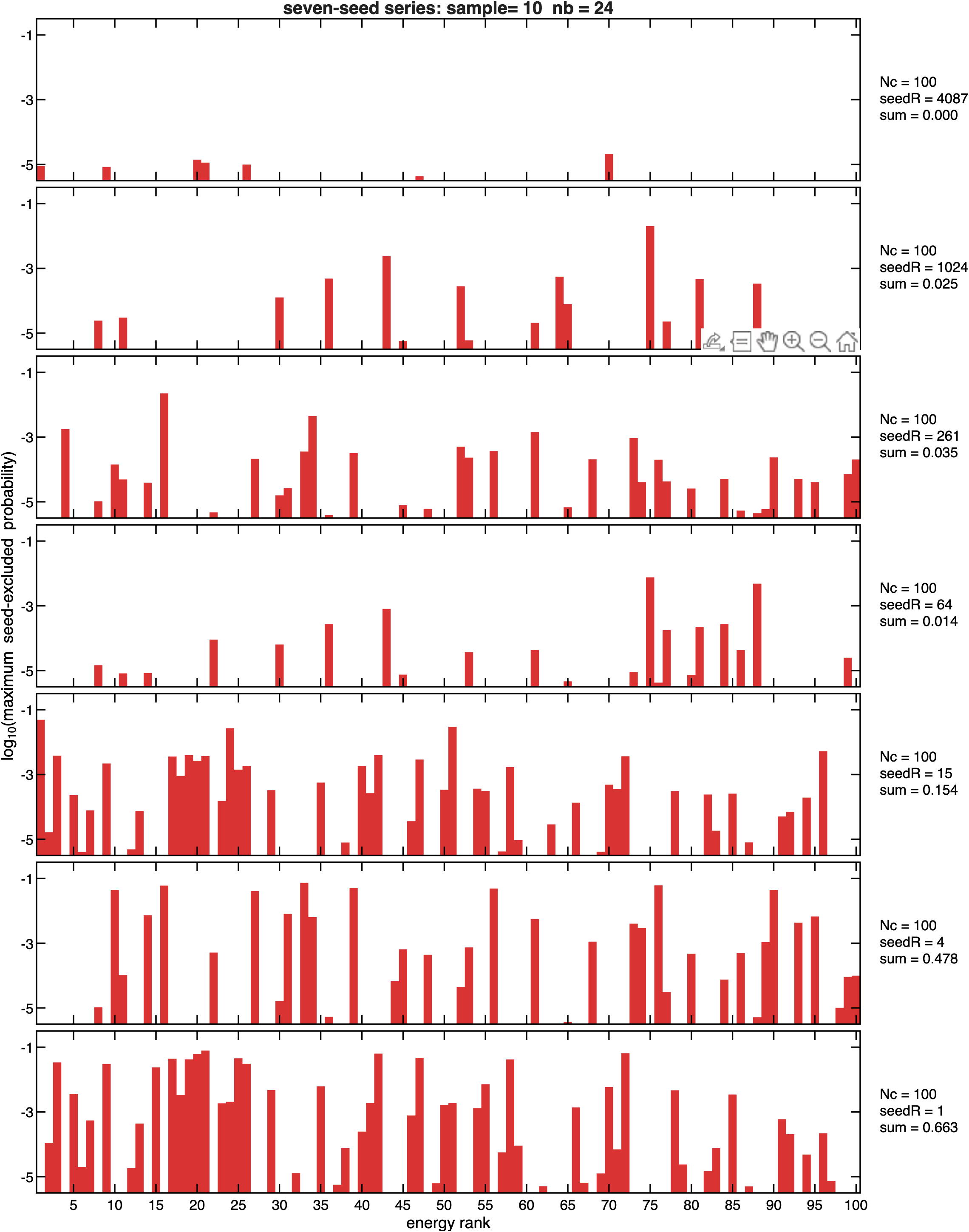}
\caption{\textbf{Dependence of top-100 maximum state probabilities on seed
rank for fixed system size.}
The same seven-seed analysis is repeated at $n_b=24$. Maximum probabilities
remain substantial for low-ranked seeds, but the number and strength of
accessible top-100 states generally decrease relative to $n_b=20$. The
irregular peak locations show that individual low-energy states are accessed
intermittently rather than through a smooth rank-dependent profile.}
\label{fig:maxTop100a2}
\end{figure}

%$$$$$$$$$$$$$$$$$$$$$$$$$$$$$$$$$$$$$$$$$$$$$$$$$$$$$$$$$$$$$$
\begin{figure}[p]
\centering
\includegraphics[
width=0.85\linewidth,
trim={0.01in 0.00in 0.00in 0.00in},
clip
]{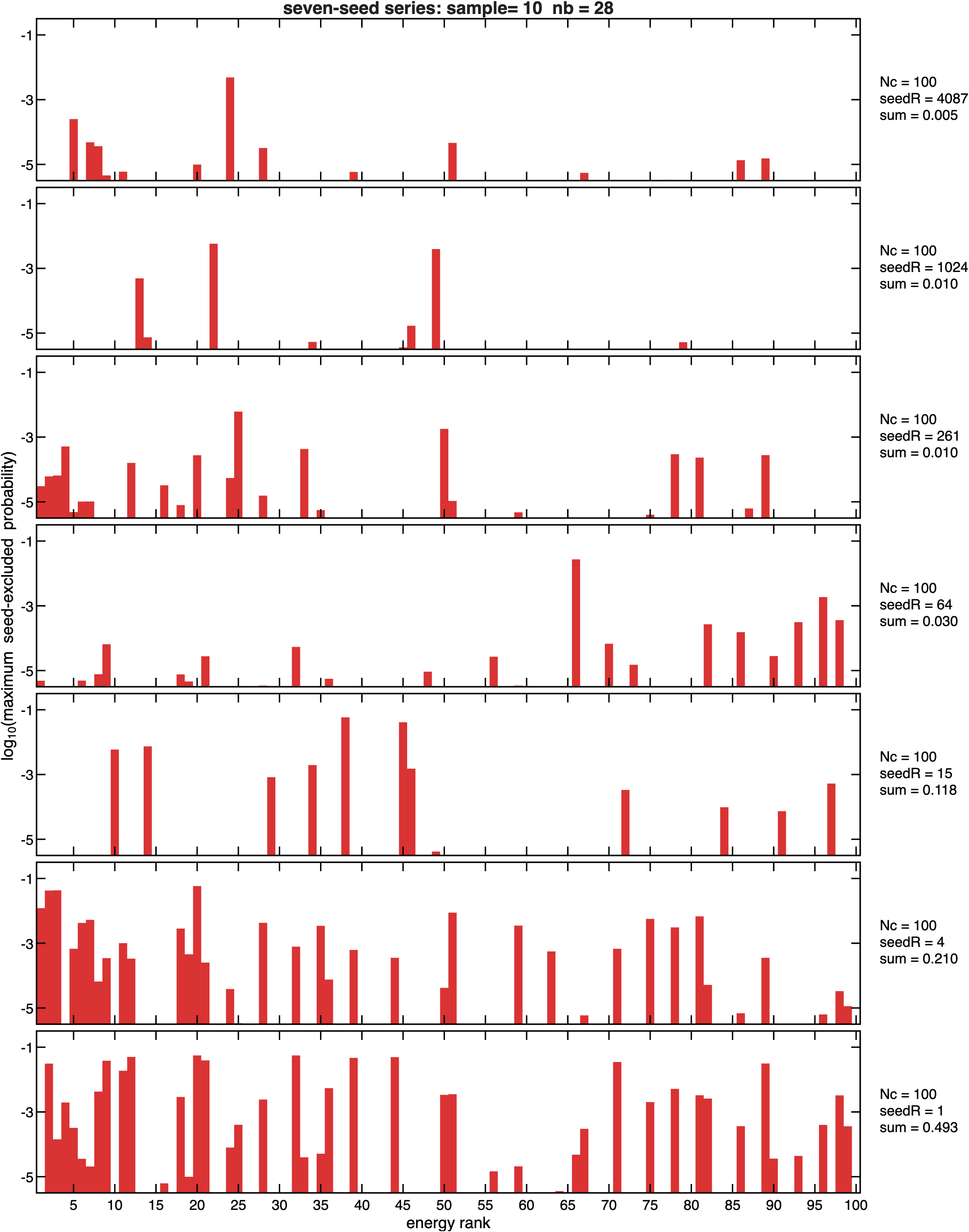}
\caption{\textbf{Dependence of top-100 maximum state probabilities on seed
rank for fixed system size.}
At $n_b=28$, strong top-100 peaks persist, particularly for seeds closest to
the ground state, but the coverage becomes increasingly sparse for seeds
further away from the ground state. Comparison with Figures~\ref{fig:maxTop100a1} and
\ref{fig:maxTop100a2} shows that the finite-shot accessibility of individual
low-energy states weakens with system size for a fixed circuit ensemble.}
\label{fig:maxTop100a3}
\end{figure}

%$$$$$$$$$$$$$$$$$$$$$$$$$$$$$$$$$$$$$$$$$$$$$$$$$$$$$$$$$$$$$$
\begin{figure}[p]
\centering
\includegraphics[
width=0.85\linewidth,
trim={0.01in 0.00in 0.00in 0.00in},
clip
]{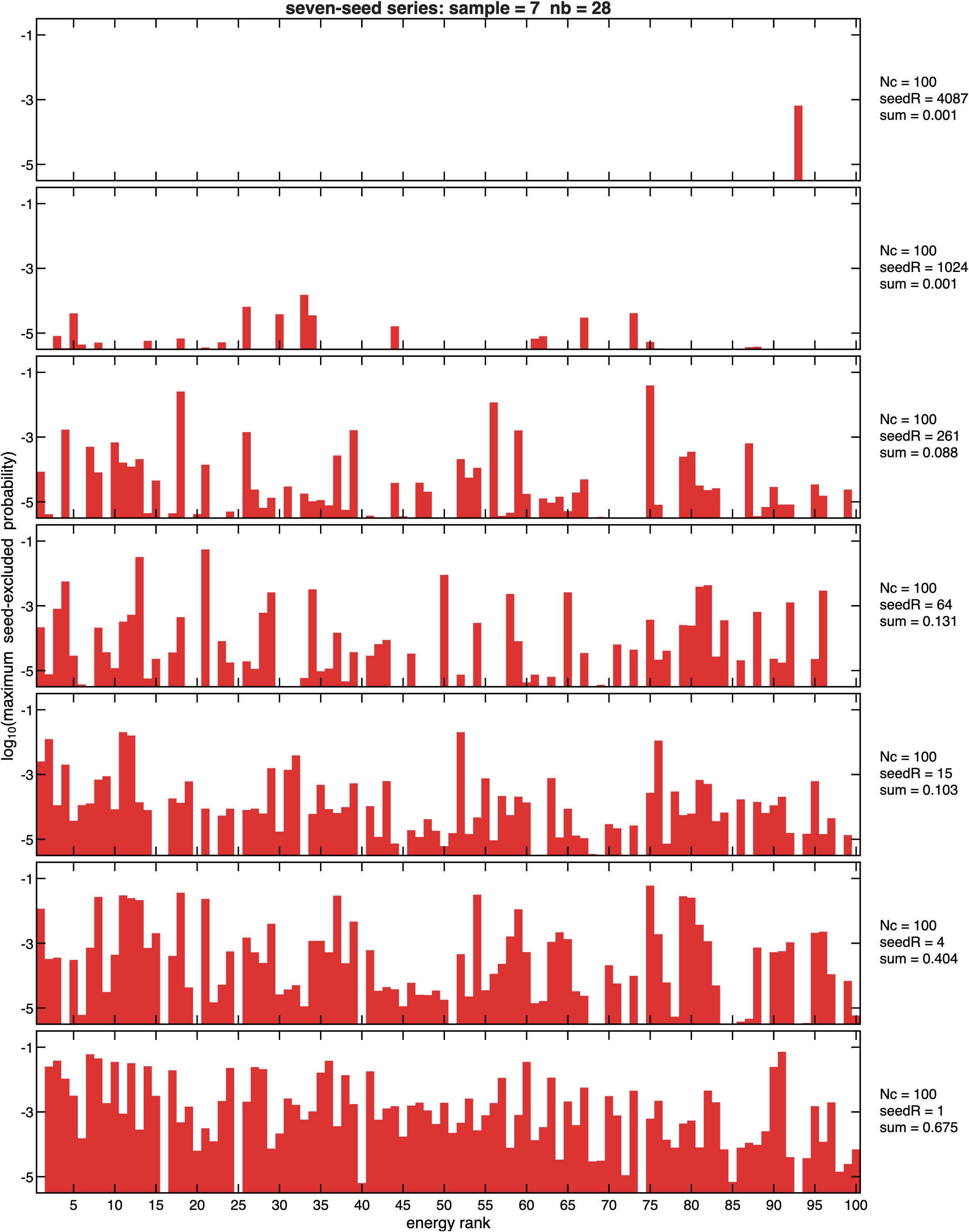}
\caption{\textbf{Seed-rank dependence for a second $n_b=28$ SK instance.}
The maximum-probability analysis is repeated for a different problem
instance. The locations and strengths of the peaks differ substantially from
Figure~\ref{fig:maxTop100a3}, while the broad tendency for lower-ranked seeds
to provide stronger top-100 access is preserved. This instance dependence
reinforces the need to sample multiple seeds and randomized circuits.}
\label{fig:maxTop100b1}
\end{figure}

%$$$$$$$$$$$$$$$$$$$$$$$$$$$$$$$$$$$$$$$$$$$$$$$$$$$$$$$$$$$$$$
\begin{figure}[p]
\centering
\includegraphics[
width=0.85\linewidth,
trim={0.01in 0.00in 0.00in 0.00in},
clip
]{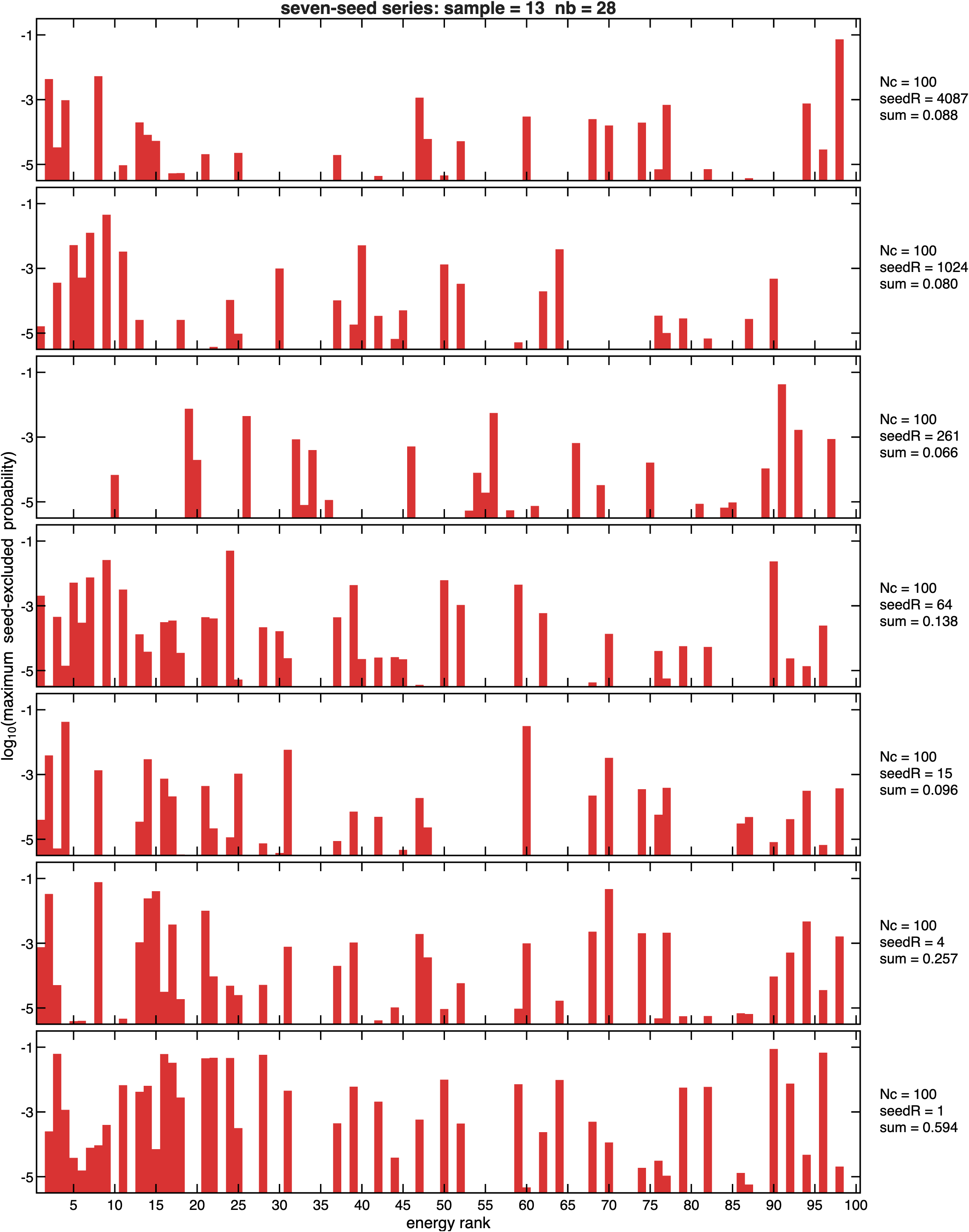}
\caption{\textbf{Seed-rank dependence for a third $n_b=28$ SK instance.}
A third instance again produces a distinct set of intermittent peaks. The
strongest maximum probabilities are not tied to fixed energy ranks across
instances, yet low-ranked seeds consistently provide broader and stronger
access to the top-100 spectrum. No single seed or state therefore controls the
observed search performance.}
\label{fig:maxTop100b2}
\end{figure}

%$$$$$$$$$$$$$$$$$$$$$$$$$$$$$$$$$$$$$$$$$$$$$$$$$$$$$$$$$$$$$$
\begin{figure}[p]
\centering
\includegraphics[
width=0.85\linewidth,
trim={0.01in 0.00in 0.00in 0.00in},
clip
]{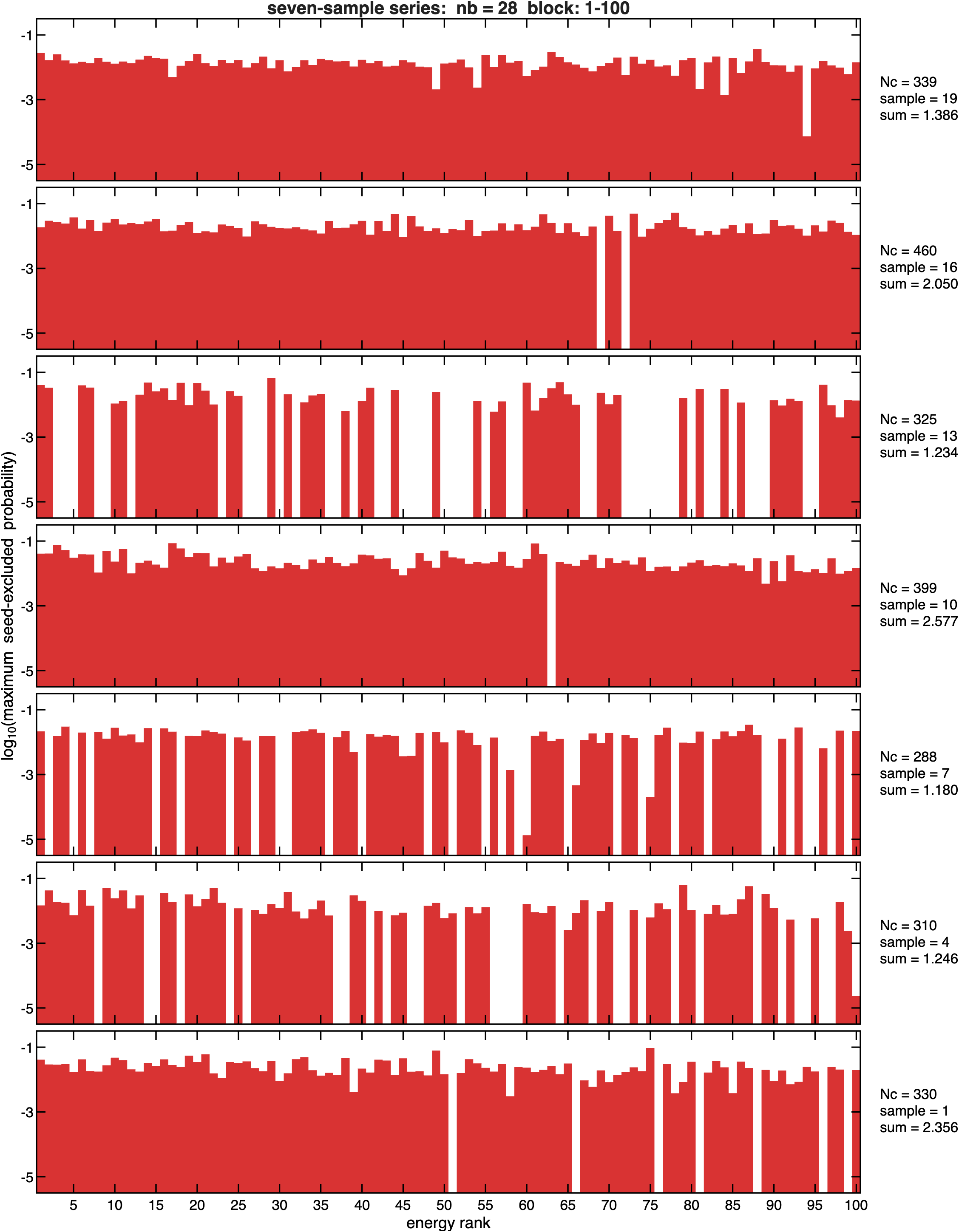}
\caption{\textbf{Top-100 maximum state probabilities using all recovered
top-100 seeds.}
Seven $n_b=28$ SK instances are shown. For each instance, the circuit ensemble
contains all search-generated seeds with ranks from 1 to 100. The maximum
probabilities provide broad and nearly uniform access across the true top-100
states, even when the ensemble-mean probabilities in Supplementary Note~5 are
much smaller. Remaining gaps are instance dependent and are more common when
fewer circuits contribute.}
\label{fig:maxTop100c1}
\end{figure}

%$$$$$$$$$$$$$$$$$$$$$$$$$$$$$$$$$$$$$$$$$$$$$$$$$$$$$$$$$$$$$$
\begin{figure}[p]
\centering
\includegraphics[
width=0.85\linewidth,
trim={0.01in 0.00in 0.00in 0.00in},
clip
]{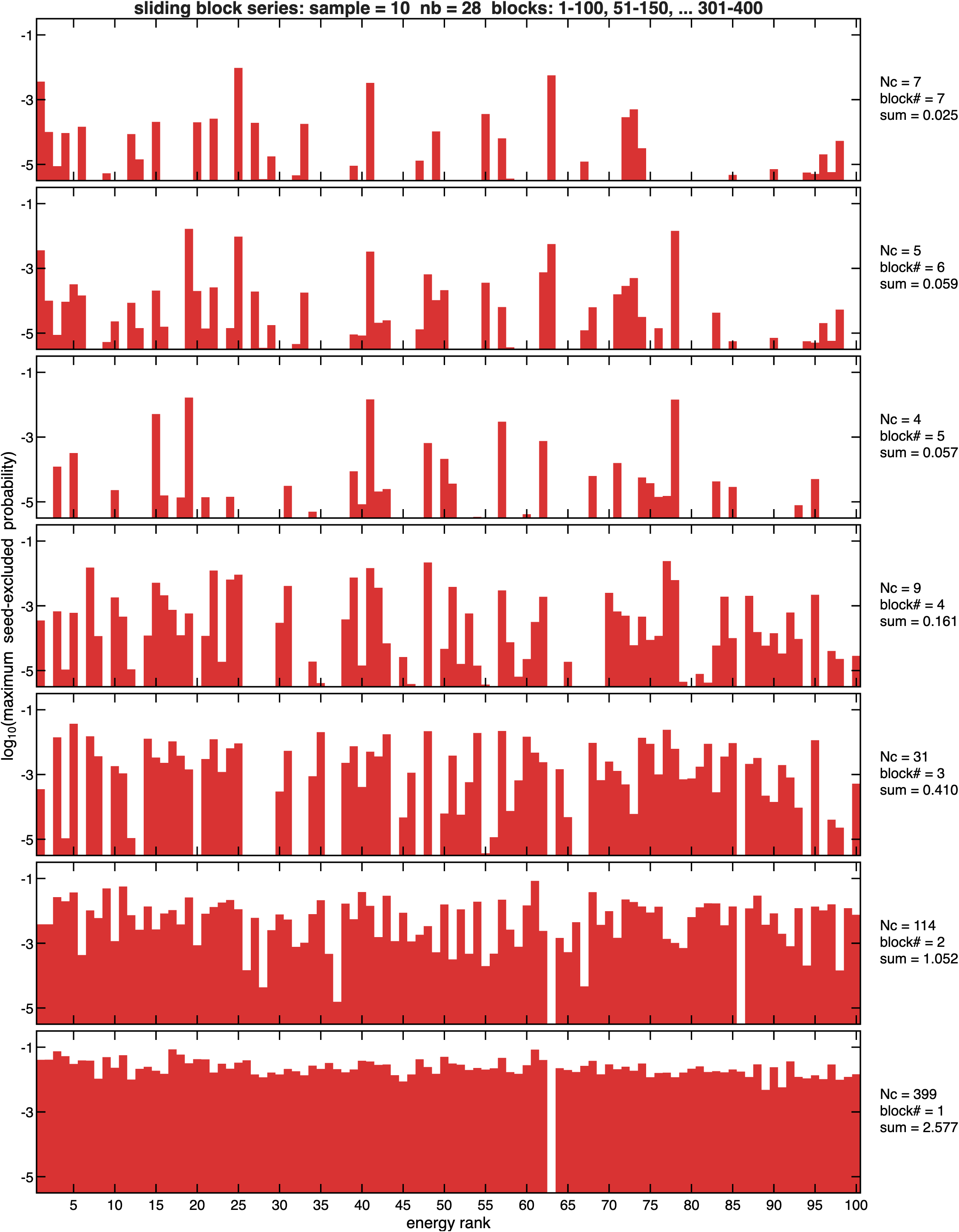}
\caption{\textbf{Dependence of top-100 maximum probability on the seed-rank
window.}
For one $n_b=28$ SK instance, circuits are grouped by overlapping 100-rank
seed windows from ranks 1--100 through 301--400 in increments of 50.
Both the number of
contributing circuits and the maximum probabilities assigned to the true
top-100 states decrease as the seed window moves away from the ground state.
The strongest and most uniform finite-shot access therefore develops after
the elite frontier accumulates seeds within or near the target low-energy
region.}
\label{fig:maxTop100d1}
\end{figure}

%$$$$$$$$$$$$$$$$$$$$$$$$$$$$$$$$$$$$$$$$$$$$$$$$$$$$$$$$$$$$$$
\begin{figure}[p]
\centering
\includegraphics[
width=0.85\linewidth,
trim={0.01in 0.00in 0.00in 0.00in},
clip
]{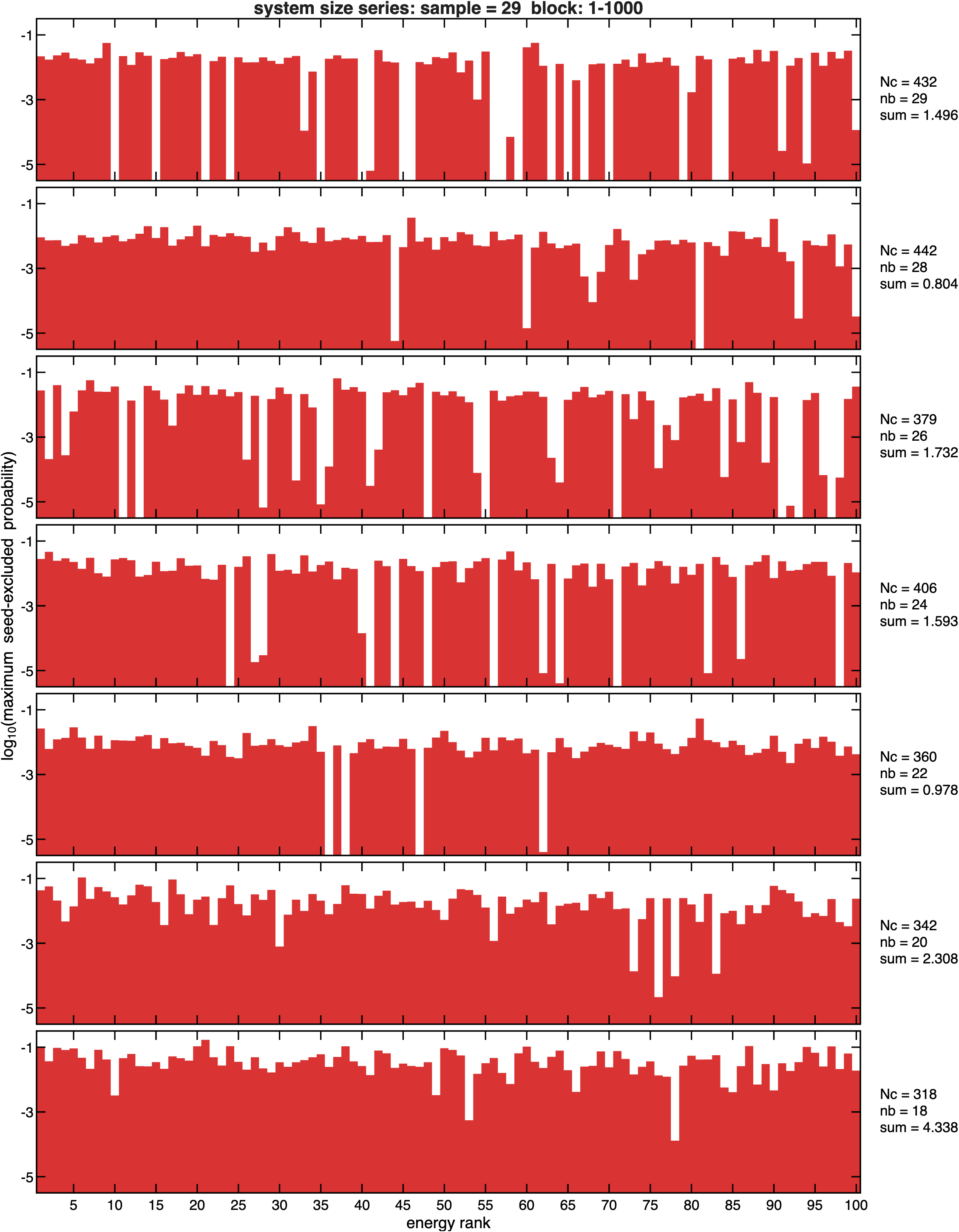}
\caption{\textbf{System-size dependence using seeds from ranks 1--1000.}
For one SK instance at each size, the circuit ensemble includes all
search-generated seeds with ranks from 1 to 1000. Maximum probabilities remain
broadly distributed across the top-100 states for $18\le n_b\le29$, showing
that individual circuits continue to generate statistically significant
low-energy peaks as the Hilbert space grows. The reported sums fluctuate
considerably, making it difficult to determine whether peak strengths and
coverage decay systematically with system size. These data suggest that larger
systems may require additional circuit realizations to maintain
comparable detection probability.}
\label{fig:maxTop100e1}
\end{figure}

%%%%%%%%%%%%%%%%%%%%%%%%%%%%%%%%%%%%%%%%%%%%%%%%%%%%%%%%%%%%%%%%%%%%%%%%%%%%%%%%%%%%%%%%
%%%%%%%%%%%%%%%%%%%%%%%%%%%%%%%%%%%%%%%%%%%%%%%%%%%%%%%%%%%%%%%%%%%%%%%%%%%%%%%%%%%%%%%%
\FloatBarrier
\suppnote{7}{Seed-excluded conditional CDF}
{supp:SeedExludedCDF}
%+++++++++++++++++++++++++++++++++++++++++++++++++++++++++++++++++++++++++++++++++++++++++++++++
\subsection*{Seed-excluded conditional cumulative distributions}
To isolate how QIPS generates proposals from a fixed seed, consider the
Sherrington--Kirkpatrick model with $n_b=20$ and seed rank $r_s=1024$.
Supplementary Figure~\ref{fig:actualCDF} shows the cumulative proposal-rank distributions
for the 96 circuits that satisfy the localization criterion. These CDFs are
constructed from the exact computational-basis probabilities of each circuit,
before removing the seed contribution. Because localized circuits retain
substantial probability at the seed state, every realization exhibits a jump
at $r_p=r_s$, whose magnitude varies across the circuit ensemble.

\begin{figure}[hbt]
    \centering
    \includegraphics[
        width=0.98\linewidth,
        trim={0.01in 0.00in 0.00in 0.00in},
        clip
    ]{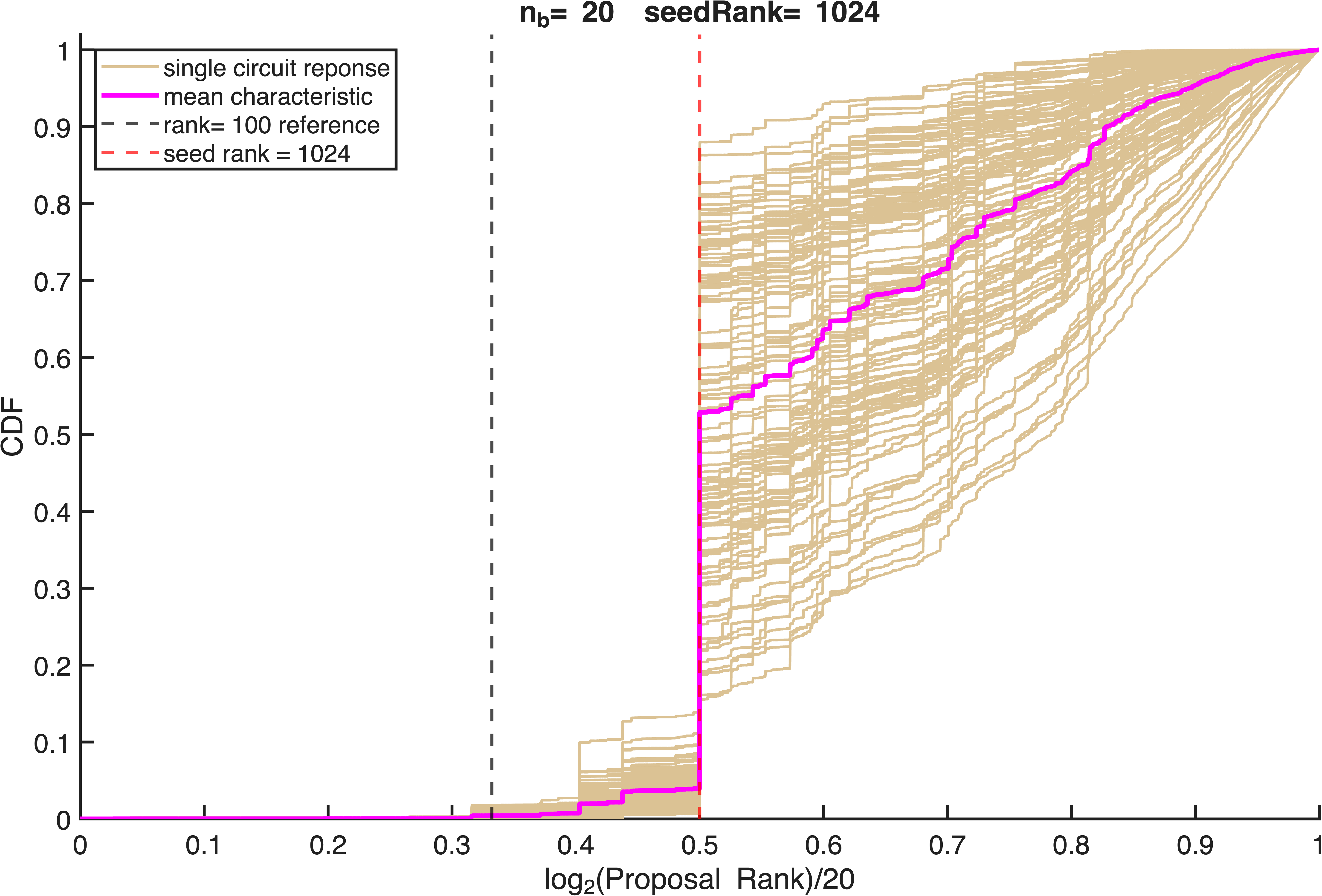}
    \caption{\textbf{Proposal-rank cumulative distributions before seed
    exclusion.}
    CDFs are shown for 96 localized circuits generated for the
    Sherrington--Kirkpatrick model with $n_b=20$ and seed rank
    $r_s=1024$. Thin tan curves denote individual circuits, and the 
    thick magenta curve denotes the ensemble mean. The jump at
    $\log_2(r_s)/n_b=0.5$ gives the probability of returning to the seed state.
    The vertical dashed lines mark rank 100 and the seed rank.}
    \label{fig:actualCDF}
\end{figure}

Repeated recovery of the seed contributes no new state to the search.
The proposal structure away from the seed is therefore characterized by the
conditional distribution
\begin{equation}
P(r_p\mid r_p\neq r_s,r_s)
=
\frac{P(r_p\mid r_s)}
     {1-p_s},
\qquad r_p\neq r_s,
\label{eq:seed_excluded_probability}
\end{equation}
where $p_s=P(r_p=r_s\mid r_s)$. Operationally, the seed probability is set
to zero and the remaining probabilities are renormalized to unity. The
resulting seed-excluded CDFs are shown in Supplementary
Figure~\ref{fig:pluckedCDF}.

Conditioning on $r_p\neq r_s$ removes repeated measurements of the seed, which 
constitute a sampling cost for QIPS. The seed-excluded CDF therefore 
characterizes the quality of non-seed proposals rather than the total search 
efficiency. Under matched end-to-end resources, the classical proposal 
mechanism remains more effective over the system sizes studied. The present 
analysis does not identify the bitstring transformations induced by the 
quantum circuit, but it shows that their rank statistics differ from those 
of the classical baseline. The classical generator uses structured bitstring 
moves that favor descent from the current seed, whereas QIPS generates 
proposals through seed-conditioned quantum interference.

\begin{figure}[hbt]
    \centering
    \includegraphics[
        width=0.98\linewidth,
        trim={0.01in 0.00in 0.00in 0.00in},
        clip
    ]{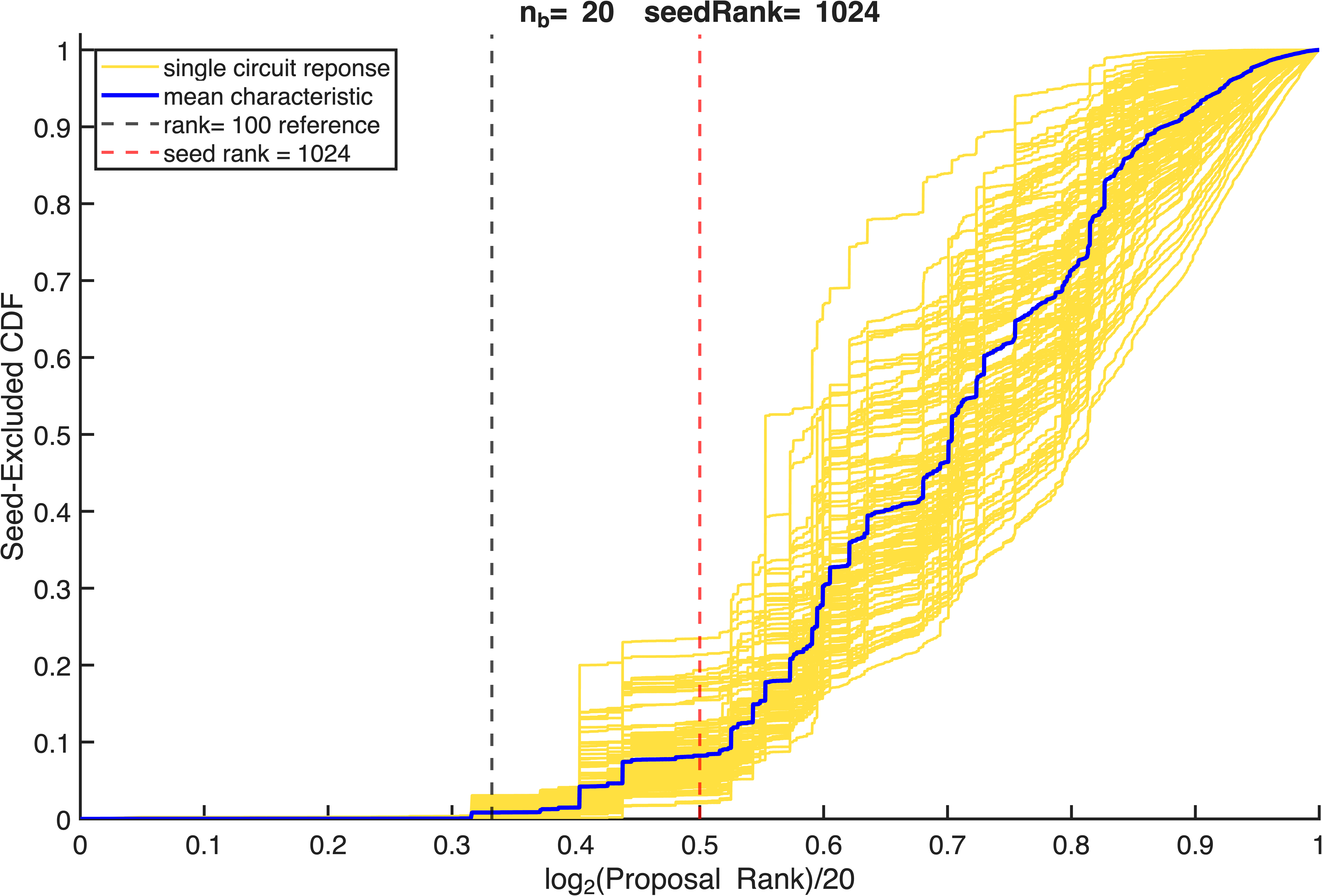}
    \caption{\textbf{Seed-excluded conditional proposal-rank distributions.}
    The seed contribution has been removed from each circuit distribution and
    the remaining conditional probability is renormalized. Thin yellow curves
    denote individual localized circuits, and the thick blue curve denotes the 
    ensemble mean. Removing the seed discontinuity exposes the productive 
    circuit-to-circuit variation and the rank structure of proposals that can 
    advance the search.}
    \label{fig:pluckedCDF}
\end{figure}

Each seed-excluded distribution is further decomposed into localized and extended 
components using a probability threshold $p_{\mathrm{loc}}$. The localized component
contains states whose probabilities exceed this threshold, whereas the remaining states
define the extended component. This partition connects naturally to finite-shot
resolution: for a fixed number of measurements, states with probabilities far
below the corresponding sampling scale will not be observed
reproducibly, thus making the concept of statistically significant probability 
peaks useful.

The extended component is accurately represented by a two-parameter probit
distribution, even at the level of individual circuit realizations. The
threshold $p_{\mathrm{loc}}$ may therefore be selected to achieve a prescribed
reconstruction tolerance: states above threshold are stored explicitly,
whereas the extended background is represented by the fitted probit form.
This localized--extended decomposition yields an ultracompact representation
of the full state probability distribution with high
reconstruction accuracy.

%%%%%%%%%%%%%%%%%%%%%%%%%%%%%%%%%%%%%%%%%%%%%%%%%%%%%%%%%%%%%%%%%%%%%%%%%%%%%%%%%%%%%%%%
%%%%%%%%%%%%%%%%%%%%%%%%%%%%%%%%%%%%%%%%%%%%%%%%%%%%%%%%%%%%%%%%%%%%%%%%%%%%%%%%%%%%%%%%
\FloatBarrier
\suppnote{8}{Quantum/classical proposal CDF comparison}
{supp:QCcdfComparison}

The seed-excluded cumulative distributions introduced in the main text are compared 
directly with those from the matched classical proposal generator in 
Supplementary Figure~\ref{fig:QvsC}. Note that Supplementary Figure~\ref{fig:QvsC}a
reproduces the corresponding main-text Figure~3c to permit side-by-side comparison.

The clearest qualitative difference appears in Supplementary
Figure~\ref{fig:QvsC}a,b. Classical proposals remain biased toward lower-energy
ranks across the full range of seed positions. QIPS instead exhibits an
approximate spectral symmetry: seeds below the median favor lower-energy
proposals, whereas seeds above the median favor higher-energy proposals. The
precise crossover was not determined, but the same qualitative behavior was
observed across all six benchmark systems. This distinction shows that the
classical and quantum proposal generators produce substantially different rank
structures. Classical proposals are therefore expected to perform better during
the early stages of a search, especially when seeds lie above the median rank.
Moreover, the CDFs indicate that as the energy rank of the seed progressively
decreases, quantum proposals become more effective than classical proposals at
generating further low-rank candidates.

\begin{figure}[!htb]
\centering
\includegraphics[
width=0.98\linewidth,
trim={0.01in 0.00in 0.00in 0.00in},
clip
]{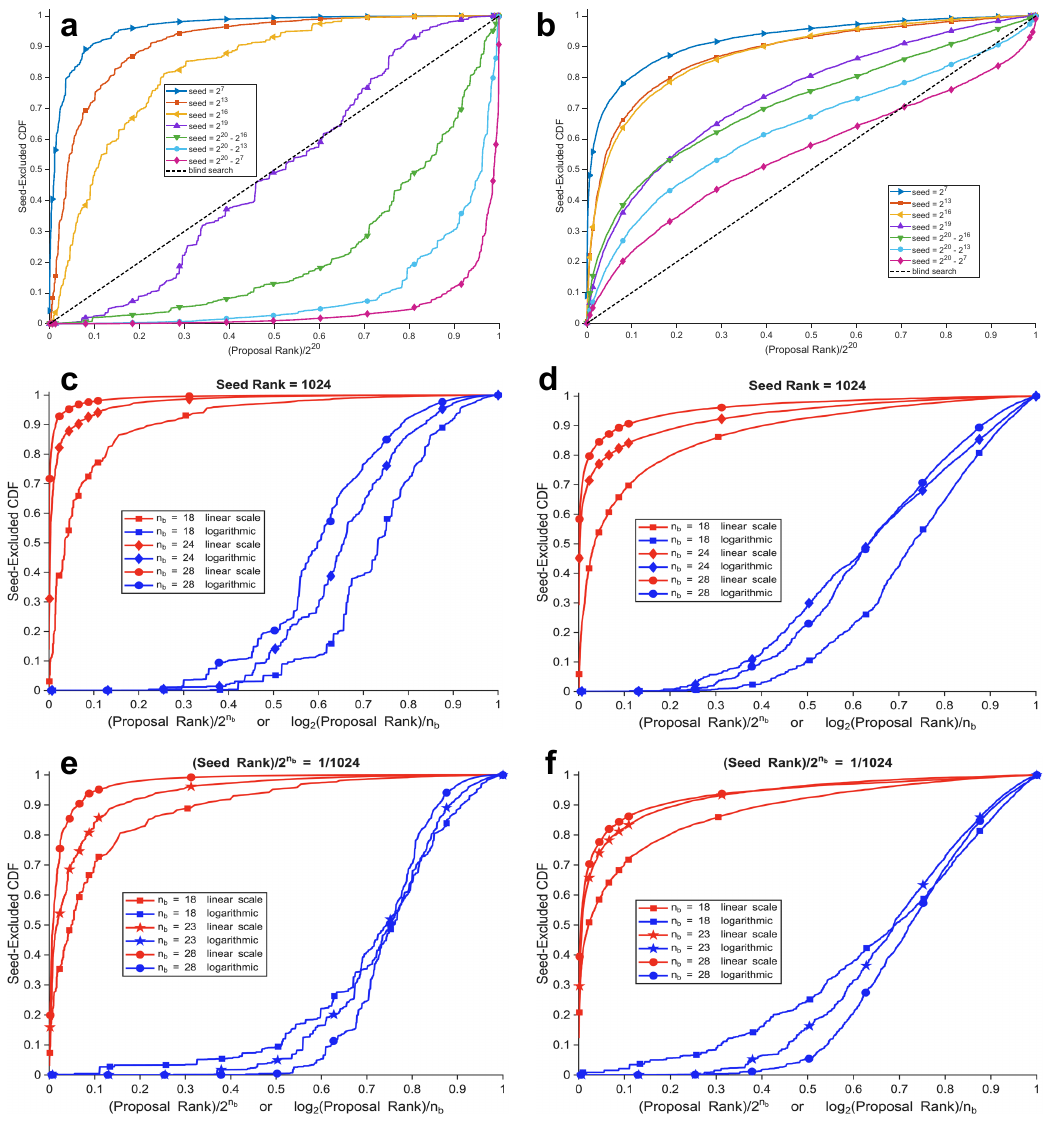}
\caption{\textbf{Seed-excluded proposal-rank CDFs for QIPS and the
matched classical proposal generator.}
\textbf{a,b}, CDFs for representative seed ranks in the
$n_b=20$ Sherrington--Kirkpatrick model, shown for QIPS and classical
proposals, respectively. The dashed diagonal denotes blind search.
\textbf{c,d}, System-size dependence at fixed seed rank
$r_s=1024$ for QIPS and classical proposals.
\textbf{e,f}, Corresponding comparison at fixed normalized seed rank
$r_s/2^{n_b}=1/1024$.
Linear and normalized logarithmic rank coordinates are shown as indicated
in the legends. All distributions are conditioned on proposing a state
distinct from the seed.}
\label{fig:QvsC}
\end{figure}

Supplementary Figure~\ref{fig:QvsC}b,c compares the system-size dependence of
the cumulative distribution functions for corresponding quantum and classical
proposal ranks at fixed seed rank. Supplementary Figure~\ref{fig:QvsC}d,e
makes the same comparison at fixed normalized seed rank, defined as
$r_s/2^{n_b}$. Because low-energy probability concentration is most clearly
resolved on a logarithmic rank scale, the cumulative distributions are shown
on both linear and logarithmic scales for fixed seed rank and fixed normalized
seed rank. Although small differences could be interpreted in favor of the quantum
proposals, this data set does not provide sufficient statistical resolution to
draw a strong conclusion about system-size dependence. Within the statistical
uncertainties, the quantum and classical proposals are comparable across all
system sizes studied here. Larger aggregated data sets, such as those analyzed
in the main text, are required to resolve statistically meaningful differences.

%%%%%%%%%%%%%%%%%%%%%%%%%%%%%%%%%%%%%%%%%%%%%%%%%%%%%%%%%%%%%%%%%%%%%%%%%%%%%%%
%%%%%%%%%%%%%%%%%%%%%%%%%%%%%%%%%%%%%%%%%%%%%%%%%%%%%%%%%%%%%%%%%%%%%%%%%%%%%%%
\FloatBarrier
\suppnote{9}{System- and target-resolved top-K benchmarks}
{supp:ResolvedBenchmarks}

\subsection*{Target-resolved benchmarks}
The aggregate benchmarks from Figure~4 of the main text are extended here
from the top-10 target to the top-1 and top-100 targets. In each figure, the
left and right columns show the matched classical and QIPS searches,
respectively, while the three rows correspond to $K=1$, 10, and 100. The
top-10 panels reproduce the main-text results to facilitate direct comparison.
Across all three targets, coverage, hit rate, and multiplicity exhibit the same
qualitative dependence on search round and system size, supporting the use of
the top-10 target for the principal analysis.

Coverage is similar for $K=1$, 10, and 100, although it decreases with
increasing system size for both proposal mechanisms. This consistency is
compatible with the outer frontier size of 100, which maintains low-energy
diversity across the same rank range. The hit rate increases strongly with
$K$, as expected because a larger target set contains more acceptable
outcomes. Mean multiplicity shows the clearest target dependence: the top-1
and top-10 curves are similar, whereas the top-100 multiplicity is lower.
Thus, repeated sampling is concentrated preferentially among the
lowest-energy states rather than distributed uniformly across the full
top-100 set.

\begin{figure}[!htb]
\centering
\includegraphics[
width=0.98\linewidth,
trim={0.01in 0.00in 0.00in 0.00in},
clip
]{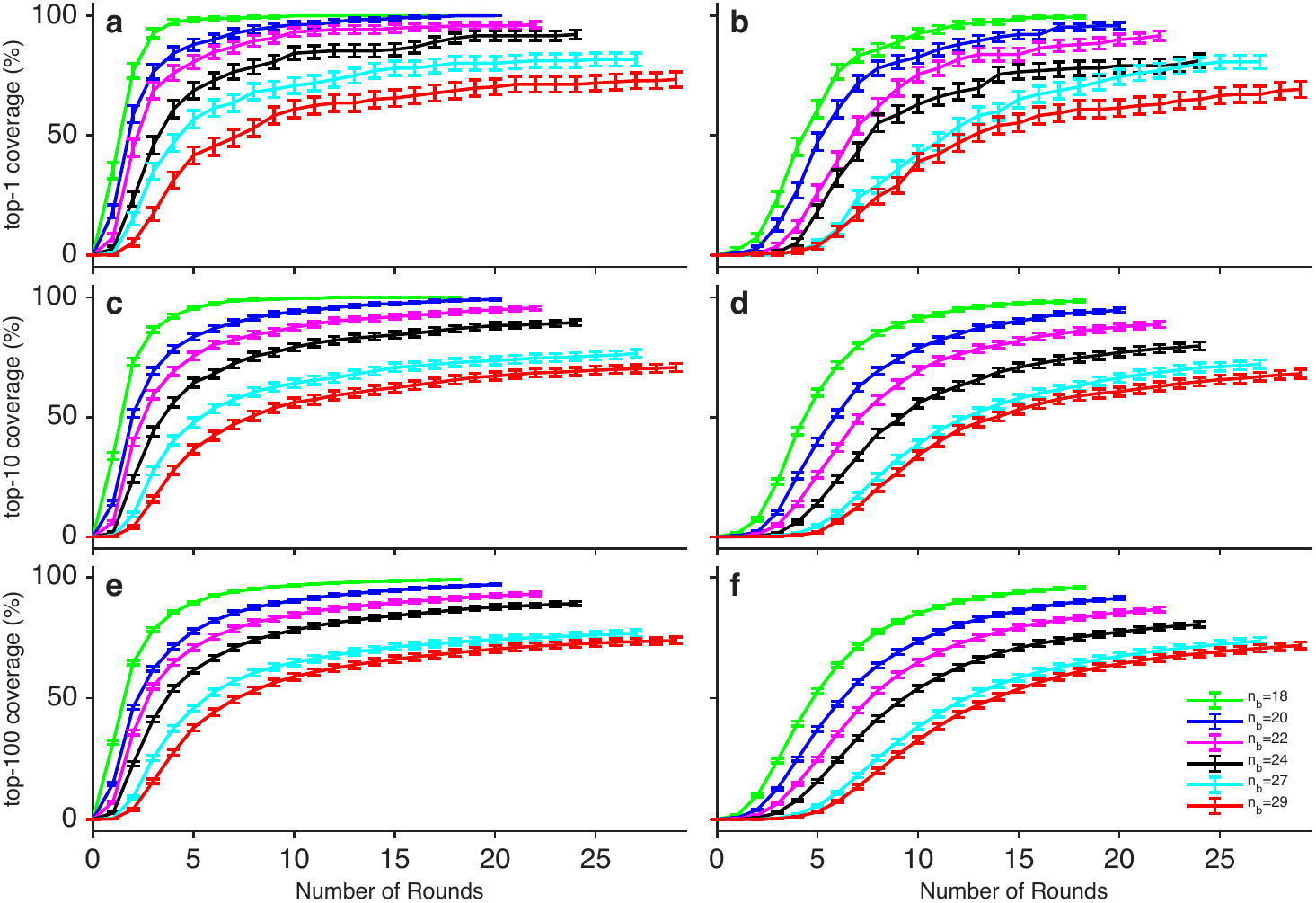}
\caption{\textbf{Target-resolved top-$K$ coverage for aggregated data.}
Classical and QIPS searches are shown in the left and right columns,
respectively. Rows correspond to $K=1$, 10, and 100. Coverage is the
fraction of distinct ground-truth top-$K$ states recovered by the search.
Curves show results for $n_b=18$, 20, 22, 24, 27, and 29, aggregated over
the six benchmark systems. Error bars denote the standard error of the mean.}
\label{fig:coverage}
\end{figure}

\begin{figure}[!htb]
\centering
\includegraphics[
width=0.98\linewidth,
trim={0.01in 0.00in 0.00in 0.00in},
clip
]{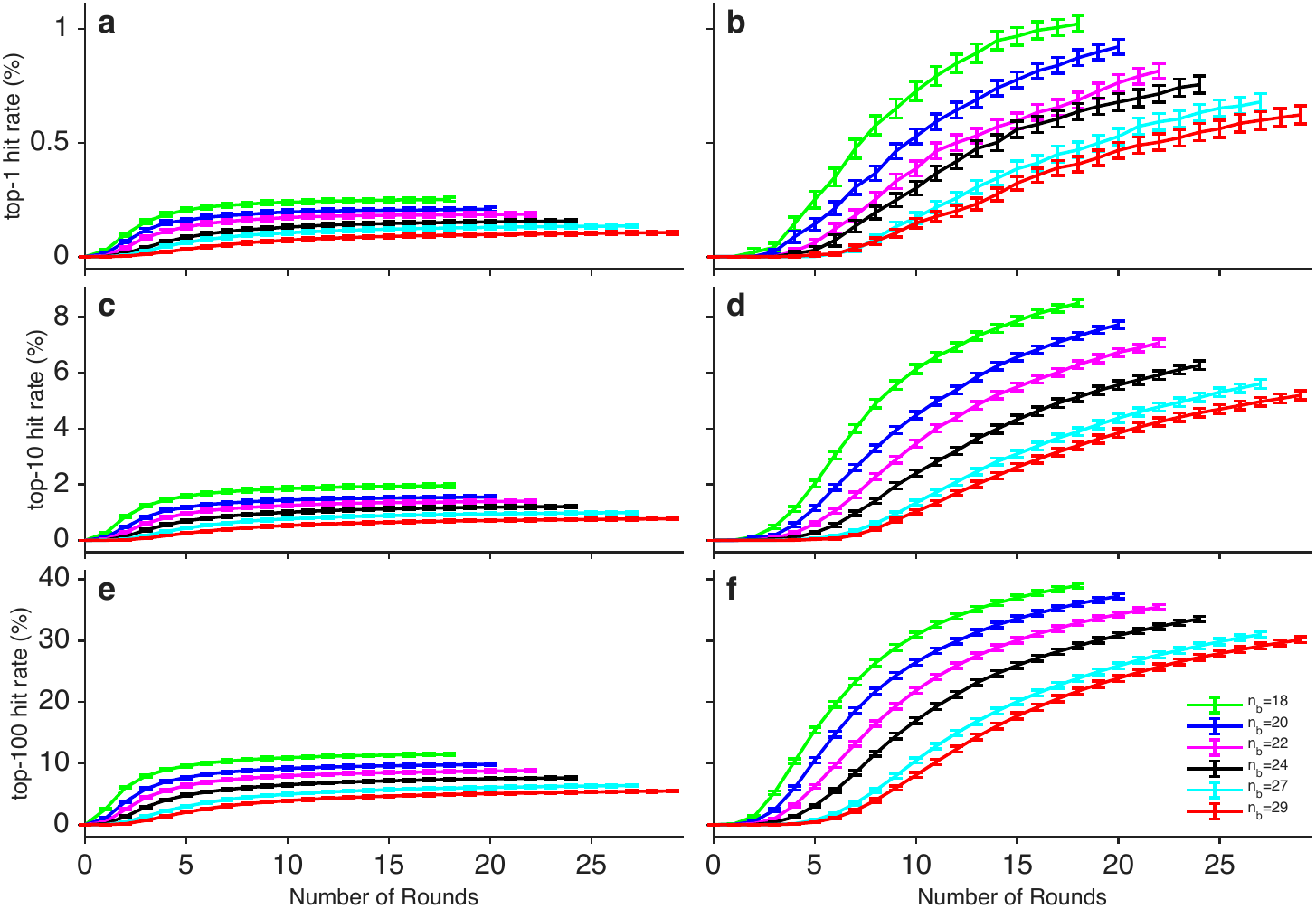}
\caption{\textbf{Target-resolved top-$K$ hit rate for aggregated data.}
Classical and QIPS searches are shown in the left and right columns,
respectively. Rows correspond to $K=1$, 10, and 100. The hit rate is the
fraction of all proposals that belong to the ground-truth top-$K$ set.
Curves show results for $n_b=18$, 20, 22, 24, 27, and 29, aggregated over
the six benchmark systems. Error bars denote the standard error of the mean.}
\label{fig:hitRate}
\end{figure}

\begin{figure}[!htb]
\centering
\includegraphics[
width=0.98\linewidth,
trim={0.01in 0.00in 0.00in 0.00in},
clip
]{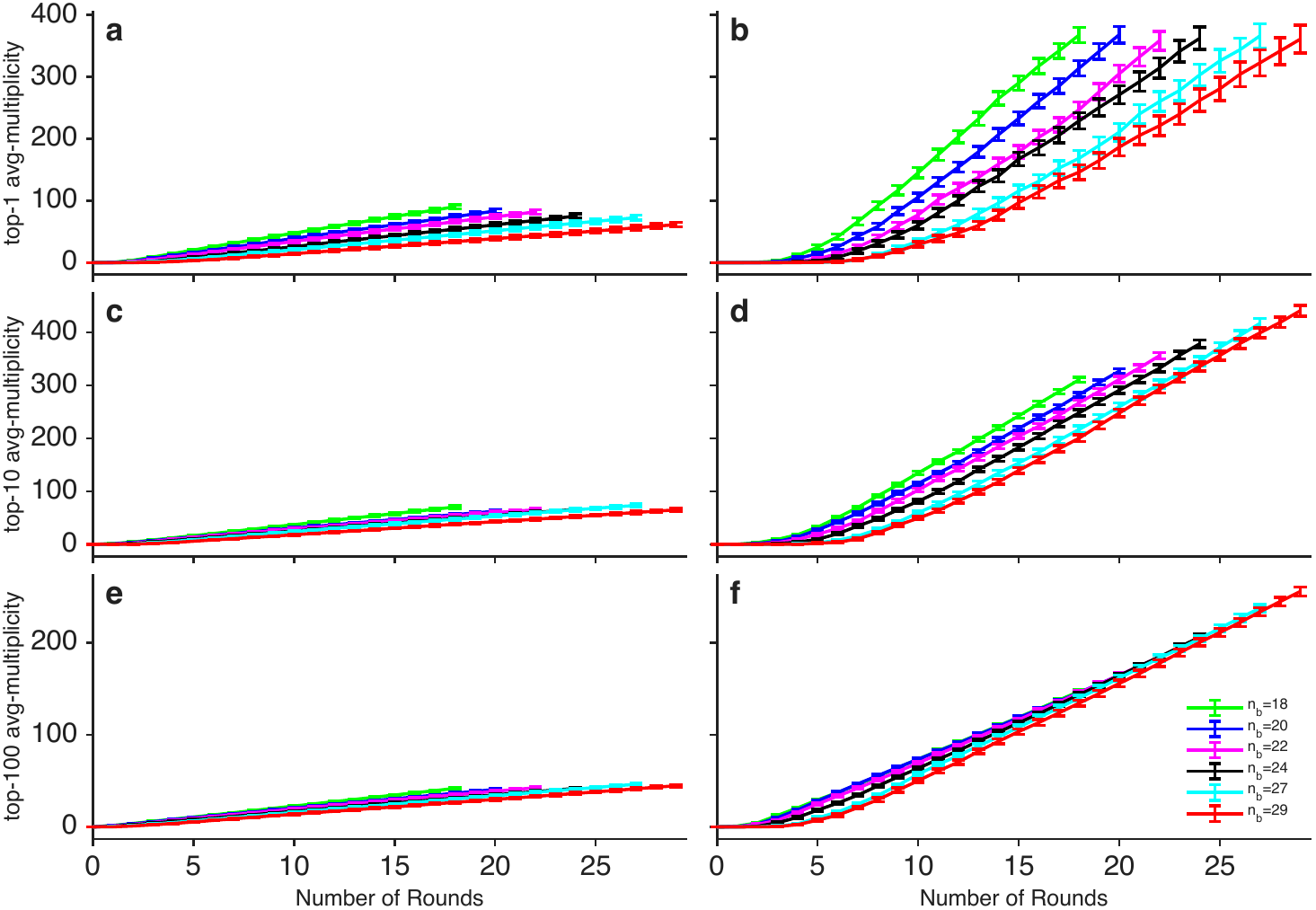}
\caption{\textbf{Target-resolved top-$K$ multiplicity for aggregated data.}
Classical and QIPS searches are shown in the left and right columns,
respectively. Rows correspond to $K=1$, 10, and 100. Mean multiplicity is
the number of top-$K$ hits divided by the number of distinct top-$K$ states
recovered. Curves show results for $n_b=18$, 20, 22, 24, 27, and 29,
aggregated over the six benchmark systems. Error bars denote the standard
error of the mean.}
\label{fig:multiplicity}
\end{figure}

\FloatBarrier
\subsection*{System-resolved benchmarks}
The three top-10 performance measures from the aggregate analysis are shown
separately for each benchmark family. Across all six systems, the classical
and QIPS searches exhibit the same qualitative trends in coverage, hit rate,
and multiplicity, although the absolute performance depends on the problem
family.

%$$$$$$$$$$$$$$$$$$$$$$$$$$$$$$$$$$$$$$$$$$$$$$$$$$$$$$$$$$$$$$
\begin{figure}[!htb]
\centering
\includegraphics[
width=0.98\linewidth,
trim={0.01in 0.00in 0.00in 0.00in},
clip
]{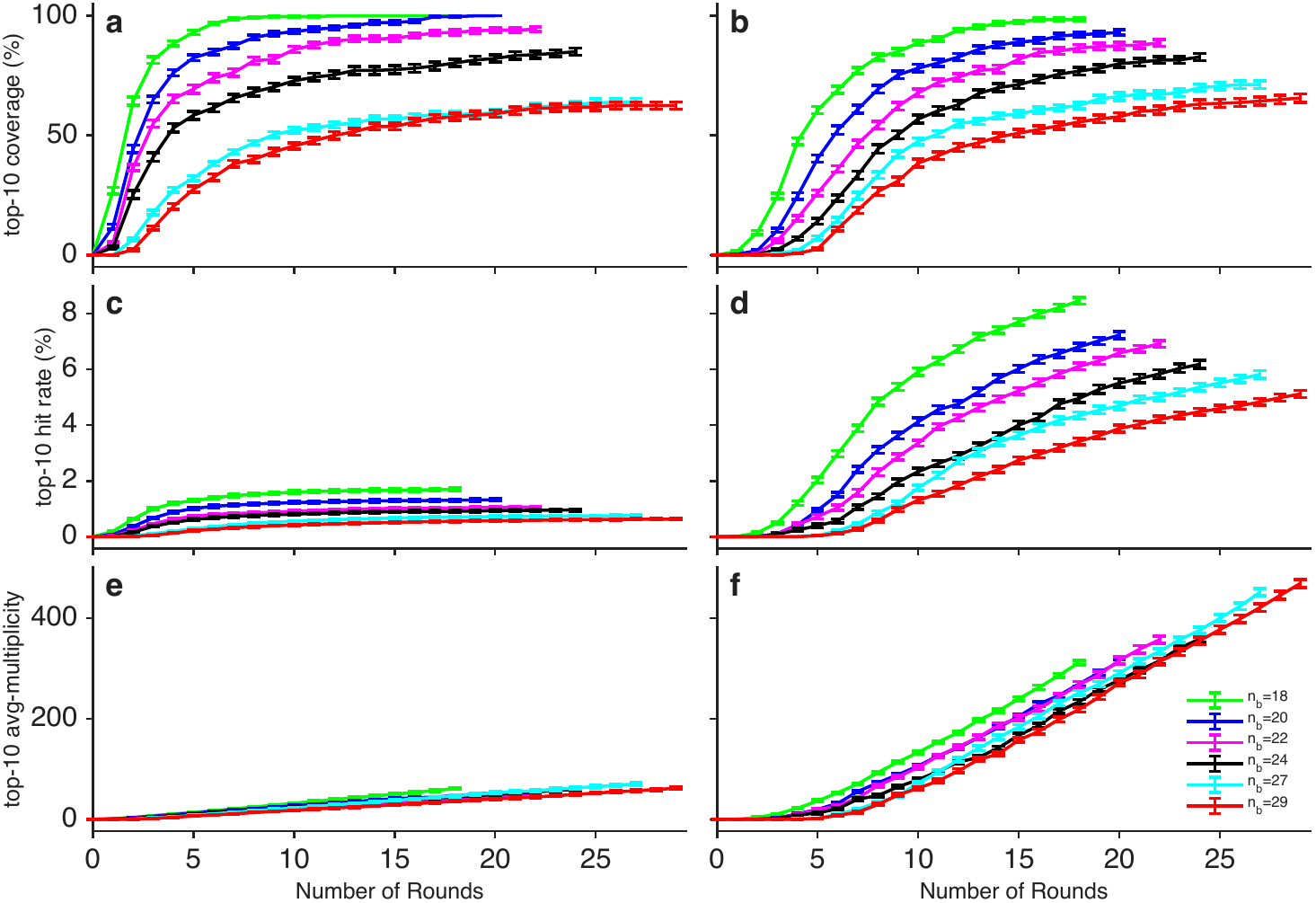}
\caption{\textbf{System-resolved top-10 benchmarks for the 3-RRG Ising model
with 70\% antiferromagnetic and 30\% ferromagnetic couplings.}
Classical and QIPS results are shown in the left and right columns,
respectively. The rows show top-10 coverage, hit rate, and mean multiplicity.
Curves correspond to $n_b=18$, 20, 22, 24, 27, and 29 and are averaged over
32 matched instances. Error bars denote the standard error of the mean.}
\label{fig:top10Sys1}
\end{figure}

%$$$$$$$$$$$$$$$$$$$$$$$$$$$$$$$$$$$$$$$$$$$$$$$$$$$$$$$$$$$$$$
\begin{figure}[!htb]
\centering
\includegraphics[
width=0.98\linewidth,
trim={0.01in 0.00in 0.00in 0.00in},
clip
]{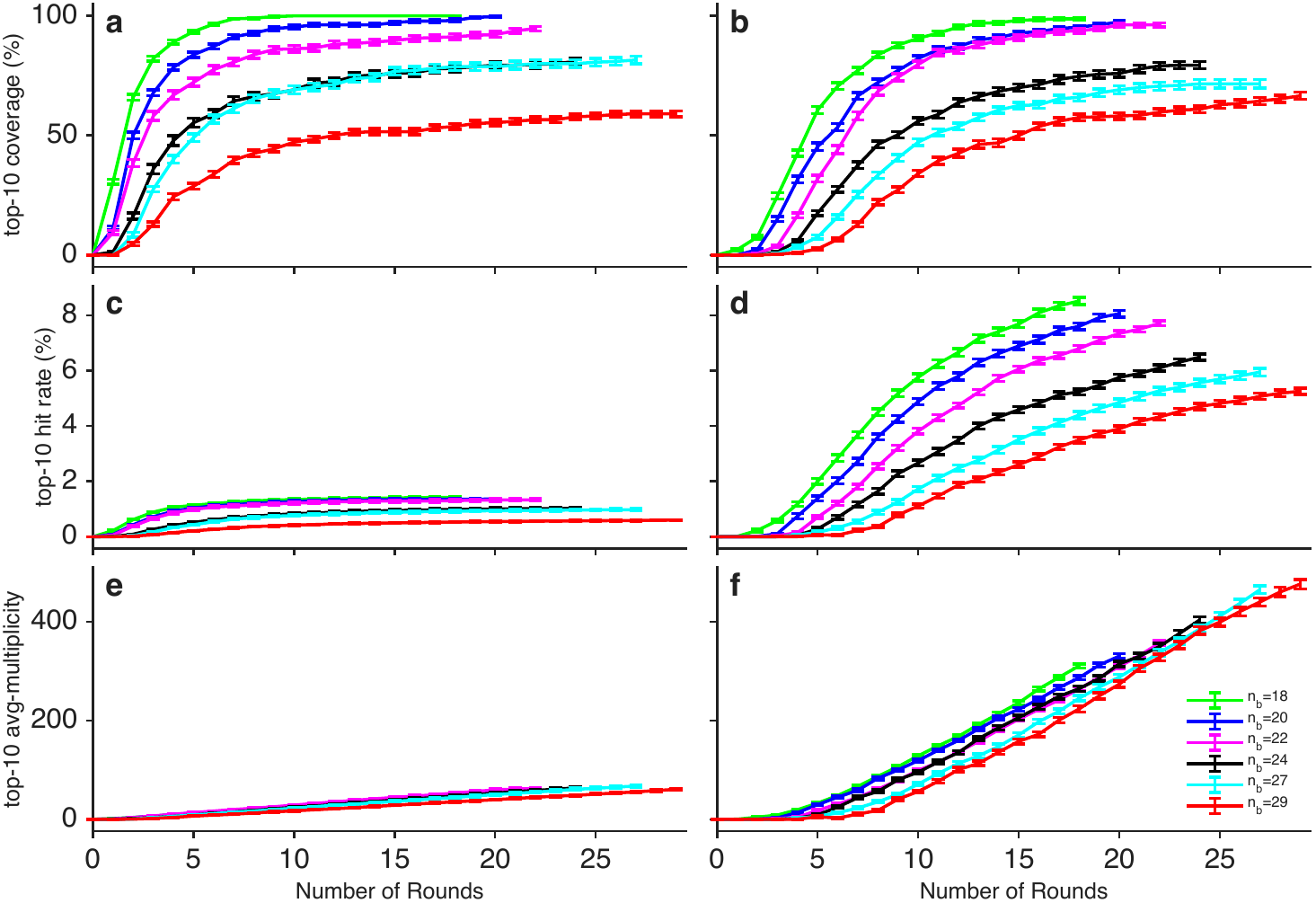}
\caption{\textbf{System-resolved top-10 benchmarks for 3-RRG MaxCut with
uniform edge weights.}
Classical and QIPS results are shown in the left and right columns,
respectively. The rows show top-10 coverage, hit rate, and mean multiplicity.
Curves correspond to $n_b=18$, 20, 22, 24, 27, and 29 and are averaged over
32 matched instances. Error bars denote the standard error of the mean.}
\label{fig:top10Sys2}
\end{figure}

%$$$$$$$$$$$$$$$$$$$$$$$$$$$$$$$$$$$$$$$$$$$$$$$$$$$$$$$$$$$$$$
\begin{figure}[!htb]
\centering
\includegraphics[
width=0.98\linewidth,
trim={0.01in 0.00in 0.00in 0.00in},
clip
]{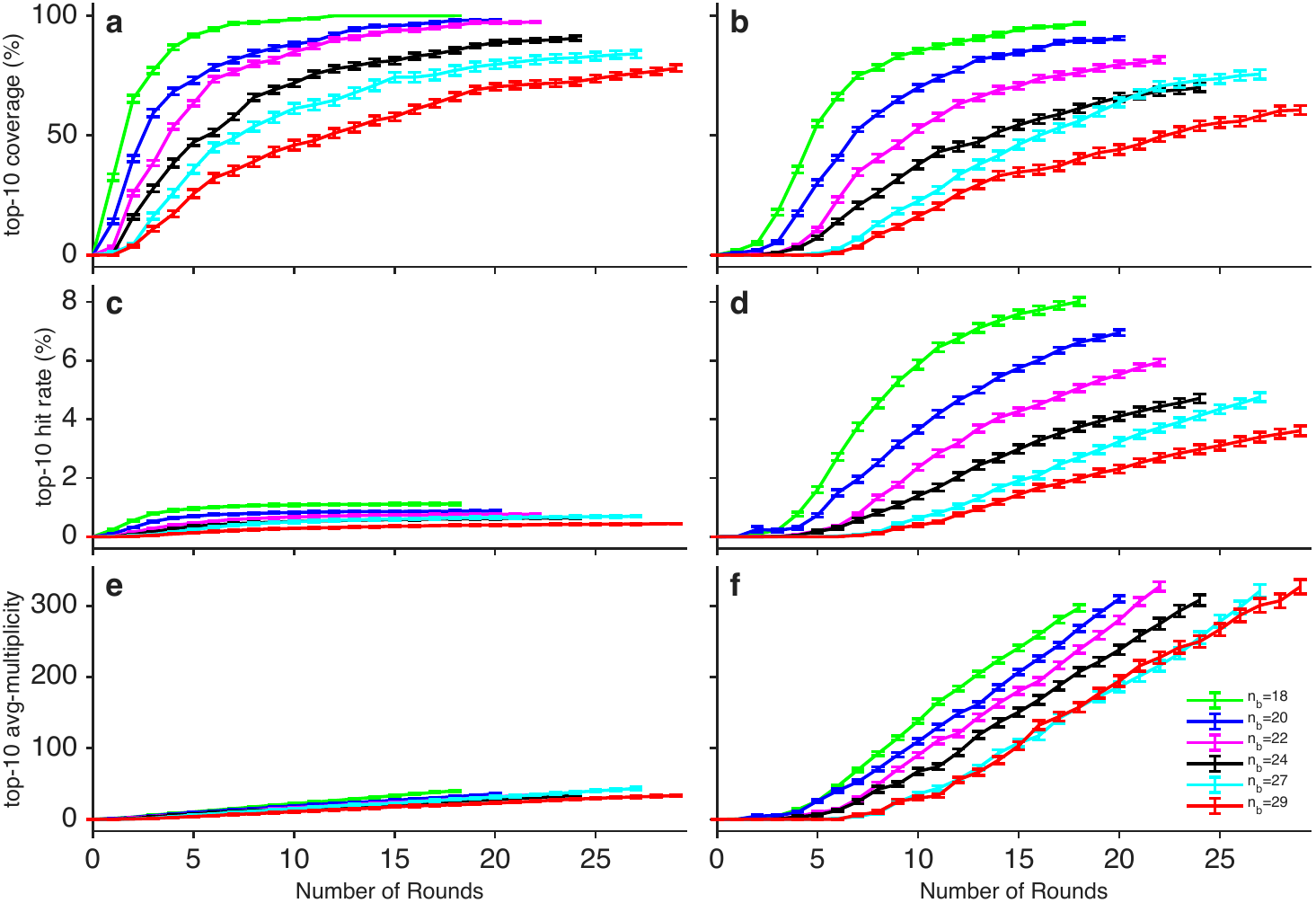}
\caption{\textbf{System-resolved top-10 benchmarks for 3-RRG maximum
independent set with uniform weights and penalties.}
Classical and QIPS results are shown in the left and right columns,
respectively. The rows show top-10 coverage, hit rate, and mean multiplicity.
Curves correspond to $n_b=18$, 20, 22, 24, 27, and 29 and are averaged over
32 matched instances. Error bars denote the standard error of the mean.}
\label{fig:top10Sys3}
\end{figure}

%$$$$$$$$$$$$$$$$$$$$$$$$$$$$$$$$$$$$$$$$$$$$$$$$$$$$$$$$$$$$$$
\begin{figure}[!htb]
\centering
\includegraphics[
width=0.98\linewidth,
trim={0.01in 0.00in 0.00in 0.00in},
clip
]{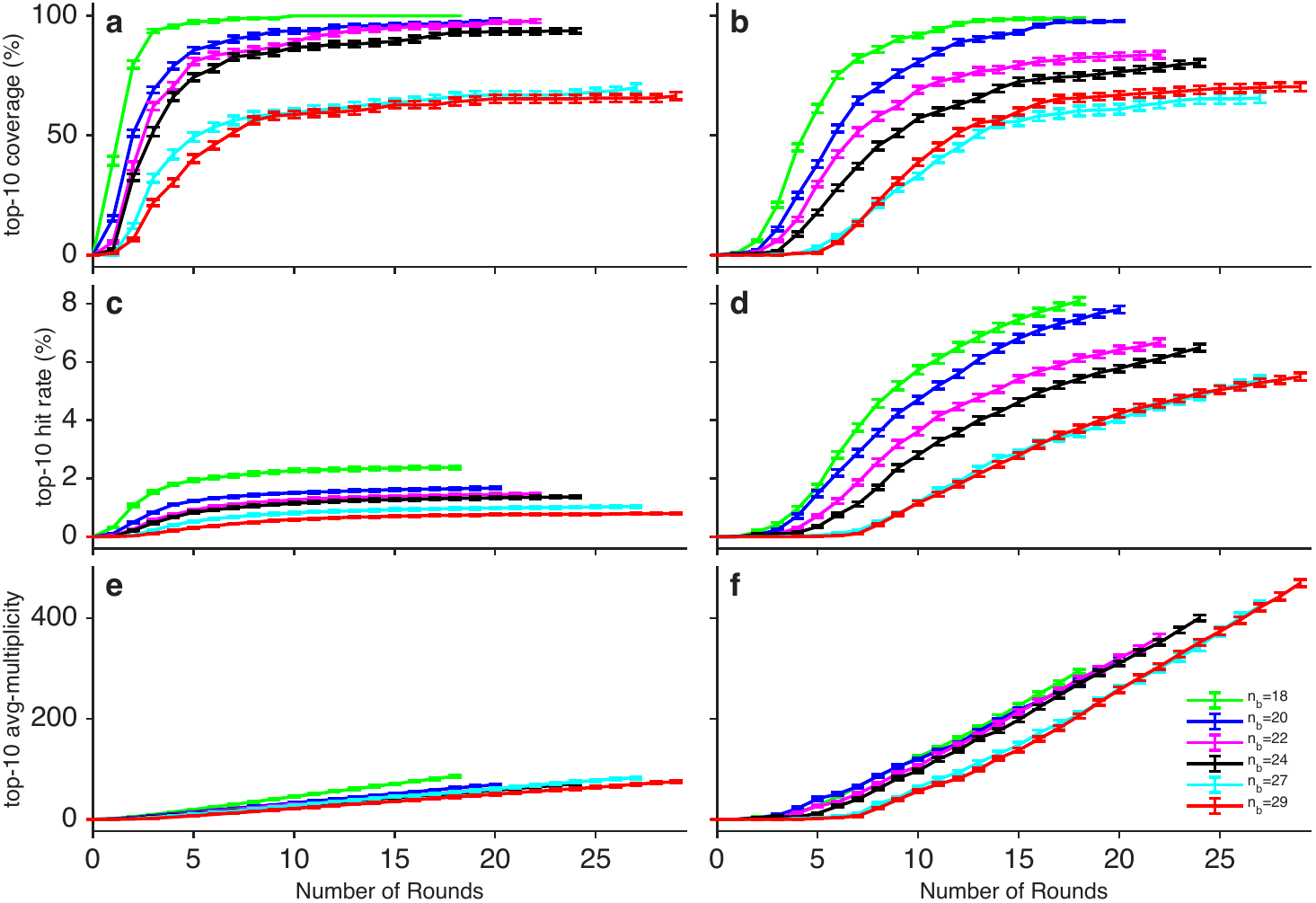}
\caption{\textbf{System-resolved top-10 benchmarks for 6-RRG MaxCut with
log-normal edge weights.}
Classical and QIPS results are shown in the left and right columns,
respectively. The rows show top-10 coverage, hit rate, and mean multiplicity.
Curves correspond to $n_b=18$, 20, 22, 24, 27, and 29 and are averaged over
32 matched instances. Error bars denote the standard error of the mean.}
\label{fig:top10Sys4}
\end{figure}

%$$$$$$$$$$$$$$$$$$$$$$$$$$$$$$$$$$$$$$$$$$$$$$$$$$$$$$$$$$$$$$
\begin{figure}[!htb]
\centering
\includegraphics[
width=0.98\linewidth,
trim={0.01in 0.00in 0.00in 0.00in},
clip
]{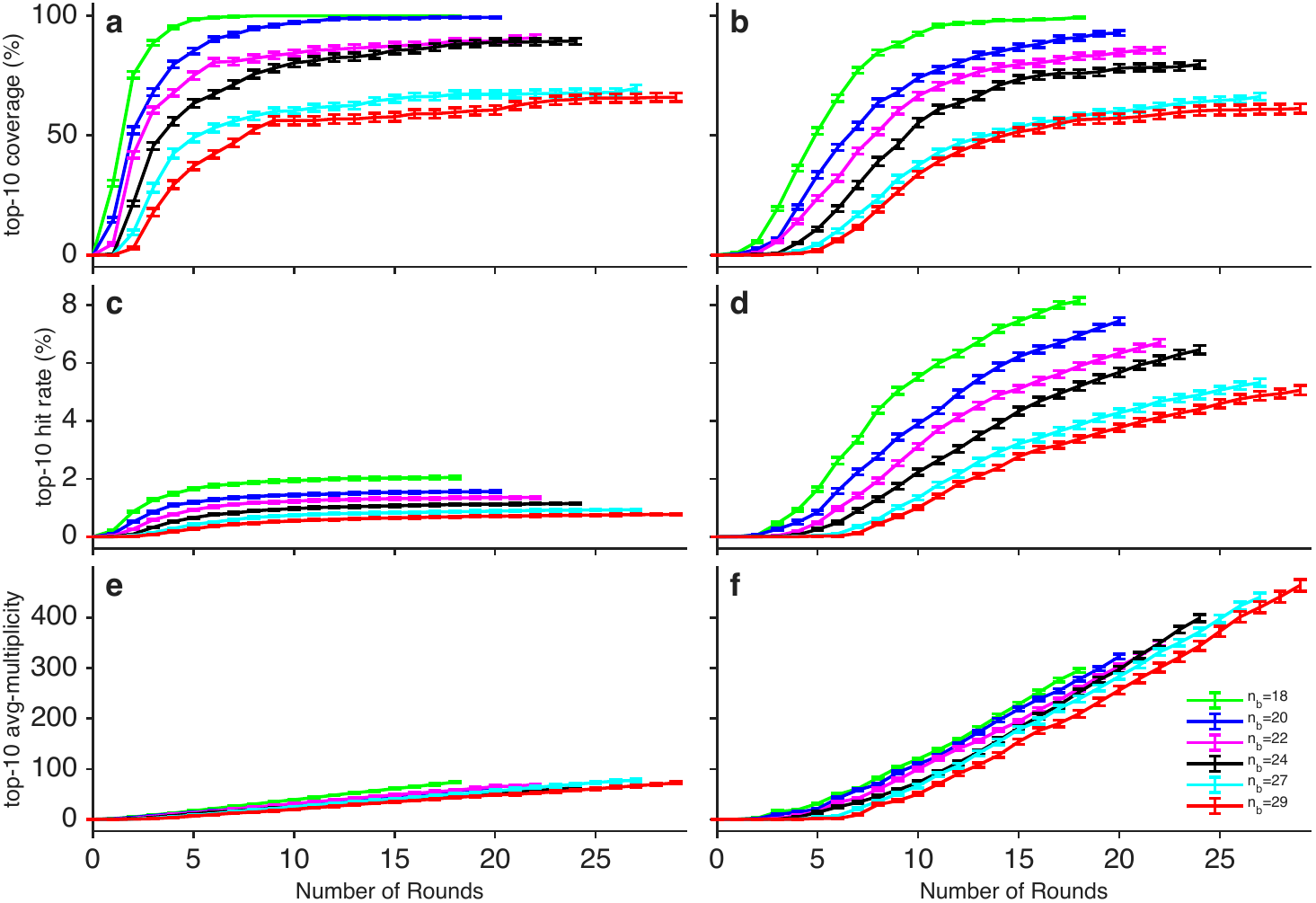}
\caption{\textbf{System-resolved top-10 benchmarks for the complete-graph
Ising model with exponential weak quenched disorder.}
Classical and QIPS results are shown in the left and right columns,
respectively. The rows show top-10 coverage, hit rate, and mean multiplicity.
Curves correspond to $n_b=18$, 20, 22, 24, 27, and 29 and are averaged over
32 matched instances. Error bars denote the standard error of the mean.}
\label{fig:top10Sys5}
\end{figure}

%$$$$$$$$$$$$$$$$$$$$$$$$$$$$$$$$$$$$$$$$$$$$$$$$$$$$$$$$$$$$$$
\begin{figure}[!htb]
\centering
\includegraphics[
width=0.98\linewidth,
trim={0.01in 0.00in 0.00in 0.00in},
clip
]{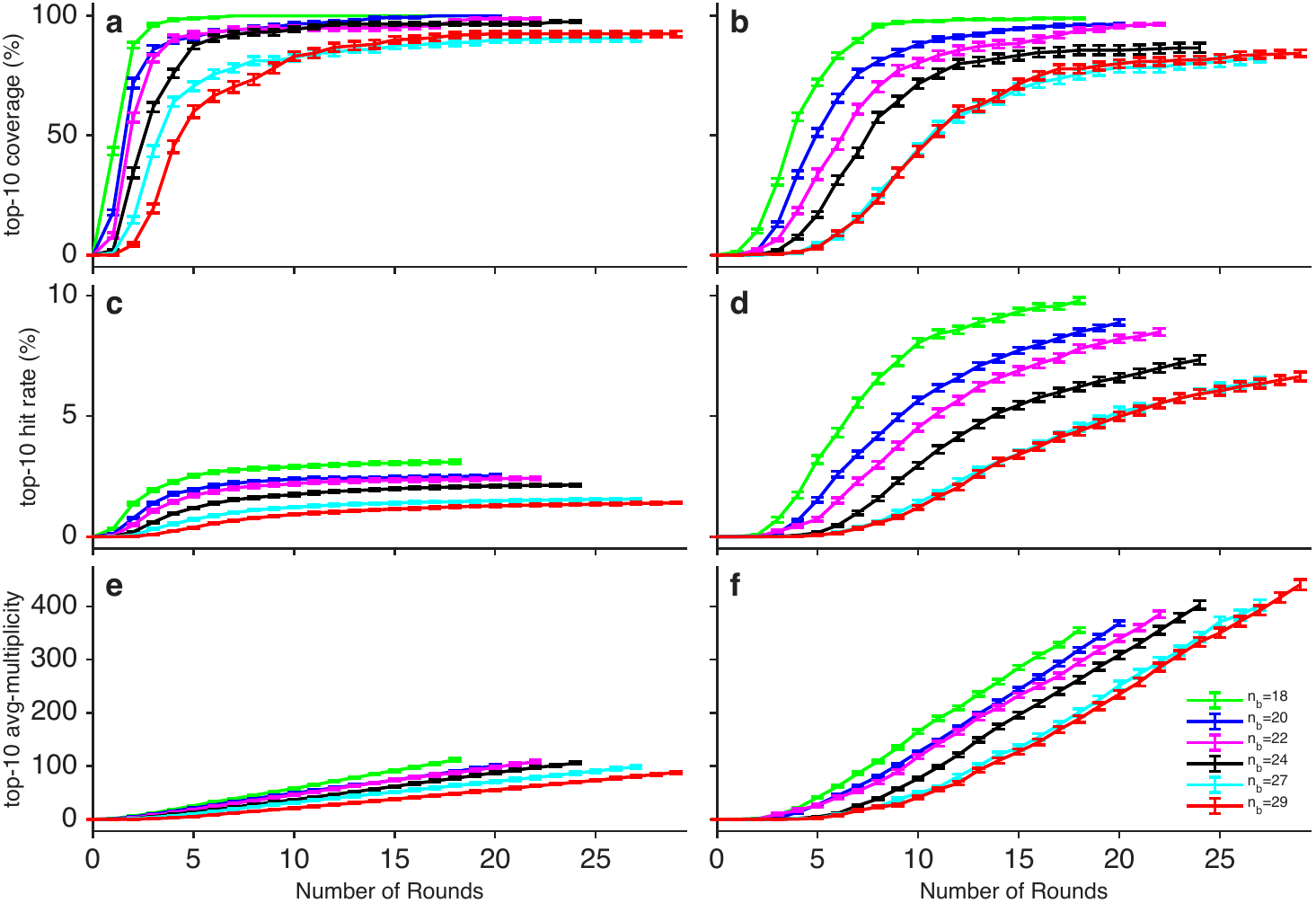}
\caption{\textbf{System-resolved top-10 benchmarks for the
Sherrington--Kirkpatrick model.}
Classical and QIPS results are shown in the left and right columns,
respectively. The rows show top-10 coverage, hit rate, and mean multiplicity.
Curves correspond to $n_b=18$, 20, 22, 24, 27, and 29 and are averaged over
32 matched instances. Error bars denote the standard error of the mean.}
\label{fig:top10Sys6}
\end{figure}

%%%%%%%%%%%%%%%%%%%%%%%%%%%%%%%%%%%%%%%%%%%%%%%%%%%%%%%%%%%%%%%%%%%%%%%%%%%%%%%
%%%%%%%%%%%%%%%%%%%%%%%%%%%%%%%%%%%%%%%%%%%%%%%%%%%%%%%%%%%%%%%%%%%%%%%%%%%%%%%
\FloatBarrier
\suppnote{10}{Circuit construction and feedback control}
{supp:CircuitFeedback}

\subsection*{Algorithmic overview and resource accounting}
Quantum Interference Proposal Search combines a classical outer search with a
quantum proposal generator. The classical component maintains an elite
frontier of low-energy states, selects reference seeds from that frontier,
evaluates newly proposed bitstrings, and updates the frontier. The quantum
component generates seed-conditioned proposal distributions through a
two-layer parameterized circuit. The measured bitstrings are used both as
optimization proposals and as finite-shot diagnostics of the interference
pattern.

The search proceeds in rounds. In each round, $N_{\mathrm{seed}}=20$ seed
states are selected from the current elite frontier. One randomized quantum
circuit is generated for each seed and measured
$N_{\mathrm{shots}}=100$ times. The production calculations use
$N_{\mathrm{round}}=n_b$ rounds, giving

\begin{equation}
N_{\mathrm{circ}}
=
N_{\mathrm{seed}}N_{\mathrm{round}}
=
20n_b
\end{equation}
quantum circuits and a total proposal budget

\begin{equation}
N_{\mathrm{prop}}
=
N_{\mathrm{shots}}N_{\mathrm{circ}}
=
2000n_b.
\label{eq:QIPS_total_budget}
\end{equation}
Repeated measurements and repeated proposals consume the same budget as new
states. The outer-loop logic and proposal budget are held fixed when quantum
and classical proposal generators are compared.

The central aim of the circuit controller is not to minimize an expectation
value of the cost Hamiltonian. Instead, it maintains an ensemble of localized
but diverse quantum interference patterns. Each pattern is centered
statistically around a selected seed while retaining enough additional
measurable outcomes to explore nearby and nonlocal regions of configuration
space. The seed energies therefore provide the principal downhill bias,
whereas the feedback controller regulates localization and diversity.

\subsection*{Cost normalization and static degeneracy breaking}

Let $E_0(s)$ denote the original QUBO or Ising energy of computational-basis
state $s$.  An extensive number of degeneracies can create practical difficulty
for a frontier because a large number of equally ranked states will belong to 
the same energy plateau. Requiring the search to enumerate an extensive degenerate 
manifold would defeat the purpose of maintaining a small elite frontier. This
is why the frontier is limited to a fixed size, and why degeneracies are 
intentionally lifted.

To define an operational ordering through such plateaus, a small static
perturbation is added to the couplings before the search begins. For edge
$(i,j)$, the disorder scale is

\begin{equation}
\sigma_{ij}^{(0)}
=
0.3\max\left(
|J_{ij}|,
\frac{\Delta E_{\mathrm{tol}}}{4}
\right),
\label{eq:static_disorder_scale}
\end{equation}
where $\Delta E_{\mathrm{tol}}=10^{-5}$ is the numerical tolerance used to
distinguish nearby energy levels. Independent Gaussian perturbations with
these standard deviations define a static disorder spectrum
$\Delta E_{\mathrm{stat}}(s)$. The operational search energy is

\begin{equation}
E(s)
=
E_0(s)
+
\epsilon_{\mathrm{stat}}\Delta E_{\mathrm{stat}}(s),
\qquad
\epsilon_{\mathrm{stat}}=10^{-9}.
\label{eq:static_degeneracy_breaking}
\end{equation}

The perturbation is too small to alter the optimization objective at the
reported numerical precision. Its role is to select a definite downhill path
among otherwise equivalent states. Performance analysis remains referenced to
the original unperturbed spectrum, for which exactly degenerate states are
treated as energetically equivalent.

The operational spectrum is converted to a dimensionless phase function. An
energy scale is estimated from 1000 uniformly sampled basis states,

\begin{equation}
E_{\mathrm{ref}}
=
\max\left[
\operatorname{median}_{s\in\mathcal S}|E(s)|,
0.1
\right],
\qquad |\mathcal S|=1000.
\label{eq:Eref}
\end{equation}
The normalized cost function used by the phase separator is

\begin{equation}
K(s)
=
\frac{2\pi E(s)}{E_{\mathrm{ref}}}.
\label{eq:normalized_cost}
\end{equation}
Only a rough common scaling is required because the phase-separator strengths
and circuit deviations are themselves randomized and regulated by feedback.

\subsection*{Canonical seed-state encoding}

Each circuit is conditioned on a reference seed bitstring

\begin{equation}
\mathbf b^{(\mathrm{seed})}
=
\left(
b_1^{(\mathrm{seed})},
\ldots,
b_{n_b}^{(\mathrm{seed})}
\right),
\qquad
b_j^{(\mathrm{seed})}\in\{0,1\}.
\end{equation}
The quantum register begins in the uniform superposition $\ket{\psi_0} = \ket{+}^{\otimes n_b}$.
The seed is not prepared directly as a computational-basis input state.
Instead, it is encoded through the canonical orientation of the second mixing
layer. The canonical angles are

\begin{align}
\bar{\delta}_1^j &= 0,
&
\bar{\theta}_1^j &= \frac{\pi}{4},
&
\bar{\phi}_1^j &= 0,
&
\bar{\alpha}_1^j &= \frac{\pi}{2},
\\
\bar{\delta}_2^j &= 0,
&
\bar{\theta}_2^j &=
\frac{\pi}{2}b_j^{(\mathrm{seed})},
&
\bar{\phi}_2^j &= 0,
&
\bar{\alpha}_2^j &= \frac{\pi}{2}.
\label{eq:canonical_angles}
\end{align}
Thus, all canonical angles are independent of the seed except
$\bar{\theta}_2^j$. The actual circuit angles are obtained by adding
stochastically generated deviations to these canonical values.

\subsection*{Two-layer quantum proposal circuit}
For qubit $j$ in layer $\ell$, define the unit vector

\begin{equation}
\mathbf n_\ell^j
=
\left(
\sin\theta_\ell^j\cos\phi_\ell^j,
\sin\theta_\ell^j\sin\phi_\ell^j,
\cos\theta_\ell^j
\right),
\end{equation}
and the corresponding single-qubit generator
$\hat B_\ell^j = \mathbf n_\ell^j\cdot\boldsymbol{\sigma}_j$. 
The single-qubit mixing rotation is

\begin{equation}
U_{B,\ell}^j
=
\exp\left(
-i\alpha_\ell^j\hat B_\ell^j
\right),
\end{equation}
and the complete mixing layer is

\begin{equation}
U_{B,\ell}
=
\prod_{j=1}^{n_b}U_{B,\ell}^j.
\end{equation}
Each phase-separator layer contains the normalized cost operator and an
optional longitudinal field,

\begin{equation}
\hat D_\ell
=
\sum_{j=1}^{n_b}\delta_\ell^j Z_j.
\end{equation}
Because $\hat K$ and $\hat D_\ell$ are diagonal in the computational basis,
they commute. The combined diagonal layer may therefore be written as

\begin{equation}
U_{K,\ell}
=
\exp\left[
-i\gamma_\ell
\left(
\hat K+\hat D_\ell
\right)
\right].
\end{equation}
The full two-layer circuit is

\begin{equation}
\ket{\psi_{\mathrm{QIP}}}
=
U_{B,2}
U_{K,2}
U_{B,1}
U_{K,1}
\ket{+}^{\otimes n_b}.
\label{eq:QIPS_circuit}
\end{equation}
The primary operator architecture is denoted
$K$--XYZ--$K$--XYZ because the mixer axes may have nonzero $x$, $y$, and
$z$ components. An ablated $K$--XZ--$K$--XZ form is obtained by setting
$\phi_\ell^j=0$. The corresponding proposal probability is

\begin{equation}
p_s
=
\left|
\braket{s|\psi_{\mathrm{QIP}}}
\right|^2.
\label{eq:QIPS_probability}
\end{equation}

\subsection*{Stochastic parameterization of angle deviations}

The circuit contains qubit-resolved angle deviations, but these deviations are
not optimized independently. Instead, their probability distributions are
controlled by six latent variables,

\begin{equation}
\left\{
a_t,
a_{u_1},
a_{u_2},
a_{v_1},
a_{v_2},
a_w
\right\}.
\end{equation}
They are mapped to bounded control variables through

\begin{align}
t &= \sin^2 a_t,
\\
u_\ell &= \sin^2 a_{u_\ell},
\\
v_\ell &= \sin^2 a_{v_\ell},
\\
w &= \sin^2 a_w.
\label{eq:latent_mapping}
\end{align}
The variable $t$ controls the overall deviation scale. The variables $u_1$
and $u_2$ regulate the coherent layer-wide displacement, while $v_1$ and
$v_2$ regulate qubit-to-qubit scatter. The variable $w$ divides the total
phase-separator strength between the two layers.

Trial latent angles are generated around the most recently accepted values.
For example,

\begin{equation}
a_t^{\mathrm{trial}}
=
a_t^{\mathrm{last}}
+
S\Delta a_t Z,
\qquad
Z\sim\mathcal N(0,1),
\label{eq:latent_trial}
\end{equation}
where $S$ is an adaptively regulated step-size factor. Analogous updates are
used for the remaining latent variables. The baseline proposal increments are

\begin{equation}
\Delta a_t=5^\circ,
\qquad
\Delta a_{u_1}
=
\Delta a_{u_2}
=
\Delta a_{v_1}
=
\Delta a_{v_2}
=
10^\circ,
\qquad
\Delta a_w=20^\circ.
\label{eq:latent_step_sizes}
\end{equation}
The initial bounded variables are approximately

\begin{equation}
t=u_1=u_2=v_1=v_2=0.3,
\qquad
w=0.7.
\end{equation}
Four angular scales define the width of the generated deviations,

\begin{equation}
a=1^\circ,
\qquad
b=4^\circ,
\qquad
c=5^\circ,
\qquad
d=20^\circ.
\end{equation}

For layer $\ell$, define

\begin{align}
A_\ell
&=
(a+b u_\ell)t,
\\
B_\ell
&=
(c+d u_\ell)t,
\\
\sigma_\ell
&=
\left[
a+(c+d)v_\ell
\right]t.
\label{eq:angle_scales}
\end{align}
For each angle family
$\chi\in\{\delta,\theta,\phi,\alpha\}$, the deviation has the form

\begin{equation}
\Delta\chi_\ell^j
=
s_{\chi,\ell}
\left(
A_\ell+B_\ell R_{\chi,\ell}
\right)
+
\eta_{\chi,\ell}^j,
\label{eq:angle_deviation}
\end{equation}
where $s_{\chi,\ell}\in\{-1,+1\}$ is chosen randomly,
$R_{\chi,\ell}$ is uniform on $[0,1]$, and
$\eta_{\chi,\ell}^j$ is Gaussian qubit-level scatter with zero mean across
the layer and standard deviation set by $\sigma_\ell$.

The total phase-separator scale is

\begin{equation}
C_\gamma
=
a+(b+c+d)t,
\end{equation}
and is divided between the two layers according to

\begin{equation}
\gamma_1
=
C_\gamma w,
\qquad
\gamma_2
=
C_\gamma(1-w).
\label{eq:gamma_partition}
\end{equation}
In the production calculations underlying all reported results, the annealing
option was disabled, so no additional dynamical scaling was applied to
$\gamma_1$ or $\gamma_2$. In addition, the two layer-specific scatter controls
were constrained by setting $v_2=v_1$.

The accepted latent variables persist as the seed changes. Because only
$\bar{\theta}_2^j$ depends on the seed, the deviation process is largely
decoupled from the canonical configuration. The retained deviations therefore
behave approximately as a stationary stochastic process, while continued
feedback allows them to re-equilibrate when the selected seed changes.

\subsection*{Dynamic cost-operator jitter}
In addition to the static perturbation used to break degeneracies, QIPS applies
a larger dynamic perturbation during circuit generation. This feature was
introduced because the dominant peaks of the quantum interference pattern for
a fixed problem instance were otherwise difficult to redistribute. Adding
random jitter to the cost operator provides a computationally inexpensive way
to diversify these peaks while preserving the underlying QUBO structure,
thereby improving the consistency and robustness of the proposal generator.

To obtain diverse jitter realizations at low computational cost, four
independent disorder spectra are precomputed and combined with evolving random
weights during the search. This produces a broad family of structured
cost-operator perturbations without regenerating the disorder from scratch for
every circuit, which ensures computational efficiency in the simulations. 

The four coupling scales are

\begin{align}
\sigma_{ij}^{(0)}
&=
0.3\max\left(
|J_{ij}|,
\frac{\Delta E_{\mathrm{tol}}}{4}
\right),
\\
\sigma_M
&=
0.3\exp\left[
\left\langle
\ln
\max\left(
|J_{ij}|,
\frac{\Delta E_{\mathrm{tol}}}{4}
\right)
\right\rangle_{ij}
\right],
\\
\sigma_{ij}^{(1)}
&=
\left(\sigma_M\right)^{1/2}
\left(\sigma_{ij}^{(0)}\right)^{1/2},
\\
\sigma_{ij}^{(2)}
&=
\left(\sigma_M\right)^{1/4}
\left(\sigma_{ij}^{(0)}\right)^{3/4}.
\label{eq:jitter_scales}
\end{align}
These define disorder spectra
$\Delta E_0(s)$, $\Delta E_1(s)$, $\Delta E_2(s)$, and
$\Delta E_M(s)$. Their weights evolve according to

\begin{equation}
w_\mu(t)
=
0.9w_\mu(t-1)
+
0.1\xi_\mu(t),
\label{eq:jitter_memory}
\end{equation}
where

\begin{equation}
\xi_\mu
=
\pm\sqrt{|Z|},
\qquad
Z\sim\mathcal N(0,1).
\end{equation}
The instantaneous jitter spectrum is

\begin{equation}
\Delta E_{\mathrm{jit}}(s,t)
=
\sum_{\mu\in\{0,1,2,M\}}
w_\mu(t)\Delta E_\mu(s).
\label{eq:jitter_spectrum}
\end{equation}
QIPS tracks exponentially smoothed frontier-contraction success rates for
jittered and unjittered circuits,

\begin{equation}
f_x(t)
=
0.95f_x(t-1)+0.05I_x(t),
\label{eq:jitter_success_rate}
\end{equation}
where $I_x=1$ when the corresponding proposal contracts the outer frontier
and is zero otherwise. The jitter amplitude is

\begin{equation}
A_{\mathrm{jit}}
=
10^{p_w},
\qquad
p_w
=
\left(2f_{\mathrm{jit}}-1\right)
\left(3-2f_0\right),
\label{eq:jitter_amplitude}
\end{equation}
where $f_{\mathrm{jit}}$ and $f_0$ are the running success rates for the
jittered and unjittered channels.

When jitter is selected, the phase function becomes

\begin{equation}
K_t(s)
=
\frac{2\pi}{E_{\mathrm{ref}}}
\left[
E(s)
+
A_{\mathrm{jit}}\Delta E_{\mathrm{jit}}(s,t)
\right].
\label{eq:jittered_cost}
\end{equation}
The choice between jittered and unjittered circuits is itself stochastic and
is biased by their running success rates. If both channels have success rates
below 0.2, the unjittered operator is selected to preserve fidelity to the
original cost function near the end of a difficult search.
The dynamic perturbation does not alter how proposal energies are evaluated.
All measured states are ranked using the operational objective $E(s)$, not the
instantaneously jittered spectrum used to generate the circuit phase.

\subsection*{Finite-shot proposal generation}

After the circuit is applied, the probability distribution in
Eq.~\eqref{eq:QIPS_probability} is measured
$N_{\mathrm{shots}}$ times. The resulting list of bitstrings forms the
proposal batch associated with the current seed.
The finite-shot record also defines two localization statistics:

\begin{equation}
n_{\mathrm{seed}}
=
\sum_{m=1}^{N_{\mathrm{shots}}}
\mathbf 1
\left[
s_m=s_{\mathrm{seed}}
\right],
\end{equation}
and

\begin{equation}
n_{\mathrm{unique}}
=
\left|
\left\{
s_m
\right\}_{m=1}^{N_{\mathrm{shots}}}
\right|.
\end{equation}
Here, $n_{\mathrm{seed}}$ is the multiplicity of the reference seed and
$n_{\mathrm{unique}}$ is the number of distinct measured bitstrings.
A useful localized quantum interference pattern should return the seed
repeatedly while also producing a relatively small number of additional 
statistically accessible states (say 5\% to 10\% of the number of shots). 
A nearly uniform distribution gives too many distinct outcomes and too 
little seed recurrence, whereas excessive concentration on the seed gives 
too few distinct proposals.

\subsection*{Efficient sampling of localized probability distributions}

Efficient simulation of localized quantum interference patterns requires
special treatment of the measurement distribution. Because the vast majority
of basis states carry negligible probability in a localized pattern, it is
unnecessary to retain and process the full array of $2^{n_b}$ probabilities
when generating finite-shot outcomes. Instead, the simulation identifies the
statistically significant component explicitly and treats the remaining
low-probability mass collectively. Most generated patterns remain at least
moderately localized because of the feedback controller, but the procedure
also accommodates strongly extended cases.

The computational task is therefore to identify the largest values of $p_s$
and define an appropriate cutoff. Directly sorting all $2^{n_b}$
probabilities becomes expensive at the largest system sizes, particularly
when broad low-probability tails contain millions of nonzero amplitudes even
though the statistically relevant component remains localized. Therefore
an accelerated sampling procedure is implemented that separates an explicitly 
retained localized component from an extended background.

By defining  $N_{\mathrm{keep}}^{\max} = 1000N_{\mathrm{shots}}$, the largest
probabilities are first identified without sorting the entire Hilbert space. A 
nominal cutoff of  $p_{\mathrm{cut}} = \frac{0.005}{N_{\mathrm{shots}}}$ is used 
with a minimum floor that retains at least 10000 states when the Hilbert space is
large.  Let $\mathcal L$ denote the retained localized set. Its explicit probability
mass is

\begin{equation}
q_{\mathrm{loc}}
=
\sum_{s\in\mathcal L}p_s.
\end{equation}
The remaining mass is

\begin{equation}
q_{\mathrm{ext}}
=
1-q_{\mathrm{loc}}.
\label{eq:extended_mass}
\end{equation}
The localized and extended shot counts are approximated by

\begin{equation}
N_{\mathrm{ext}}
=
\left\lceil
N_{\mathrm{shots}}q_{\mathrm{ext}}
\right\rceil,
\qquad
N_{\mathrm{loc}}
=
N_{\mathrm{shots}}-N_{\mathrm{ext}}.
\label{eq:shot_partition}
\end{equation}
The $N_{\mathrm{loc}}$ localized measurements are sampled from the retained
states according to their relative probabilities. The
$N_{\mathrm{ext}}$ extended measurements are drawn uniformly from the full
computational basis as an approximation to the diffuse residual component.

Importantly, the extended probability mass is not transferred to the
localized states. The retained probabilities are renormalized only for the
conditional task of drawing the localized subset of shots, while the number
of such shots is reduced according to $q_{\mathrm{ext}}$. The procedure
therefore avoids artificially enhancing the overall sampling frequency of the
localized component.
This protocol was implemented before the extended component was found to retain
a stronger-than-uniform bias toward low-energy states, as described by the
probit model. Treating the extended background as uniform therefore
underestimates its low-energy contribution and provides a conservative
approximation to ideal quantum sampling.

After most simulations were complete, the extended component of the quantum
interference pattern was found to retain reproducible structure described by a
probit model. The code was therefore augmented with an ultracompressed storage
scheme that reconstructs the full probability distribution $p_s$ to a
user-specified accuracy. This procedure is used only when saving probability
records for subsequent analysis and does not affect the proposals generated
during the search.

\subsection*{Pseudo-energy objective for interference localization}

Each finite-shot record is assigned a pseudo-energy that depends only on
$n_{\mathrm{seed}}$ and $n_{\mathrm{unique}}$,

\begin{equation}
E_{\mathrm{QIP}}
=
E_{\mathrm{DS}}(n_{\mathrm{seed}})
+
E_{\mathrm{NU}}(n_{\mathrm{unique}}).
\label{eq:QIP_pseudo_energy}
\end{equation}
The subscripts DS and NU refer to the seed-state degeneracy and number of
unique outcomes, respectively. The objective contains no direct energy-based
reward for the measured bitstrings. The pseudo-energy landscape used in this 
work is shown in Supplementary Figure~\ref{fig:QIPobjective}. Its role is to 
maintain a statistically useful degree of localization.

\begin{figure}[!h]
    \centering
    \includegraphics[
        width=0.92\linewidth,
        trim={0.01in 0.00in 0.00in 0.00in},
        clip
    ]{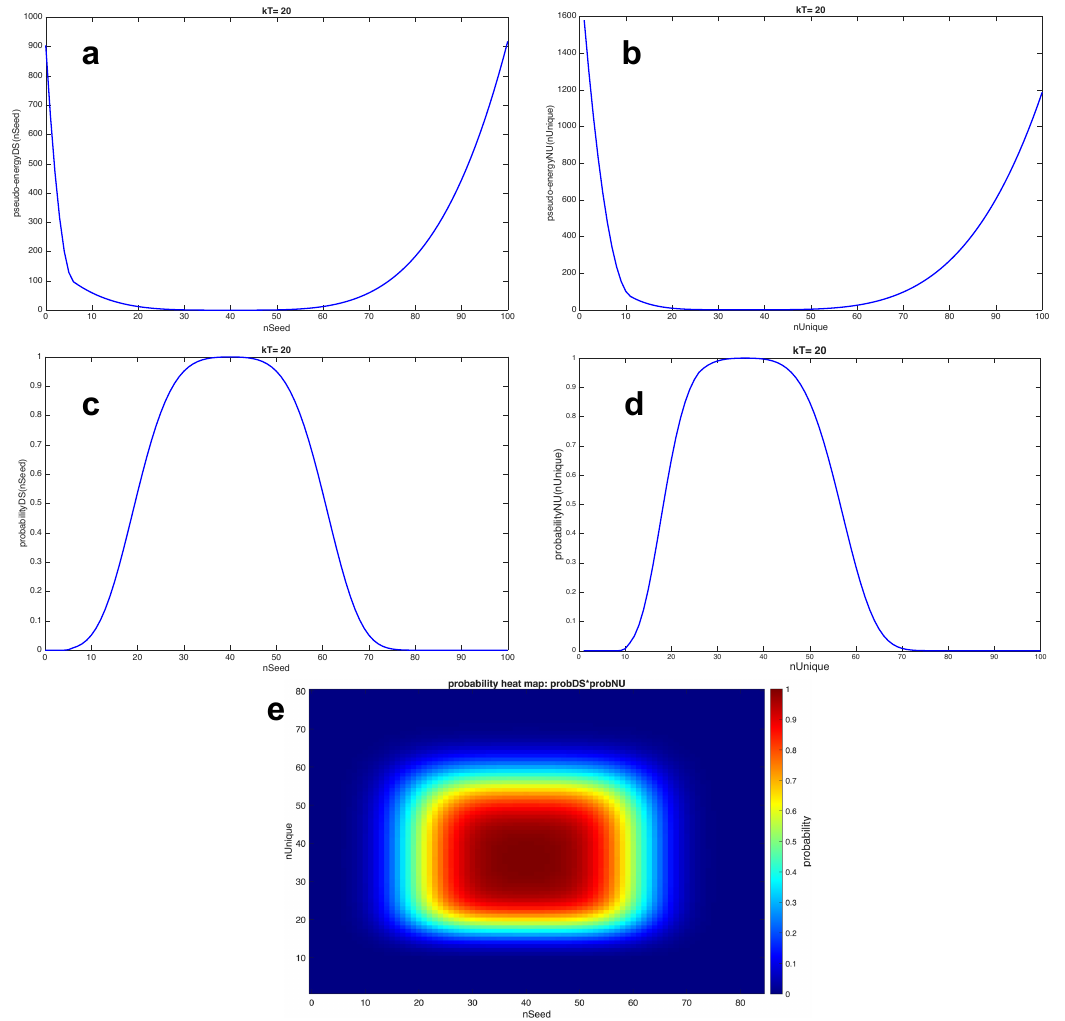}
    \caption{\textbf{Pseudo-energy objective used to regulate localized
    quantum interference patterns.}
    \textbf{a,b}, Pseudo-energy contributions associated with the measured
    seed multiplicity $n_{\mathrm{seed}}$ and number of distinct outcomes
    $n_{\mathrm{unique}}$, respectively, for
    $N_{\mathrm{shots}}=100$.
    \textbf{c,d}, Corresponding Boltzmann weights
    $\exp[-E_{\mathrm{DS}}/kT]$ and
    $\exp[-E_{\mathrm{NU}}/kT]$ for $kT=20$.
    \textbf{e}, Product of the two Boltzmann factors, showing the joint
    preference for intermediate seed multiplicity and a finite number of
    distinct outcomes. The objective suppresses both delocalized patterns
    and excessive concentration on the seed. Although the landscape is shown
    for 100 shots, the axes may be interpreted as percentages for other shot
    counts because the construction scales with $N_{\mathrm{shots}}$.}
    \label{fig:QIPobjective}
\end{figure}

The preferred seed-multiplicity range is

\begin{equation}
n_{\mathrm{seed}}^{\min}
=
\left\lceil
0.1N_{\mathrm{shots}}
\right\rceil,
\qquad
n_{\mathrm{seed}}^{\max}
=
\left\lceil
0.7N_{\mathrm{shots}}
\right\rceil,
\end{equation}
with midpoint

\begin{equation}
n_{\mathrm{seed}}^{\mathrm{ave}}
=
\frac{
n_{\mathrm{seed}}^{\min}
+
n_{\mathrm{seed}}^{\max}
}{2}.
\end{equation}
The preferred unique-outcome range is

\begin{align}
n_{\mathrm{unique}}^{\min}
&=
1+
\left\lceil
\frac{
N_{\mathrm{shots}}
-
n_{\mathrm{seed}}^{\max}
}{3}
\right\rceil,
\\
n_{\mathrm{unique}}^{\max}
&=
1+
\left\lceil
\frac{
N_{\mathrm{shots}}
-
n_{\mathrm{seed}}^{\min}
}{1.5}
\right\rceil,
\end{align}
with midpoint

\begin{equation}
n_{\mathrm{unique}}^{\mathrm{ave}}
=
\frac{
n_{\mathrm{unique}}^{\min}
+
n_{\mathrm{unique}}^{\max}
}{2}.
\end{equation}
The factors 3 and 1.5 correspond to target average multiplicities among
nonseed states. Define

\begin{equation}
\Delta_{\mathrm{seed}}
=
\frac{
n_{\mathrm{seed}}
-
n_{\mathrm{seed}}^{\mathrm{ave}}
}{
N_{\mathrm{shots}}
},
\end{equation}
and

\begin{equation}
\Delta_{\mathrm{unique}}
=
\frac{
n_{\mathrm{unique}}
-
n_{\mathrm{unique}}^{\mathrm{ave}}
}{
N_{\mathrm{shots}}
}.
\end{equation}
The central pseudo-energy wells are

\begin{equation}
E_{\mathrm{DS}}^{(0)}
=
30\Delta_{\mathrm{seed}}^2
+
7000\Delta_{\mathrm{seed}}^4,
\end{equation}
and

\begin{equation}
E_{\mathrm{NU}}^{(0)}
=
30\Delta_{\mathrm{unique}}^2
+
7000\Delta_{\mathrm{unique}}^4.
\end{equation}
Additional penalties suppress severely underlocalized records,

\begin{equation}
P_{\mathrm{DS}}
=
\begin{cases}
20(n_{\mathrm{seed}}-6)^2,
&n_{\mathrm{seed}}\leq6,
\\
0,
&n_{\mathrm{seed}}>6,
\end{cases}
\end{equation}
and

\begin{equation}
P_{\mathrm{NU}}
=
\begin{cases}
12(n_{\mathrm{unique}}-11)^2,
&n_{\mathrm{unique}}\leq11,
\\
0,
&n_{\mathrm{unique}}>11.
\end{cases}
\end{equation}
The unique-outcome contribution is further reweighted by

\begin{equation}
B_{\mathrm{NU}}
=
\max\left(
1,
-10\Delta_{\mathrm{unique}}
\right).
\end{equation}
The final terms are

\begin{align}
E_{\mathrm{DS}}
&=
E_{\mathrm{DS}}^{(0)}
+
P_{\mathrm{DS}},
\\
E_{\mathrm{NU}}
&=
B_{\mathrm{NU}}E_{\mathrm{NU}}^{(0)}
+
P_{\mathrm{NU}}.
\label{eq:pseudo_energy_terms}
\end{align}
The numerical coefficients define an algorithmic scale rather than a physical
energy. They were designed relative to the fixed Monte Carlo temperature
$kT=20$.

\FloatBarrier

\subsection*{Metropolis feedback and qualification-rate control}

The latent deviation parameters are updated using a Metropolis acceptance
rule. Let $E_{\mathrm{QIP}}^{\mathrm{last}}$ denote the pseudo-energy of the
most recently accepted parameter set. It is initialized to a sufficiently
large value so that the first trial is accepted. A trial is always accepted when
$ E_{\mathrm{QIP}}^{\mathrm{trial}} < E_{\mathrm{QIP}}^{\mathrm{last}}$, 
otherwise, it is accepted with probability

\begin{equation}
P_{\mathrm{acc}}
=
\exp\left[
-\frac{
E_{\mathrm{QIP}}^{\mathrm{trial}}
-
E_{\mathrm{QIP}}^{\mathrm{last}}
}{kT}
\right].
\label{eq:QIP_MC_acceptance}
\end{equation}
Although a simulated annealing protocol was implemented, tested and found useful, the 
production calculations use a fixed value $kT=20$.

With the Monte Carlo (MC) simulation in place, it is worth noting that the 
accepted QIP score need not decrease monotonically. Uphill moves preserve 
diversity among localized patterns and the MC method generally prevents any
possible collapse onto one angle-deviation set. When a trial is accepted, 
its six latent angles become the retained state of the Metropolis process.

In addition to the continuous pseudo-energy, a circuit is classified as
qualified when

\begin{equation}
n_{\mathrm{seed}}
\geq
n_{\mathrm{seed}}^{\min} \quad \mbox{and}  \quad 
n_{\mathrm{unique}}^{\min}
\leq
n_{\mathrm{unique}}
\leq
n_{\mathrm{unique}}^{\max}.
\label{eq:QIP_qualification}
\end{equation}
The exponentially smoothed qualification rate is

\begin{equation}
f_{\mathrm{QIP}}(t)
=
0.95f_{\mathrm{QIP}}(t-1)
+
0.05I_{\mathrm{QIP}}(t),
\label{eq:qualification_rate}
\end{equation}
where $I_{\mathrm{QIP}}=1$ for a qualified pattern and zero otherwise. The
target rate is set to $f_{\mathrm{QIP}}^{\mathrm{target}} = 0.8$.
The latent proposal step scale is regulated according to

\begin{equation}
S_{t+1}
=
S_t
+
\left(
f_{\mathrm{QIP}}
-
f_{\mathrm{QIP}}^{\mathrm{target}}
\right)
5^\circ
\exp(0.2Z),
\qquad
Z\sim\mathcal N(0,1),
\label{eq:step_scale_feedback}
\end{equation}
with $0.08\leq S_t\leq2$.
If too many patterns qualify, the trial steps become larger and increase
diversity. If too few qualify, the steps contract.

When the qualification rate falls more than 0.2 below its target, an
additional correction acts on the overall deviation scale. If too many
distinct states are measured, the retained value of $a_t$ is contracted. 
If too few are measured, it is shifted partially toward $\pi/2$, thereby
increasing $t=\sin^2a_t$. The phase-partition variable is also weakly 
regulated. If $w<0.2$, the retained latent angle $a_w$ is shifted toward 
$\pi/4$, preventing one of the two phase-separator layers from remaining 
nearly inactive for an extended period.

Tests in which the angle deviations were held fixed revealed a strong
persistence of localization across different seed states. Deviation sets that
produced localized interference patterns for one seed generally did so for
others, whereas poorly localized sets tended to remain ineffective. This
suggests that, for a given problem instance, families of seed-conditioned
circuits may share recurring connected structures of high-probability peaks,
resembling attractor-like organization in parameter space.

A systematic characterization of this structure was beyond the scope of the
present work. Instead, QIPS uses a pragmatic diversification strategy in which
the deviation parameters are treated as an approximately stationary stochastic
process and are continually perturbed, including through cost-operator jitter,
to avoid repeatedly sampling the same sparse set of peaks.

\subsection*{Elite frontier and exponential seed selection}

All measured states are merged with the previously retained states, duplicate
bitstrings are removed, and the resulting list is sorted by operational
energy. The elite frontier is then truncated to

\begin{equation}
N_F
=
D\max\left(
N_{\mathrm{seed}},
N_{\mathrm{target}}
\right),
\label{eq:frontier_size}
\end{equation}
where $D=5$, $N_{\mathrm{seed}}=20$, and $N_{\mathrm{target}}=10$. Therefore, 
$N_F=100$. This strict state-count cutoff is important when large degenerate 
manifolds are present. The small static perturbation in 
Eq.~\eqref{eq:static_degeneracy_breaking} defines the ordering used to choose
which 100 states remain active.

Seeds are selected from the frontier using an exponentially decaying rank
bias. Let the frontier states be ordered from lowest to highest operational
energy and indexed by $j=1,\ldots,N_F$. The geometric parameter is chosen so
that the nominal probability of drawing beyond the active frontier is

\begin{equation}
q_{\mathrm{tail}}=0.05.
\end{equation}
Thus,

\begin{equation}
p
=
1-q_{\mathrm{tail}}^{1/N_F},
\end{equation}
and a candidate index is generated as

\begin{equation}
J
=
\left\lfloor
\frac{
\ln(1-U)
}{
\ln(1-p)
}
\right\rfloor
+1,
\qquad
U\sim\mathrm{Uniform}(0,1).
\label{eq:seed_geometric}
\end{equation}
If $J>N_F$, the index is replaced by a uniform draw from
$\{1,\ldots,N_F\}$. The resulting distribution combines a low-rank
exponential preference with a 5\% exploration tail.

The 20 seeds selected for a round are drawn independently and therefore with
replacement. A highly weighted seed may be used more than once. The selected
seeds are subsequently processed from higher to lower energy, but this
ordering does not alter their selection probabilities.

\subsection*{Frontier contraction and resource-limited search}

Let $\mathcal F_t$ denote the active frontier after proposal step $t$, and let
$r(s)$ be the total ordering rank obtained after static degeneracy breaking.
The outer-frontier rank is

\begin{equation}
R_F(t)
=
\max_{s\in\mathcal F_t}r(s).
\end{equation}
A proposal step is classified as successful when

\begin{equation}
R_F(t)
<
R_F(t-1).
\label{eq:frontier_contraction}
\end{equation}
Thus, success is defined by contraction of the complete elite frontier, not
only by discovery of a new absolute minimum. This criterion rewards progress
in filling and improving the entire low-energy set.
The contraction indicator is used to update progress statistics and the
relative success rates of jittered and unjittered proposals. Earlier versions
used a moving energy threshold, but frontier contraction was found to be more
robust across different benchmark families and is used for all reported
production data.

The benchmark calculations terminate after the fixed resource allocation in
Eq.~\eqref{eq:QIPS_total_budget}. The code also supports optional early
termination after repeated unsuccessful rounds or after ground-state
discovery, but these modes are not used for the matched fixed-budget
comparisons.

\subsection*{Matched classical proposal generator}

The primary classical baseline uses Explore Random Bit Flips, a generic
kick-and-repair proposal generator operating under the same finite proposal
budget. Starting from a selected seed, the method samples a Hamming-distance kick
$K$. The short-range component uses

\begin{equation}
K\in
\left\{
1,\ldots,\min(8,n_b)
\right\},
\end{equation}
with weights

\begin{equation}
P_{\mathrm{short}}(K)
\propto
\exp[-0.45(K-1)].
\end{equation}
The long-range component is approximately uniform over

\begin{equation}
K
\in
\left[
\left\lceil0.25n_b\right\rceil,
\left\lceil0.50n_b\right\rceil
\right].
\end{equation}
The mixture weights are

\begin{equation}
P(\mathrm{short})=0.85,
\qquad
P(\mathrm{long})=0.15.
\end{equation}
After flipping exactly $K$ bits, the resulting state is subjected to at most
four greedy one-bit repair steps. At each repair step, all one-bit neighbors
are examined, and the best improving move is applied. If the repaired state
improves upon the current anchor, it becomes the new anchor for subsequent
kicks.

The classical budget counts all attempted kicked states and repair neighbors,
including repeated states. Previously evaluated energies may be retrieved
from memory without recomputation, and selecting an already known anchor is
not charged as a new proposal. This convention gives the classical method a
small practical advantage while preserving matched proposal counts.

A blind-search baseline is obtained by drawing bitstrings uniformly from the
full Hilbert space. All classical and quantum proposal generators use the same
outer frontier, seed-selection logic, and stopping conditions.

\subsection*{Empirical classical proposal distribution}

For comparisons based on proposal-rank cumulative distributions, the classical
proposal generator is represented by its empirical finite-shot proposal
distribution. Each classical proposal batch contains
$N_{\mathrm{shots}}=100$ attempted bitstrings. The empirical probability
assigned to state $s$ is

\begin{equation}
\hat p_s^{(\mathrm{cl})}
=
\frac{n_s}{N_{\mathrm{shots}}},
\end{equation}
where $n_s$ is the number of times state $s$ appears in the batch. Thus, a
state proposed once has probability $0.01$, a state proposed twice has
probability $0.02$, and states not proposed in the batch have probability zero.
Repeated classical proposals are therefore retained as multiplicities rather
than collapsed to a binary visited/unvisited indicator.

This construction makes the classical proposal distribution an empirical
finite-shot distribution, directly analogous to the measurement record produced
by a quantum circuit. It also means that classical proposal distributions are
sparse by construction: only the proposed states need to be stored. Under this
representation, classical batches are localized in Hilbert space because their
support contains at most $N_{\mathrm{shots}}$ distinct states. The distinction
from QIPS is not whether the empirical distribution is sparse, but how the
support and multiplicities are generated. Classical proposals arise from the
kick-and-repair mechanism, whereas QIPS proposals arise from sampling a
seed-conditioned quantum interference pattern.

\subsection*{Hybrid modes and ablation controls}

The implementation supports several proposal modes:

\begin{enumerate}[itemsep=1pt, topsep=3pt, parsep=0pt, partopsep=0pt]
\item quantum proposals only;
\item classical kick-and-repair proposals only;
\item blind uniform proposals;
\item quantum proposals followed by classical proposals;
\item classical proposals followed by quantum proposals;
\item interleaved classical and quantum proposals.
\end{enumerate}

\vspace{0.30 cm}
\noindent The principal quantum ablations include:

\begin{enumerate}[itemsep=1pt, topsep=3pt, parsep=0pt, partopsep=0pt]
\item disabling the cost phase separator;
\item replacing the XYZ mixers by XZ mixers;
\item disabling the longitudinal random fields;
\item disabling dynamic cost-operator jitter;
\item shuffling the correspondence between computational-basis states and
energy values;
\item holding the seed fixed;
\item holding the angle deviations fixed;
\item disabling accelerated sparse probability sampling.
\end{enumerate}

\vspace{0.30 cm}
The energy-label shuffling control destroys the quadratic correspondence
between nearby computational-basis states and the cost spectrum while
preserving the same set of energies. Under this control, the structured QIPS
proposal mechanism degrades toward blind sampling, showing that its
effectiveness depends on the relation between circuit interference and QUBO
structure. The implementation also includes placeholders for hardware-noise
models, but no hardware noise was used in the reported simulations, and such
effects are beyond the scope of the present work.

\subsection*{Algorithm summary}

The complete QIPS procedure may be summarized as follows:

\vspace{0.30 cm}
\begin{enumerate}[itemsep=1pt, topsep=3pt, parsep=0pt, partopsep=0pt]
\item Construct the QUBO or Ising energy spectrum and add a negligible static
perturbation to define an ordering through exact degeneracies.

\item Initialize the elite frontier from randomly sampled states.

\item Select 20 seed states from the frontier using the exponential
rank-biased distribution in Eq.~\eqref{eq:seed_geometric}.

\item For each seed, construct its canonical two-layer circuit and sample a
trial set of latent deviation parameters around the most recently accepted
values.

\item Optionally perturb the cost phase by the adaptive jitter spectrum.

\item Evaluate the two-layer circuit and generate 100 measured bitstrings.

\item Merge the measured states into the elite list and truncate the frontier
to its 100 lowest operational-energy states.

\item Calculate $n_{\mathrm{seed}}$, $n_{\mathrm{unique}}$, and the
pseudo-energy in Eq.~\eqref{eq:QIP_pseudo_energy}.

\item Accept or reject the trial latent parameters using the Metropolis rule
in Eq.~\eqref{eq:QIP_MC_acceptance}.

\item Update the qualification-rate controller, jitter success rates, and
frontier-contraction statistics.

\item Repeat for all seeds and rounds until the fixed proposal budget is
exhausted.
\end{enumerate}

\vspace{0.25 cm}
\noindent The algorithm therefore separates three roles. The canonical angles encode
the selected seed, the randomized deviations generate a diverse ensemble of
localized interference patterns, and the classical frontier supplies the
downhill search bias by preferentially reusing low-energy states.

%%%%% \end{document}

\endgroup

\end{document}